\documentclass{aastex7}
\usepackage[T1]{fontenc}
\usepackage{amsmath}	% Advanced maths commands
\usepackage{amssymb}	% Extra maths symbols
\usepackage{bm}		% Bold maths symbols, including upright Greek
\usepackage{threeparttable}

\begin{document}

%\title{Three-dimensional radiation and relativistic magnetohydrodynamics of magnetically arrested disk accretion flows around active galactic nuclei}
%\title{Radiation and Relativistic Magnetohydrodynamics of Magnetically Arrested Disks in AGN: Insights from 3D Global Simulations}
\title{Three-dimensional Global Relativistic Radiation Magnetohydrodynamics of Magnetically Arrested Disk Accretion Flows in AGNs}

%%%%%%%%%%%%%%%%%%%%%%%%%%%%%%%% Authors %%%%%%%%%%%%%%%%%%%%%%%%%%%%%%%%%%%%%%%%%%%%%%%%%%%%%%%%%%

\author[0000-0002-3672-6271]{Ramiz Aktar}
\affiliation{Department of Physics and Institute of Astronomy, National Tsing Hua University, 30013 Hsinchu, Taiwan}
\affiliation{Institute of Astronomy and Astrophysics, Academia Sinica, Taipei 10617, Taiwan, R.O.C.}
\email[show]{ramizaktar@gmail.com}  

\author[0000-0002-1473-9880]{Kuo-Chuan Pan}
\affiliation{Department of Physics and Institute of Astronomy, National Tsing Hua University, 30013 Hsinchu, Taiwan}
\affiliation{Center for Theory and Computation, National Tsing Hua University, Hsinchu 30013, Taiwan}
\affiliation{Physics Division, National Center for Theoretical Sciences, National Taiwan University, Taipei 10617, Taiwan}
\email{kuochuan.pan@gapp.nthu.edu.tw}

\author{Toru Okuda}
\affiliation{Hakodate Campus, Hokkaido University of Education, Hachiman-Cho 1-2, Hakodate
040-8567, Japan}
\email{bbnbh669@ybb.ne.jp}

%%%%%%%%%%%%%%%%%%%%%%%%%%%%%%%%%%%%%%%%%%%%%%%%%%%%%%%%%%%%%%%%%%%%%%%%%%%%%%%%%%%%%%%%%%%%%%%%%%


%%%%%%%%%%%%%%%%%%%%%%%%%%%%%%%% Abstract %%%%%%%%%%%%%%%%%%%%%%%%%%%%%%%%%%%%%%%%%%%%%%%%%%%%%%%%%

\begin{abstract}

We perform three-dimensional radiation-relativistic magnetohydrodynamic (3D Rad-RMHD) simulations of accretion flows around spinning active galactic nuclei (AGNs). Our study focuses on the magnetically arrested disk (MAD) state, adopting a single-temperature model that includes bremsstrahlung opacity as the sole radiation process while varying the black hole spin from non-spinning to rapidly spinning cases. We find that the MAD state persists across all spin values, as demonstrated by the normalized magnetic flux at the horizon and the physically motivated spatially averaged plasma beta. The overall flow dynamics remain qualitatively similar for all spin models in 3D flow, suggesting that black hole spin has minimal influence on the accretion dynamics.
In addition, we conduct post-processing using a two-temperature model to calculate the luminosities from synchrotron and bremsstrahlung radiation. We find that the total radiation luminosity is significantly higher than the luminosities from synchrotron and bremsstrahlung. This finding highlights the influence of radiation on the dynamics of the accretion flow. Our analysis shows that the electron temperature is significantly high in the jet region, regardless of spin. We further find that the temporal evolution of both radiative and synchrotron luminosities exhibits qualitatively similar behavior across all spin values. Finally, our results indicate that black hole spin has minimal impact on the spectral energy distribution (SED) in MAD state accretion flows.

\end{abstract}

%%%%%%%%%%%%%%%%%%%%%%%%%%%%%%%%%%%%%%%%%%%%%%%%%%%%%%%%%%%%%%%%%%%%%%%%%%%%%%%%%%%%%%%%%%%%%%%%%%


%\keywords{\uat{Galaxies}{573} --- \uat{Cosmology}{343} --- \uat{High Energy astrophysics}{739} --- \uat{Interstellar medium}{847} --- \uat{Stellar astronomy}{1583} --- \uat{Solar physics}{1476}}

%%%%%%%%%%%%%%%%%%%%%%%%%%%%%%%%  Introduction %%%%%%%%%%%%%%%%%%%%%%%%%%%%%%%%%%%%%%%%%%%%%%%%%%%%

\section{Introduction} \label{sec:intro}

Active Galactic Nucleus (AGN) activity is often linked to powerful, collimated jets. In the case of M87, highly relativistic jets have been directly observed \citep{Junor-etal-99, Ly-etal-07, Hada-etal-16, Cui-etal-23, Goddi-etal-25}. Recently, the Event Horizon Telescope (EHT) Collaboration published a polarized image of M87, which demonstrates that the structured magnetic field near the event horizon is responsible for synchrotron emission \citep{Event-Horizon-etal-2021}. In contrast, Sgr A* does not show a persistent jet signature; instead, it exhibits episodic X-ray and near-infrared flares, detected approximately on a daily basis \citep{Baganoff-etal-2001, Genzel-etal-2003, Meyer-etal-2008, Degenaar-etal-2013, Neilsen-etal-2013, Neilsen-etal-2015, Ponti-etal-2015, Fazio-etal-2018, Boyce-etal-2019}. In this context, \citet{GRAVITY-Collaboration-18} reported the detection of near-infrared flares originating within about \(10r_g\) of the event horizon, where \(r_g\) represents the Schwarzschild radius. These flares were observed to move at approximately 0.3 times the speed of light and exhibited a continuous rotation of the polarization angle, which is consistent with the orbital motion of hot plasma in a strong poloidal magnetic field. This observational evidence confirms that the magnetic field plays a crucial role in producing jets and flares in AGNs.

\citet{Narayan-etal-03} first identified a significant magnetic state of accretion flow known as the ``magnetically arrested disk'' (MAD). The fundamental physics of the MAD state is that in a highly magnetized accretion flow, magnetic flux accumulates near the black hole's event horizon. This accumulation of magnetic flux causes the magnetic pressure to balance the ram pressure within the disk, which in turn hinders further mass accretion \citep{Narayan-etal-03, Igumenshchev-08, Tchekhovskoy-etal11}. Over the years, several independent simulation studies have investigated the MAD state in accretion flows around black holes \citep{McKinney-etal-12, Narayan-etal12, Dihingia-etal21, Mizuno-etal-21, Chatterjee-Narayan-22, Fromm-etal-22, Janiuk-James-22, Dhang-etal23, Jiang-etal-23, Aktar-etal24a, Aktar-etal24b, Zhang-etal-24, Aktar-etal25}. The EHT Collaboration has recently achieved a major breakthrough by imaging the supermassive black holes in M87 and the Galactic Center, Sgr A* \citep{Event-Horizon-etal-2019, Event-Horizon-etal-2022a}. Horizon-scale observations, when compared with synthetic radio images from GRMHD simulations, provide compelling evidence that strongly ordered magnetic fields are a hallmark of accretion in the MAD regime \citep{Event-Horizon-etal-2021, Event-Horizon-etal-2024}. Independent evidence for such a state has also been provided by the GRAVITY Collaboration, whose high-angular-resolution near-infrared observations of flaring events in Sgr A* reveal signatures consistent with plasma dynamics governed by a MAD configuration \citep{GRAVITY-Collaboration-18, GRAVITY-Collaboration-20}. Moreover, the MAD state is of particular importance because it enables efficient jet launching and facilitates the extraction of black hole spin energy observed in AGNs.

The radiation mechanism plays a crucial role in the behavior of accretion flows as mass accretion rates approach near Eddington levels (\(\gtrsim 10^{-2} \dot{M}_{\rm Edd}\)) or exceed Eddington rates (i.e., $> \dot{M}_{\rm Edd}$) \citep{Elvis-etal-94, Ohsunga-etal-05, Davis-Laor-11, Done-etal-12, Sadowski-etal-14, McKinney-etal-14, Netzer-15, Sadowski-Narayan-16}. In contrast, the radiation process can generally be disregarded in flows with low accretion rates, specifically in sub-Eddington conditions (i.e., $<< 10^{-2} \dot{M}_{\rm Edd}$). This is particularly relevant for low-luminosity active galactic nuclei (LLAGNs) and radiatively inefficient accretion flows (RIAFs) \citep{Narayan-etal95, Yuan-Narayan14, White-etal-20, Dhang-etal23, Aktar-etal24a}. Moreover, the implementation of radiation mechanism in the simulation model around black holes considering both optically thin and thick in the flow is quite challenging. In this regards, \citet{Ohsunga-etal-05} initiated to implement radiation mechanism in accretion flow in a global simulation models around black holes. Following that, there are several independent simulation models have been developed to incorporate radiation process around black holes \citep{Ohsuga-etal-09, Ohsuga-Mineshige11, Sadowski-etal-13, Sadowski-etal-14, McKinney-etal-14, Takahashi-etal-16, Okuda-etal22, Okuda-etal23, Aktar-etal24b}. Simulations of radiation-dominated accretion flows have been conducted, particularly focusing on the highly magnetized MAD state \citep{Sadowski-etal-13, Sadowski-etal-15, Morales-Teixeira-etal-18, Yao-etal-21, Curd-Narayan23, Aktar-etal24b}. Recently, \citet{Aktar-etal24b} performed a comparative study between MAD and ``standard and normal evolution (SANE)'' based on their two-dimensional (2D), global, radiation, relativistic, magnetohydrodynamics (Rad-RMHD) model around spinning AGN. They observed no characteristic difference of spectral energy distribution (SED) between MAD and SANE state. Further, \citet{Aktar-etal25} examined the comparison between two-dimensional (2D) and three-dimensional (3D) flow in MAD state considering resistive flow around black holes. They showed that dynamical parameters are quite different for 3D models compared 2D models when the flow enters into the MAD state. In this study, our primary focus is to assess how the spin of a black hole affects the 3D radiation magnetohydrodynamic flows, as well as to analyze the resulting luminosities and spectral energy distributions.

We organize the paper as follows. In Section \ref{simulation-scheme}, we present the description of the numerical model and governing equations. In  Section \ref{results}, we discuss the simulation results of our model in detail. Finally,  we draw the concluding remarks and summary in Section \ref{conclusion}.

%%%%%%%%%%%%%%%%%%%%%%%%%%%%%%%%%%%%%%%%%%%%%%%%%%%%%%%%%%%%%%%%%%%%%%%%%%%%%%%%%%%%%%%%%%%%%%%%%%%

\setcounter{footnote}{0}

%%%%%%%%%%%%%%%%%%%%%%%%%%%%%%%%%%%%%%%%%%%%%%%%%%
\section{Simulation Setup}
\label{simulation-scheme}
%%%%%%%%%%%%%%%%%%%%%%%%%%%%%%%%%%%%%%%%%%%%%%%%%%

We investigate three-dimensional (3D) simulations of radiation and relativistic magnetohydrodynamics (3D-Rad-RMHD) using the numerical simulation code PLUTO\footnote{\url{http://plutocode.ph.unito.it}} \citep{Migone-etal07, Fuksman-Mignone-19}. In our analysis, we adopt a unit system where \( G = M_{\rm BH} = c = 1 \). Here, \( G \) represents the gravitational constant, \( M_{\rm BH} \) is the mass of the black hole, and \( c \) is the speed of light. As a result, we measure distance, velocity, and time in terms of \( r_g = \frac{GM_{\rm BH}}{c^2} \), \( c \), and \( t_g = \frac{GM_{\rm BH}}{c^3} \), respectively. In this work, we apply the ideal special relativistic magnetohydrodynamics (RMHD) equations while neglecting explicit resistivity in the flow.

%%%%%%%%%%%%%%%%%%%%%%%%%%%%%%%%%%%%%%%%%%%%%%%%%%
\subsection{Governing Equations for 3D-Rad-RMHD}
%%%%%%%%%%%%%%%%%%%%%%%%%%%%%%%%%%%%%%%%%%%%%%%%%%

In this section, we present the governing equations of ideal RMHD for the interaction between matter and electromagnetic (EM) fields in cylindrical coordinates \((r, \phi, z)\). The governing equations are as follows \citep{Fuksman-Mignone-19, Aktar-etal24b}:

\begin{align}
& \label{governing_eq_1} \frac{\partial (\rho \Gamma)}{\partial t} + \nabla \cdot (\rho \Gamma \bm{v}) = 0, \\
& \frac{\partial \varepsilon}{\partial t} + \nabla \cdot (\bm{m} - \rho \Gamma \bm{v}) = G^0, \\
& \frac{\partial \bm{m}}{\partial t} + \nabla \cdot (\rho h \Gamma^2 \bm{v} \bm{v} - \bm{B} \bm{B} - \bm{E} \bm{E}) + \nabla P_t = \bm{G}, \\
& \frac{\partial \bm{B}}{\partial t} + \nabla \times \bm{E} = 0, \\
& \frac{\partial E_{\rm rad}}{\partial t} + \nabla \cdot \bm{F}_{\rm rad} = -G^0, \\
& \text{and} \\
& \label{governing_eq_7} \frac{\partial \bm{F}_{\rm rad}}{\partial t} + \nabla \cdot \bm{P}_{\rm rad} = -\bm{G},
\end{align}

where \(\rho\), \(\bm{v}\), \(\bm{m}\), \(h\), \(\Gamma\), and \(\bm{B}\) represent the mass density, fluid velocity, momentum density, specific enthalpy, Lorentz factor, and mean magnetic field, respectively. The electric field \(\bm{E}\) is given by \(\bm{E} = -\bm{v} \times \bm{B}\). Additionally, \(E_{\rm rad}\), \(\bm{F}_{\rm rad}\), and \(\bm{P}_{\rm rad}\) denote the radiation energy density, radiation flux, and pressure tensor as moments of the radiation field, respectively. In this context, the total pressure, momentum density, and energy density of matter and electromagnetic (EM) fields are expressed as follows:
\begin{align}
& P_t = P_{\rm gas} + \frac{E^2 + B^2}{2},\\
& \bm{m} = \rho h \Gamma^2 \bm{v} + \bm{E} \times \bm{B},\\
& \text{and}\\
& \varepsilon = \rho h \Gamma^2 - P_{\rm gas} - \rho \Gamma + \frac{E^2 + B^2}{2}.
\end{align}
Additionally, the radiation-matter interaction terms, denoted as $(G^0, \bm{G})$, are given by:
\begin{align}
(G^0, \bm{G})_{\rm comov} = \rho[\kappa (E_{\rm rad} - a_R T^4), (\kappa + \sigma) \bm{F}_{\rm rad}]_{\rm comov},
\end{align}
In this equation, all fields are measured in the fluid's comoving frame. Here, $\kappa$ and $\sigma$ represent the frequency-averaged absorption and scattering opacities, respectively. In this work, we adopt the free-free absorption opacity defined as: \(\kappa = \rho T^{-7/2}~{\rm cm}^2~{\rm g}^{-1} \) and the scattering opacity is taken as: \( \sigma = 0.4~{\rm cm}^2~{\rm g}^{-1} \). In these equations, $T$ denotes the temperature of the fluid, and $a_R$ is the radiation density constant. The radiation transfer equations are solved based on the gray approximation and by employing the M1 closure \citep{Fuksman-Mignone-19}.

%%%%%%%%%%%%%%%%%%%%%%%%%%%%%%%%%%%%%%%%%%%%%%%%%%%%%%%%%%%%%%%%%%%%%%%%%
\subsection{Expression for closure relation of EoS and radiation fields}
%%%%%%%%%%%%%%%%%%%%%%%%%%%%%%%%%%%%%%%%%%%%%%%%%%%%%%%%%%%%%%%%%%%%%%%%%

To close the system of equations (\ref{governing_eq_1}-\ref{governing_eq_7}), additional sets of relations are required. One of these is the equation of state (EoS), which provides a connection between thermodynamic quantities. In this work, we consider a constant-\(\gamma\) EoS, defined as follows:
\begin{align} 
\label{EoS_eq}
h = 1 + \frac{\gamma}{\gamma -1} \Theta,
\end{align}
where \(\Theta = \frac{P_{\rm gas}}{\rho}\) and \(\gamma\) is the adiabatic index. 

Furthermore, another closure relation is necessary for the radiation fields, which relates the pressure tensor \(P_{\rm rad}^{ij}\), the energy density \(E_{\rm rad}\), and the radiation flux \(\bm{F}_{\rm rad}\). The closure relation is given by \citep{Fuksman-Mignone-19, Aktar-etal24b}:
\begin{align}
& P_{\rm rad}^{ij} = D^{ij} E_{\rm rad}, \\
& D^{ij} = \frac{1-\xi}{2} \delta^{ij} + \frac{3 \xi -1}{2} n^i n^j,\\
& \xi = \frac{3 + 4 f^2}{5 + 2 \sqrt{4 - 3 f^2}},
\end{align}
where \(\bm{n} = \bm{F}_{\rm rad}/|\bm{F}_{\rm rad}|\) and \(f = |\bm{F}_{\rm rad}|/E_{\rm rad}\). Additionally, \(\delta^{ij}\) is the Kronecker delta.

%%%%%%%%%%%%%%%%%%%%%%%%%%%%%%%%%%%%%%%%%%%%%%%%%%%%%%
\subsection{Gravitational potential of the black hole}
%%%%%%%%%%%%%%%%%%%%%%%%%%%%%%%%%%%%%%%%%%%%%%%%%%%%%%
\label{Kerr_pot}

In this study, we model the gravitational effects around a spinning black hole using the effective Kerr potential proposed by \citet{Dihingia-etal18b}. The effective Kerr potential is expressed as follows \citep{Dihingia-etal18b, Aktar-etal24a, Aktar-etal24b, Aktar-etal25}:
\begin{align} \label{DDMC18_pot}
\Phi^{\rm eff}\left(r,z,a_k, \lambda\right)= \frac{1}{2} \ln \left(\frac{A(2R - \Sigma) r^2 - 4 a_k^2 r^4}{\Sigma \lambda \left(\Sigma \lambda R^2 + 4 a_k r^2 R - 2 \lambda R^3\right) - A \Sigma r^2 }\right),
\end{align}
where \( R = \sqrt{r^2 + z^2} \) is the spherical radial distance, \( \Delta = a_k^2 + R^2 - 2R \), \( \Sigma = \frac{a_k^2 z^2}{R^2} + R^2 \), and \( A = \left(a_k^2 + R^2\right)^2 - \frac{a_k^2 r^2 \Delta}{R^2} \). In this context, \( \lambda \) represents the specific flow angular momentum, and \( a_k \) denotes the dimensionless spin of the black hole, defined as \( a_k = \frac{J}{M_{\rm BH}} \), where \( J \) is the angular momentum of the black hole. 
To derive the Keplerian angular momentum, we evaluate the potential in the equatorial plane (\( z \rightarrow 0 \)) using the formula \( \lambda_K = \sqrt{r^3 \frac{\partial \Phi^{\rm eff}}{\partial r}\big|_{\lambda \rightarrow 0}} \). The angular frequency is then given by \( \Omega = \lambda / r^2 \). We incorporate the effective Kerr potential in the PLUTO code, following the methodology outlined in our previous works \citet{Aktar-etal24a, Aktar-etal24b, Aktar-etal25}.

It is important to note that \citet{Dihingia-etal18b} conducted various comparative studies between general relativity (GR) and the effective Kerr potential through analytical methods. They demonstrated that the transonic properties obtained using the effective Kerr potential are in excellent agreement with those derived from GR in the semi-relativistic regime around Kerr black holes. Consequently, our numerical approach effectively retains the key space-time characteristics around black holes by employing the effective Kerr potential. Furthermore, our simulation model is capable of achieving higher spatial resolutions compared to GRMHD simulations for a given computational resource \citep{Aktar-etal24a, Aktar-etal24b, Aktar-etal25}.

%%%%%%%%%%%%%%%%%%%%%%%%%%%%%%%%%%%%%%%%%%%%%%%%%%
\subsection{Initial equilibrium Torus and atmospheric condition}
\label{torus_set_up}
%%%%%%%%%%%%%%%%%%%%%%%%%%%%%%%%%%%%%%%%%%%%%%%%%%

The initial equilibrium torus is constructed using the Newtonian analog of relativistic tori as described by \cite{Abramowicz-etal78}. In this study, we adopt the same formalism to establish the initial equilibrium torus outlined by \citet{Aktar-etal24a, Aktar-etal24b, Aktar-etal25}. The density distribution of the torus is derived by considering a constant angular momentum flow \citep{Matsumoto-etal96, Hawley-00, Kuwabara-etal05, Aktar-etal24a, Aktar-etal24b}:
\begin{align}
\Phi^{\rm eff}\left(r,z, a_k, \lambda\right) + \frac{\gamma}{\gamma -1}\frac{P_{\rm gas}}{\rho}={\cal C},
\end{align}
where `$\mathcal{C}$' is the integration constant and is evaluated under the condition of zero gas pressure \((P_{\rm gas} \rightarrow 0)\) at the location \(r = r_{\rm min}\) in the equatorial plane. Here, \(r_{\rm min}\) denotes the inner edge location of the torus. Consequently, the density distribution can be expressed using the adiabatic equation of state, \(P_{\rm gas} = K \rho^{\gamma}\), as follows:
\begin{align}
\rho=\left[\frac{\gamma-1}{K\gamma}\left({\cal C}-\Phi^{\rm eff}\left(r,z, a_k, \lambda\right)\right)\right]^{\frac{1}{\gamma -1}},   
\end{align}
where \(K\) is determined based on the maximum density condition \((\rho_{\rm max})\) at the location \(r = r_{\rm max}\) in the equatorial plane. It is given by:
\begin{align}
K=\frac{\gamma - 1}{\gamma}\left[\mathcal{C}-\Phi^{\rm eff}\left(r_{\rm max},0, a_k, \lambda\right)\right]\frac{1}{\rho_{\rm max}^{\gamma-1}}.   
\end{align}
Here, the term \(\Phi^{\rm eff}\) refers to the effective Kerr potential as mentioned in Equation \ref{DDMC18_pot}.

In this work, we assume that the atmosphere evolves freely and that the initial density and pressure in the area surrounding the torus are very low \citep{Aktar-etal25}. The density and pressure distributions outside the torus are defined as \(\rho_{\rm atm} = \rho_{\rm floor} r^{-3/2}\) and \(P_{\rm atm} = P_{\rm floor} r^{-5/2}\), with the floor values set to be extremely small (see section \ref{floor-condition}). These atmospheric conditions are commonly considered in various GRMHD simulation codes \citep{Gammie-etal03, Porth-etal-16, White-etal-2016, Liska-etal-2018}.

%%%%%%%%%%%%%%%%%%%%%%%%%%%%%%%%%%%%%%%%%%%%%%%%%%%%%%%%%%%%%%%%%%%%%%%%%
%                        Table 1
%%%%%%%%%%%%%%%%%%%%%%%%%%%%%%%%%%%%%%%%%%%%%%%%%%%%%%%%%%%%%%%%%%%%%%%%%
\begin{table}
%\begin{minipage}{180mm}
%\ContinuedFloat
\caption{Conversion of code and C.G.S units used in this paper 
\label{Table-1}}
 \begin{tabular}{@{}c c c   } 
 \hline
 Units &  code units   &   C.G.S values   \\ 
 \hline  
 Density  &   $\rho_0 = 1 \times 10^{-10}$ &  $\rho_0 = 1 \times 10^{-10}$ g cm$^{-3}$  \\
 Length   &  $ r_g = GM_{\rm BH}/c^2$      &  $r_g = 1.485 \times 10^{13}$ cm   \\
 Velocity &  $c$ &   $c = 2.998 \times 10^{10}$ cm s$^{-1}$ \\
 Time     &  $t_g = GM_{\rm BH}/c^3$  &  $t_g = 4.953 \times 10^{2}$ s \\
 Magnetic field &   $B = \frac{B_0}{c \sqrt{4 \pi \rho_0}}$  & $B_0 = 1.063 \times 10^6$  G \\
 Mass     &   $M_{\rm BH} = 10^8 M_{\odot}^*$ &    \\
 Eddington luminosity  & $L_{\rm Edd} = 4 \pi GcM_{\rm BH}/\kappa_{\rm es}^{**}$ & $1.25 \times 10^{46}$ erg $s^{-1}$  \\
 Eddington accretion rate & $\dot{M}_{\rm Edd} = L_{\rm Edd}/c^2$   &  $1.40 \times 10^{25}$ g $s^{-1}$     \\
 \hline
$*$ $M_{\odot}$ is the mass of the sun. \\
$**\kappa_{\rm es} = 0.4$ cm$^2$ g$^{-1}$ is the electron scattering opacity.  \\
%$**$ $r_g = \frac{GM_{\rm BH}}{c^2}$% = Schwarzschild radius.

 \end{tabular}
 %\end{minipage}
\end{table}

%====================================================================

%%%%%%%%%%%%%%%%%%%%%%%%%%%%%%%%%%%%%%%%%%%%%%%%%%%%%%%%%%%%%%%%%%%%%%%%%
%                        Table 2
%%%%%%%%%%%%%%%%%%%%%%%%%%%%%%%%%%%%%%%%%%%%%%%%%%%%%%%%%%%%%%%%%%%%%%%%%

\begin{table*}[htbp]
\centering
\begin{threeparttable}
\caption{Simulation models.}
\label{Table-2}

\begin{tabular}{@{}c c c c c c c c c c @{}}
\hline
Model\tnote{$a$} & $r_{\rm min}$\tnote{b} & $r_{\rm max}$\tnote{c} & $\rho_{\rm max}$\tnote{d} & $\beta_0$\tnote{e}  & $\lambda$\tnote{f} & $E_{\rm rad}$\tnote{g} & $a_k$\tnote{h} & $(n_r \times n_\phi \times n_z)$\tnote{i}  \\
      & ($r_g$) & ($r_g$) & (g cm$^{-3}$)  &  & ($r_gc$) & (erg cm$^{-3}$) & & & \\
\hline
$a0p00$  & 20 & 40 & $10^{-10}$ & 10 & 6.25 & $5 \times 10^{-10}$ & 0.0 & $(920 \times 92 \times 920)$  \\
$a0p20$  & '' & '' & ''         & '' & ''   & ''                  & 0.20 & ''  \\
$a0p50$  & '' & '' & ''         & '' & ''   & ''                  & 0.50 & ''  \\
$a0p80$  & '' & '' & ''         & '' & ''   & ''                  & 0.80 & ''  \\
$a0p98$  & '' & '' & ''         & '' & ''   & ''                  & 0.98 & ''  \\ 

\hline
\end{tabular}

\begin{tablenotes}
\footnotesize
\item[] 
$(a):$ Model names. 
$(b):$ inner edge of the initial equilibrium torus.
$(c):$ location of the maximum pressure of the initial equilibrium torus.
$(d):$ maximum density at $r_{\rm max}$.
$(e):$ initial plasma beta parameter.
$(f):$ specific flow angular momentum. 
$(g):$ initial radiation energy density.
$(h):$ dimensionless spin parameter of the black hole.
$(i):$ resolution of the simulation.
\end{tablenotes}

\end{threeparttable}
    
\end{table*}
%%%%%%%%%%%%%%%%%%%%%%%%%%%%%%%%%%%%%%%%%%%%%%%%%%%%%%%%%%%%%%%%%%%%%%%%%%

%%%%%%%%%%%%%%%%%%%%%%%%%%%%%%%%%%%%%%%%%%%%%%%%%%
\subsection{Configuration of initial magnetic field in the torus}
\label{Mag_config}
%%%%%%%%%%%%%%%%%%%%%%%%%%%%%%%%%%%%%%%%%%%%%%%%%%

In this work, we initialize a poloidal magnetic field by introducing a purely toroidal component of the vector potential \citep{Hawley-etal02}. The expression for the toroidal component of the vector potential is given by \citep{Hawley-etal02, Aktar-etal24a, Aktar-etal24b, Aktar-etal25}:
\begin{align}
A_\phi = B_0 [\rho(r,\phi,z) - \rho_{\rm min}],
\end{align}
where \(B_0\) represents the normalized initial magnetic field strength and \(\rho_{\rm min}\) is the minimum density within the torus. The initial magnetic field strength is characterized by the ratio of gas pressure to magnetic pressure, known as the plasma beta parameter, which is defined as \(\beta_0 = \frac{2 P_{\rm gas}}{B_0^2}\). Additionally, we define the magnetization parameter as \(\sigma_{\rm M} = \frac{B^2}{\rho}\), which represents the ratio of magnetic energy to rest mass energy in the flow \citep{Dihingia-etal21, Dihingia-etal22, Dhang-etal23, Curd-Narayan23, Aktar-etal24a, Aktar-etal24b, Aktar-etal25}. This magnetization parameter provides a measure of how strongly the accreting plasma is magnetized, effectively comparing the magnetic energy content of the flow to its rest-mass energy. In this study, we set the initial plasma beta parameter to $\beta_0 = 10$ for all cases.

%%%%%%%%%%%%%%%%%%%%%%%%%%%%%%%%%%%%%%%%%%%%%%%%%%
\subsection{Initial and boundary conditions}
\label{boundary_cond}
%%%%%%%%%%%%%%%%%%%%%%%%%%%%%%%%%%%%%%%%%%%%%%%%%%

To simulate radiative relativistic magnetohydrodynamic (Rad-RMHD) accretion flows, we employ the HLL Riemann solver along with second-order linear interpolation for spatial discretization. Additionally, we implement the second-order Runge-Kutta method for time integration. To ensure the divergence-free condition, given by \(\nabla \cdot \bm{B} = 0\), we utilize the hyperbolic divergence cleaning method \citep{Migone-etal07}. In our simulation, the computational domain extends radially from $1.5 r_g \lesssim r \lesssim 200 r_g$, vertically from $-100 r_g \lesssim z \lesssim 100 r_g$, and azimuthally from $0 \lesssim \phi \lesssim 2 \pi$. We use a spherical radius \(R_{\rm in} = \sqrt{r^2 + z^2}\) to represent the spherical region of a black hole in our model. The inner boundary conditions are set as absorbing conditions around the black hole \citep{Okuda-etal19, Okuda-etal22, Okuda-etal23, Aktar-etal24a, Aktar-etal24b, Aktar-etal25}, with the inner boundary located at \(R_{\rm in} = 2.5 r_g\), which serves as the event horizon in our model. At the inner boundary, where \(R \leq R_{\rm in}\), the radial velocity is assigned the free-fall value, while the remaining variables are interpolated from the outer region near the boundary. All other boundaries are configured with outflow boundary conditions \citep{Aktar-etal24a, Aktar-etal24b, Aktar-etal25}. For our simulations, we consider a resolution for the 3D models of \((n_r \times n_\phi \times n_z) = (920 \times 92 \times 920)\). We maintain uniform grid sizes in the \(r\), \(\phi\), and \(z\) directions for our studies. In this paper, we select a high resolution in the \(r\) and \(z\) directions while opting for a comparatively lower resolution in the \(\phi\) direction. This approach allows us to generate results within an optimized time frame using the Rad-RMHD module in 3D. However, the comparatively lower resolution in the \(\phi\) direction is unlikely to impact the overall flow dynamics.

In this study, we establish the initial equilibrium torus with the inner edge positioned at \(r_{\rm min} = 20 r_g\) and the maximum pressure located at \(r_{\rm max} = 40 r_g\). It is generally agreed that these dimensions are appropriate for the initial equilibrium torus in a MAD state \citep{Wong-etal-21, Fromm-etal-22}. We set the minimum density of the torus to \(\rho_{\rm min} = 0.1 \rho_{\rm max}\), where \(\rho_{\rm max}\) represents the maximum density of the torus. For our simulation, we designate \(\rho_{\rm max} = \rho_0\), with \(\rho_0\) being the reference density. We fix the reference density at \(\rho_0 = 10^{-10} \, \text{g} \, \text{cm}^{-3}\), and the constant adiabatic index is set to \(\gamma = \frac{4}{3}\). Additionally, we consider the radiation energy density, which is set at \(E_{\rm rad} = 5 \times 10^{-10} \, \text{erg} \, \text{cm}^{-3}\). For our representative case, we choose the mass of the black hole to be \(M_{\rm BH} = 10^8 M_{\odot}\). In our simulation model, we convert the code units to C.G.S. units as shown in Table \ref{Table-1}.

%%%%%%%%%%%%%%%%%%%%%%%%%%%%%%%%%%%%%%%%%%%%%%%%%%
\subsection{Floor and ceiling conditions}
%%%%%%%%%%%%%%%%%%%%%%%%%%%%%%%%%%%%%%%%%%%%%%%%%%
\label{floor-condition}

To avoid encountering negative density and pressure in supersonic and highly magnetized flows, particularly near the event horizon, we consider very low minimum values for both density and pressure. We set the floor density at \(\rho_{\text{floor}} = 10^{-6} \rho_0\) and the floor pressure at \(P_{\text{floor}} = 10^{-8} P_0\), where $\rho_0$ and $P_0$ are the reference density and pressure, respectively. Moreover, we impose a ceiling condition on magnetization parameter, limiting it to \(\sigma_{\text{M}} < 100\) in our simulation model, which is a typical consideration in GRMHD simulations \citep{Tchekhovskoy-etal11, Porth-etal-16, Dihingia-etal21, Narayan-etal-22, Chatterjee-Narayan-22, Dhang-etal23, Jiang-etal-23, Dhruv-etal-2025}.

%%%%%%%%%%%%%%%%%%%%%%%%%%%%%%%%%%%%%%%%%%%%%%%%%%
\subsection{Magnetic state of the flow}
\label{magnetic_state}
%%%%%%%%%%%%%%%%%%%%%%%%%%%%%%%%%%%%%%%%%%%%%%%%%%

To examine the magnetic state, we calculate the magnetic flux accumulated at the horizon. We define normalized magnetic flux threading the black hole horizon, denoted as \(\dot{\phi}_{\rm acc}\), which is commonly referred to as the MAD parameter. The MAD parameter is defined in Gaussian units for a 3D model in cylindrical coordinates \citep{Tchekhovskoy-etal11, Narayan-etal12, Dihingia-etal21, Dhang-etal23, Aktar-etal24a, Aktar-etal24b, Aktar-etal25} as follows:
\begin{align}
\dot{\phi}_{\rm acc} = \frac{\sqrt{4 \pi}}{2 \sqrt{\dot{M}_{\rm acc}}} \int^{2 \pi}_{0} \int^{z_{\rm in}}_{-z_{\rm in}} |B_r|_{R = R_{\rm in}} \, r\, dz \, d\phi,
\end{align}
where \(R_{\rm in} = \sqrt{r^2 + z^2_{\rm in}}\) represents the inner boundary or the horizon for our model, and \(B_r\) is the radial component of the magnetic field. In this study, we define the normalized magnetic flux in Gaussian units by including the factor of \(\sqrt{4 \pi}\) \citep{Narayan-etal-22, Aktar-etal25}. The corresponding mass accretion rate, \(\dot{M}_{\rm acc}\), is given by:
\begin{align}
\dot{M}_{\rm acc} = - \int^{2 \pi}_{0} \int^{z_{\rm in}}_{-z_{\rm in}} \rho (r,z, \phi) \, r \, v_r \, dz \, d\phi,
\end{align}
where the negative sign indicates the inward direction of the accretion flow. We also calculate the density-weighted spatial average of the plasma beta parameter \(\beta_{\rm ave} = \frac{\int \rho \beta dV}{\int \rho dV}\) across the computational domain to identify the magnetic state of the flow \citep{Aktar-etal25}.

The accretion state transitions to the MAD state when the normalized magnetic flux \(\dot{\phi}_{\rm acc}\) reaches a critical value. It is generally accepted that the MAD state is achieved when this critical value satisfies \(\dot{\phi}_{\rm acc} \gtrsim 50\) in Gaussian units \citep{Tchekhovskoy-etal11, Narayan-etal12, Dihingia-etal21, Jiang-etal-23, Aktar-etal24b}. However, this fiduciary value is somewhat arbitrary and does not have a solid physical foundation. Recently, \citet{Aktar-etal25} proposed a physically motivated method to examine the magnetic state by analyzing the spatial average of the plasma beta parameter. They observed that the flow enters the MAD state when \(\beta_{\rm ave} \lesssim 1\), meaning that the magnetic pressure is comparable to or greater than the gas pressure in the flow.

%%%%%%%%%%%%%%%%%%%%%%%%%%%%%%%%%%%%%%%%%%%%%%%%%%%
%%                        Figure 1
%%%%%%%%%%%%%%%%%%%%%%%%%%%%%%%%%%%%%%%%%%%%%%%%%%%
\begin{figure*}
	\begin{center}
        \includegraphics[width=0.40\textwidth]{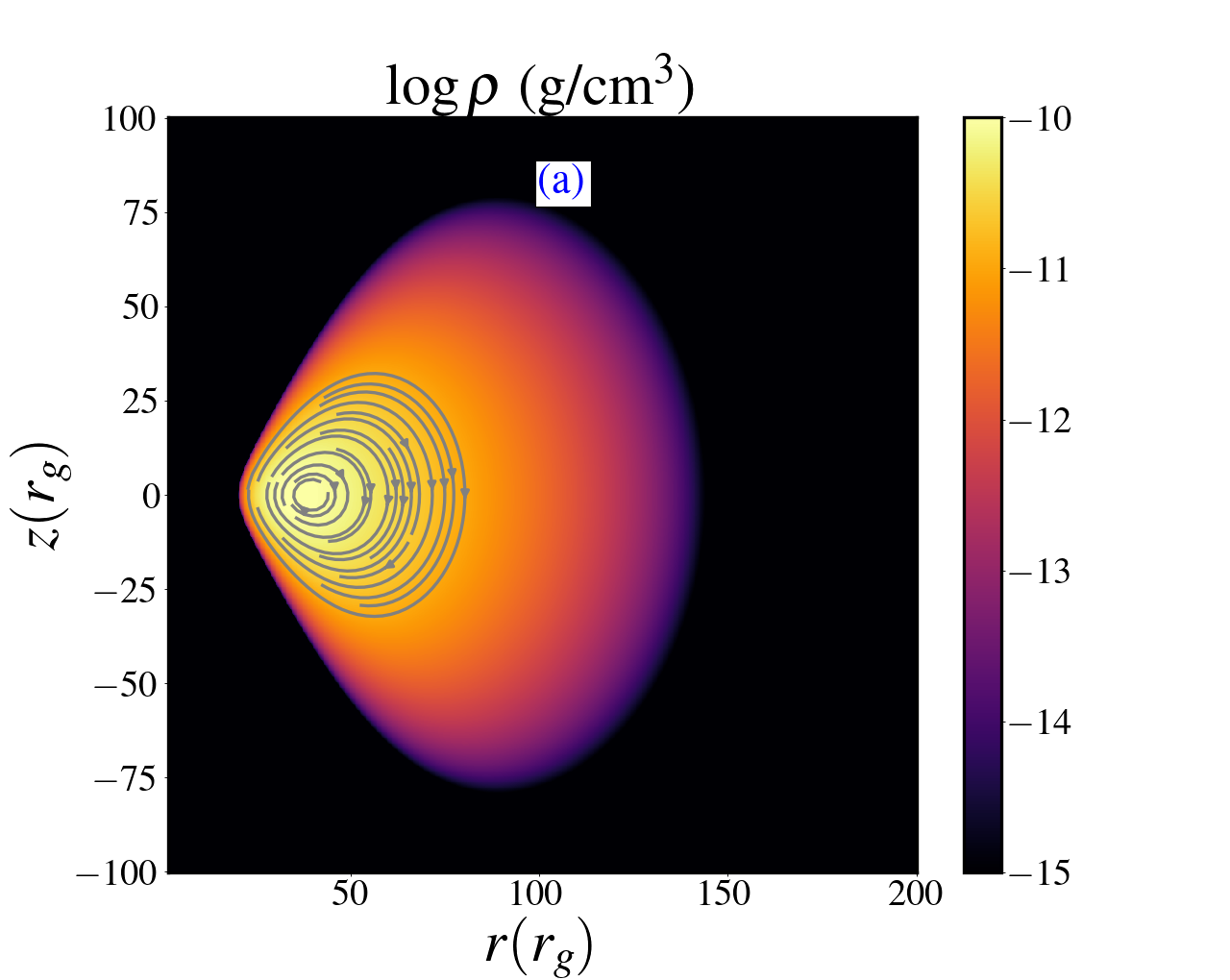} 
        \includegraphics[width=0.35\textwidth]{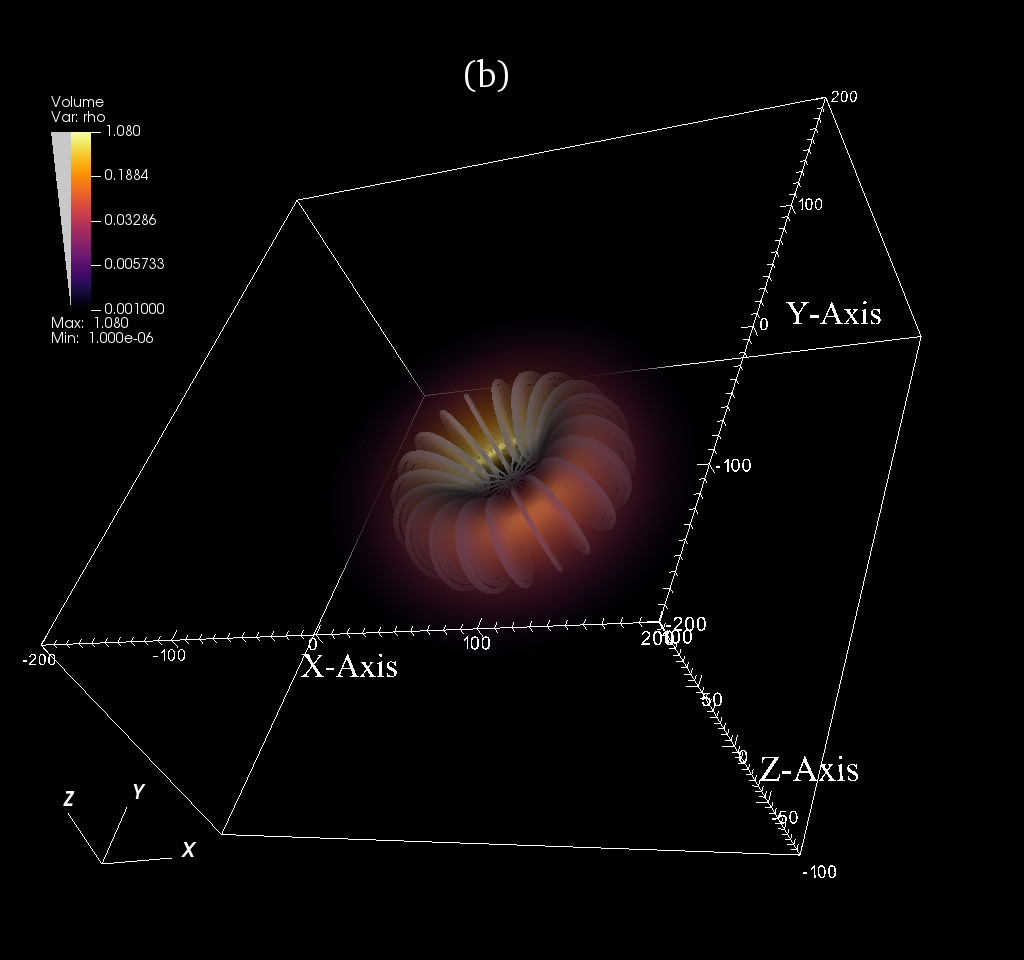} 
	\end{center}
	\caption{The distribution of density of the initial equilibrium torus $(a)$ for 2D slice in ($r-z$) plane for 3D model, and $(b)$ volume rendering plot at $t = 0~t_g$. The gray lines illustrate the magnetic field lines. Here, we fix spin of the black hole as $a_k = 0.98$.}
	\label{Figure_1}
\end{figure*}
%%%%%%%%%%%%%%%%%%%%%%%%%%%%%%%%%%%%%%%%%%%%%%%%%%%%

%%%%%%%%%%%%%%%%%%%%%%%%%%%%%%%%%%%%%%%%%%%%%%%%%%%
%%                        Figure 2
%%%%%%%%%%%%%%%%%%%%%%%%%%%%%%%%%%%%%%%%%%%%%%%%%%%
\begin{figure}
	\begin{center}
        \includegraphics[width=0.49\textwidth]{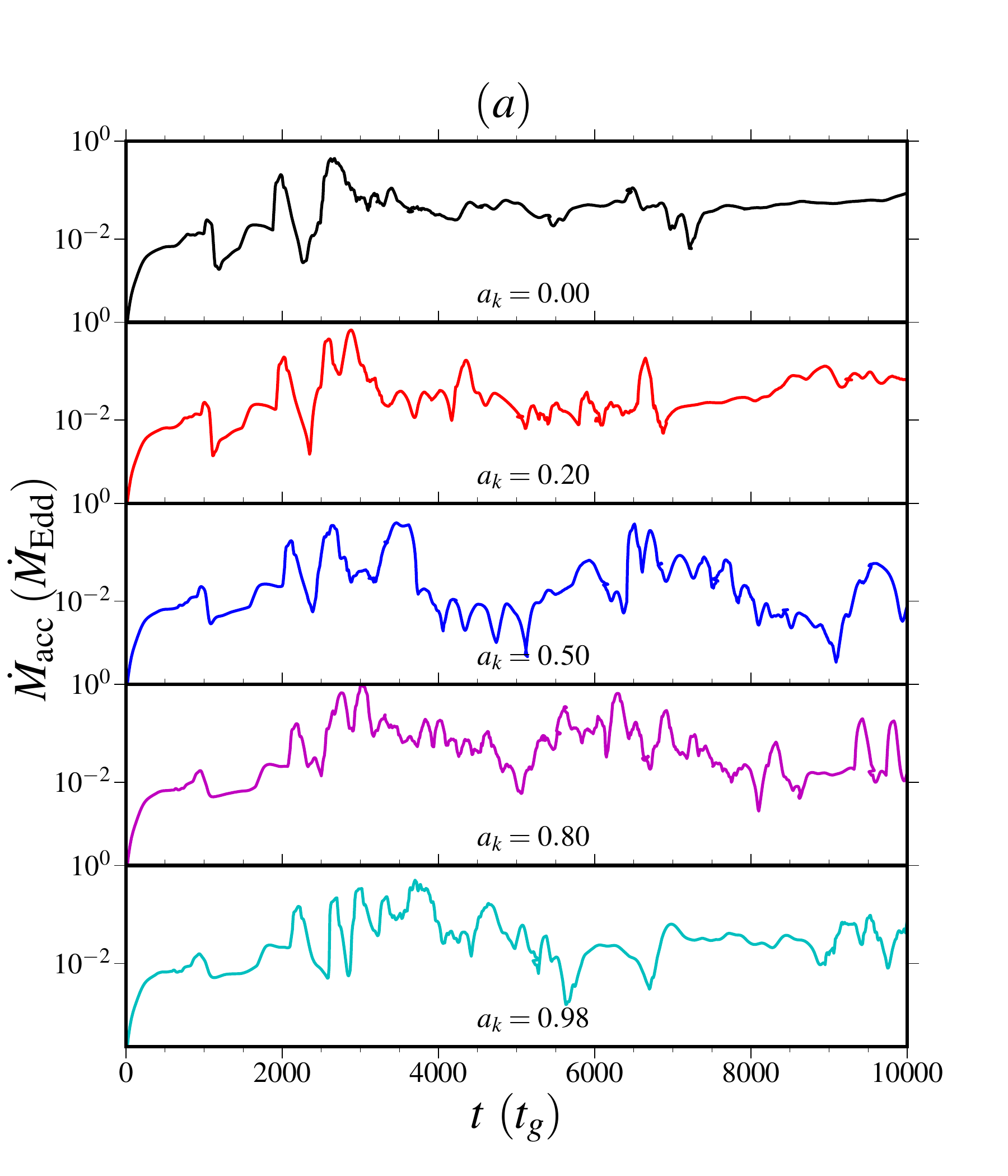} 
	      \includegraphics[width=0.49\textwidth]{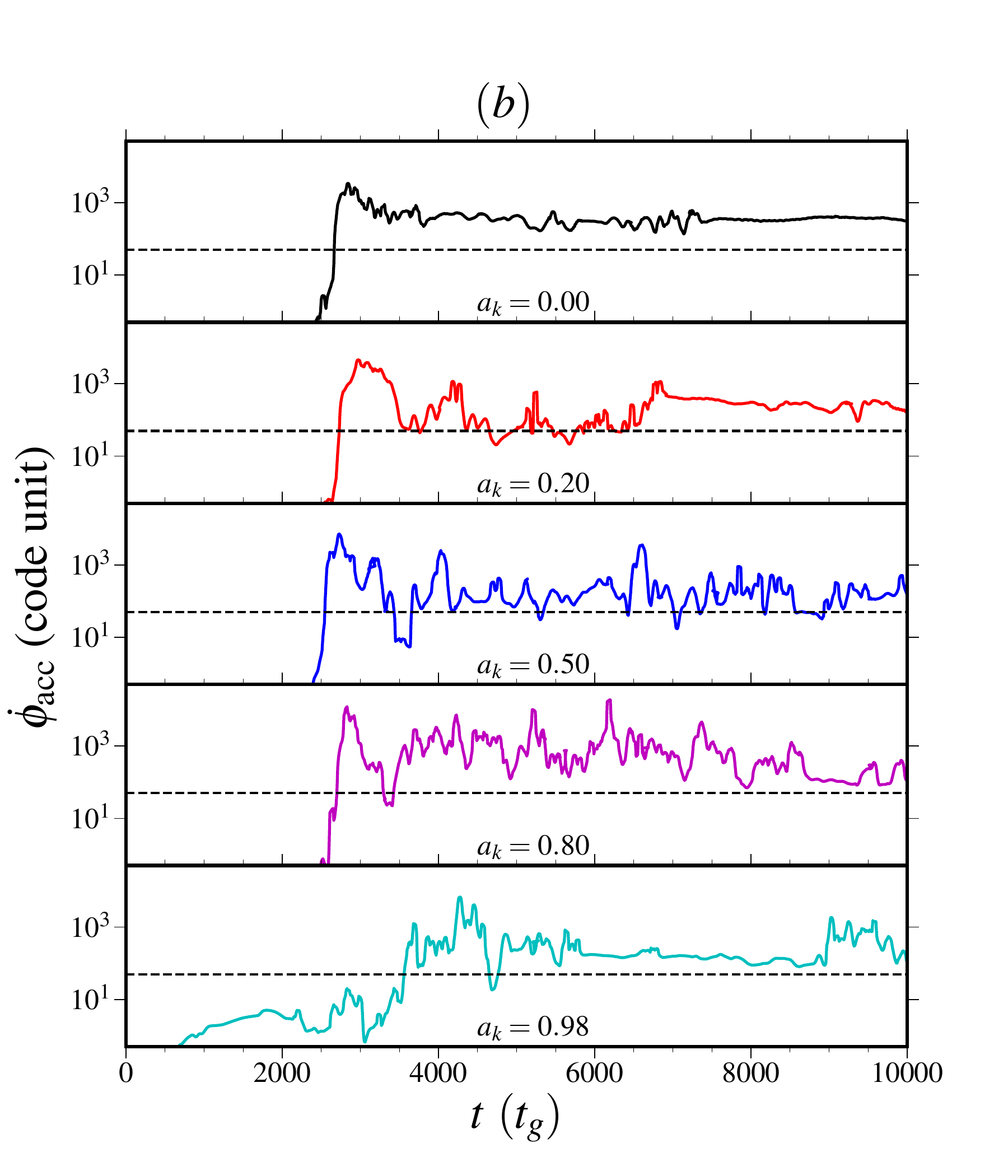} 
        \vskip -3.0mm
        \includegraphics[width=0.49\textwidth]{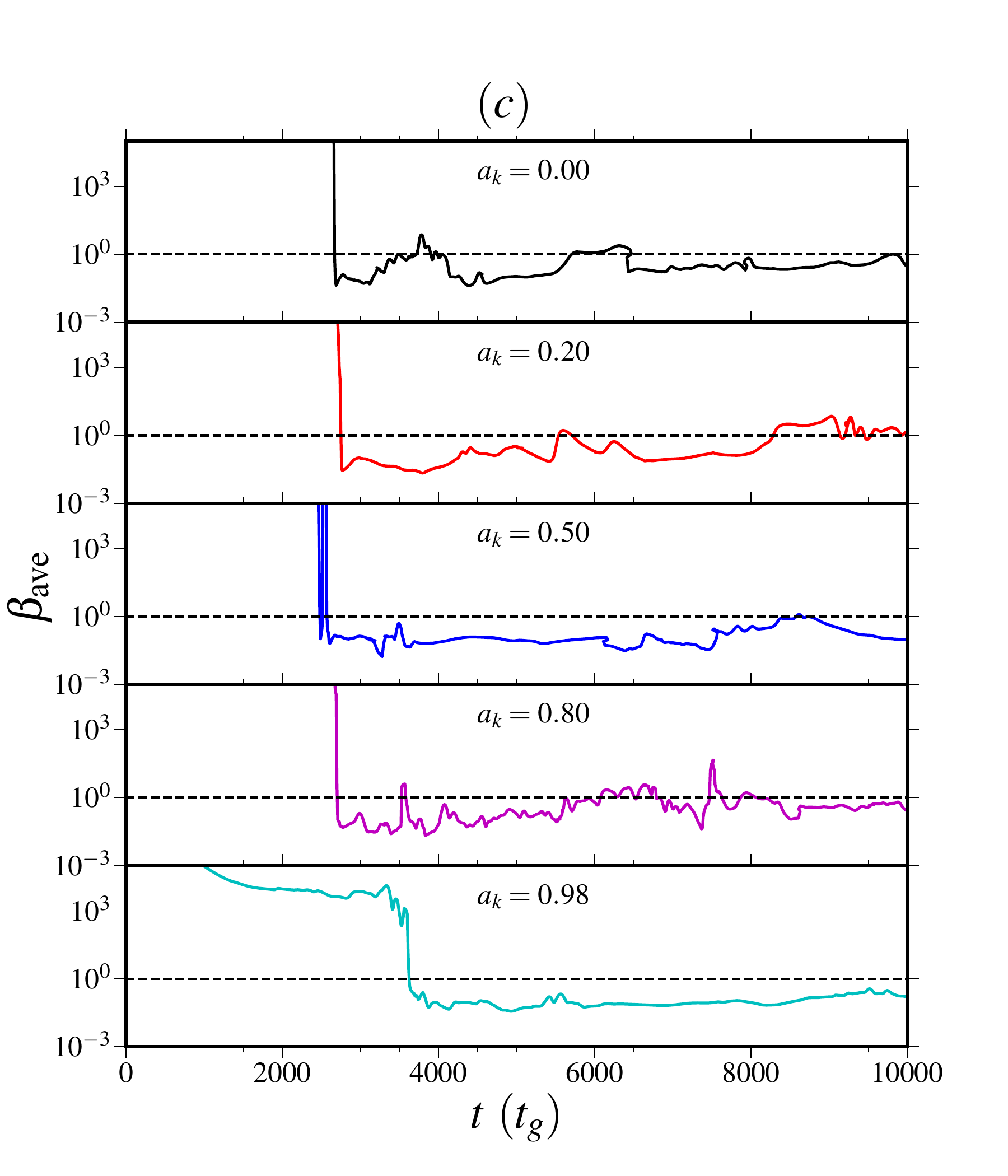} 
	\end{center}
	\caption{Comparison of temporal evolution of $(a)$: mass accretion rate ($\dot{M}_{\rm acc}$) in Eddington units ($\dot{M}_{\rm Edd}$), $(b)$: normalized magnetic ﬂux ($\dot{\phi}_{\rm acc}$) accumulated at the black hole horizon and ($c$): spatial average plasma beta ($\beta_{\rm ave}$) with the simulation time for different black hole spin. Here, we consider the spin values as $a_k = 0.0, 0.20, 0.50, 0.80$, and $0.98$. See the text for details.}
	\label{Figure_2}
\end{figure}
%%%%%%%%%%%%%%%%%%%%%%%%%%%%%%%%%%%%%%%%%%%%%%%%%%%%

%%%%%%%%%%%%%%%%%%%%%%%%%%%%%%%%%%%%%%%%%%%%%%%%%%%
%%                        Figure 3
%%%%%%%%%%%%%%%%%%%%%%%%%%%%%%%%%%%%%%%%%%%%%%%%%%%
\begin{figure*}
	\begin{center}
        \hskip -2.5mm
        \includegraphics[width=0.20\textwidth]{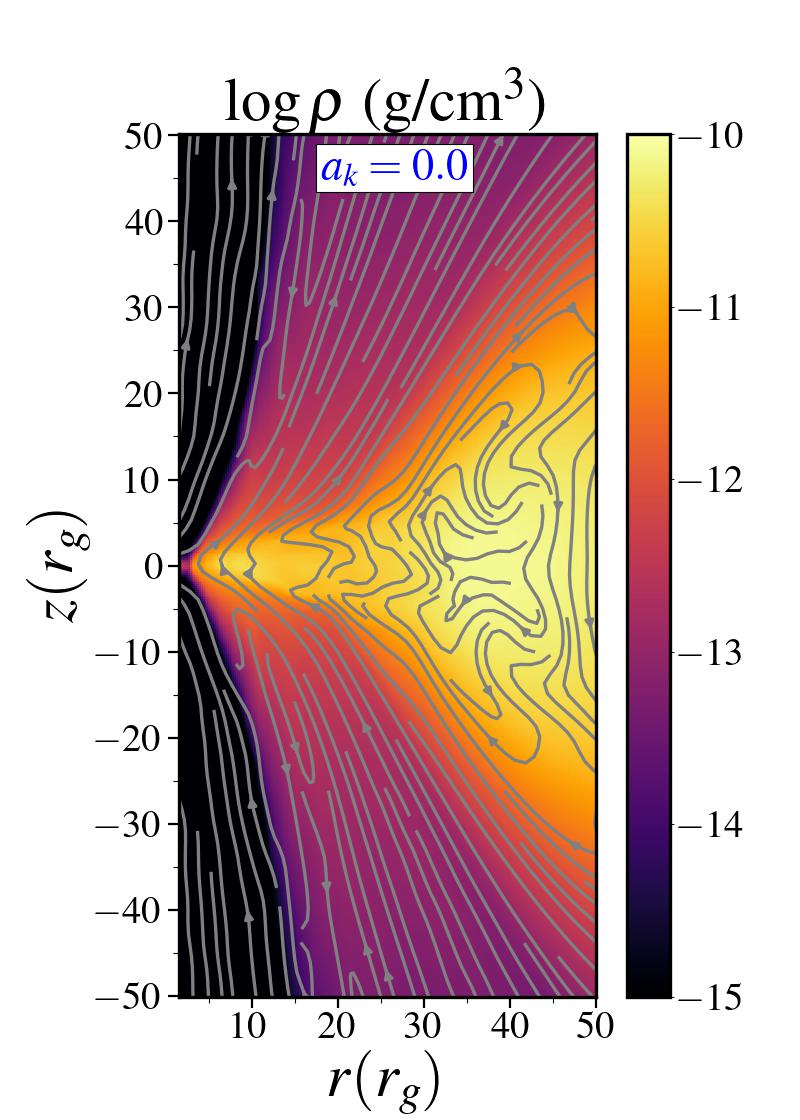} 
        \hskip -2.5mm
        \includegraphics[width=0.20\textwidth]{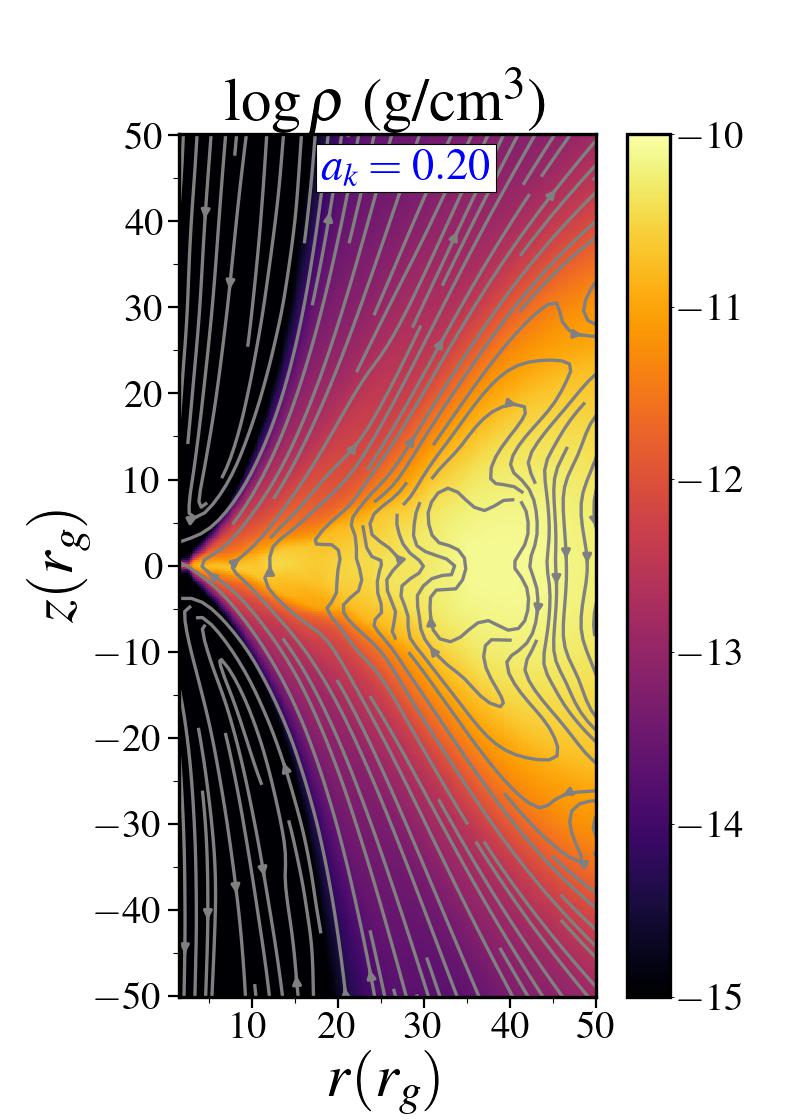} 
        \hskip -2.5mm
	\includegraphics[width=0.20\textwidth]{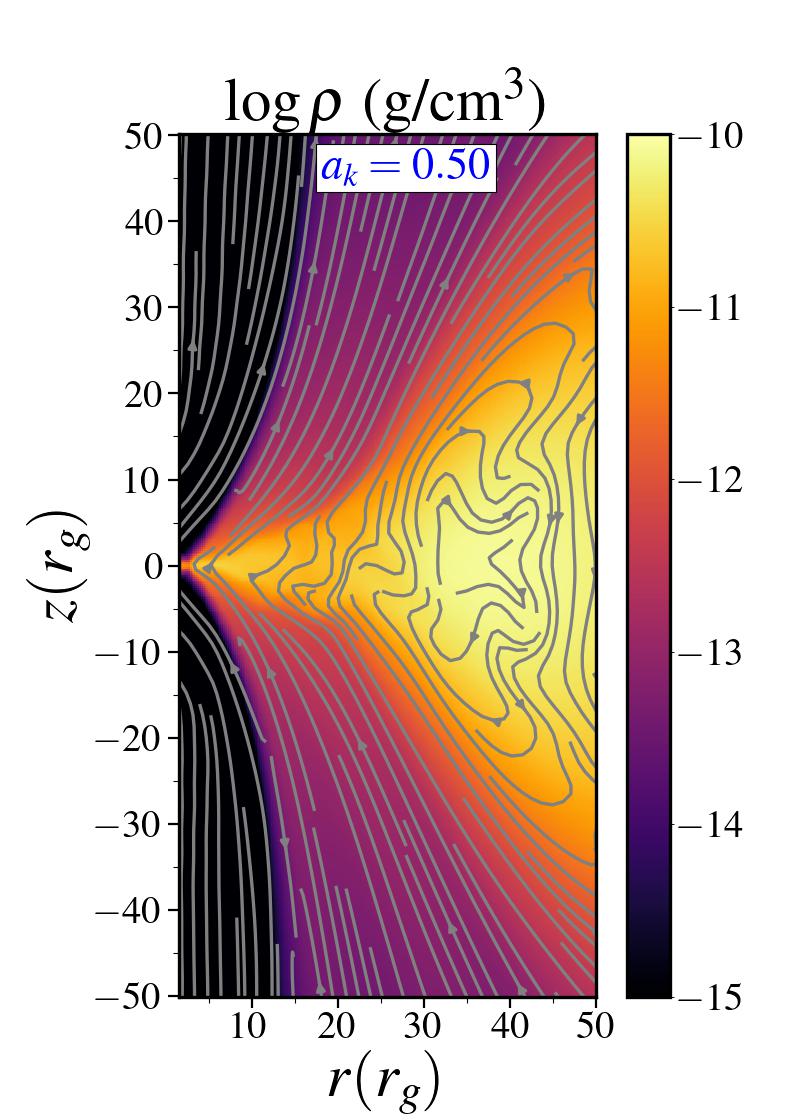} 
        \hskip -2.5mm
        \includegraphics[width=0.20\textwidth]{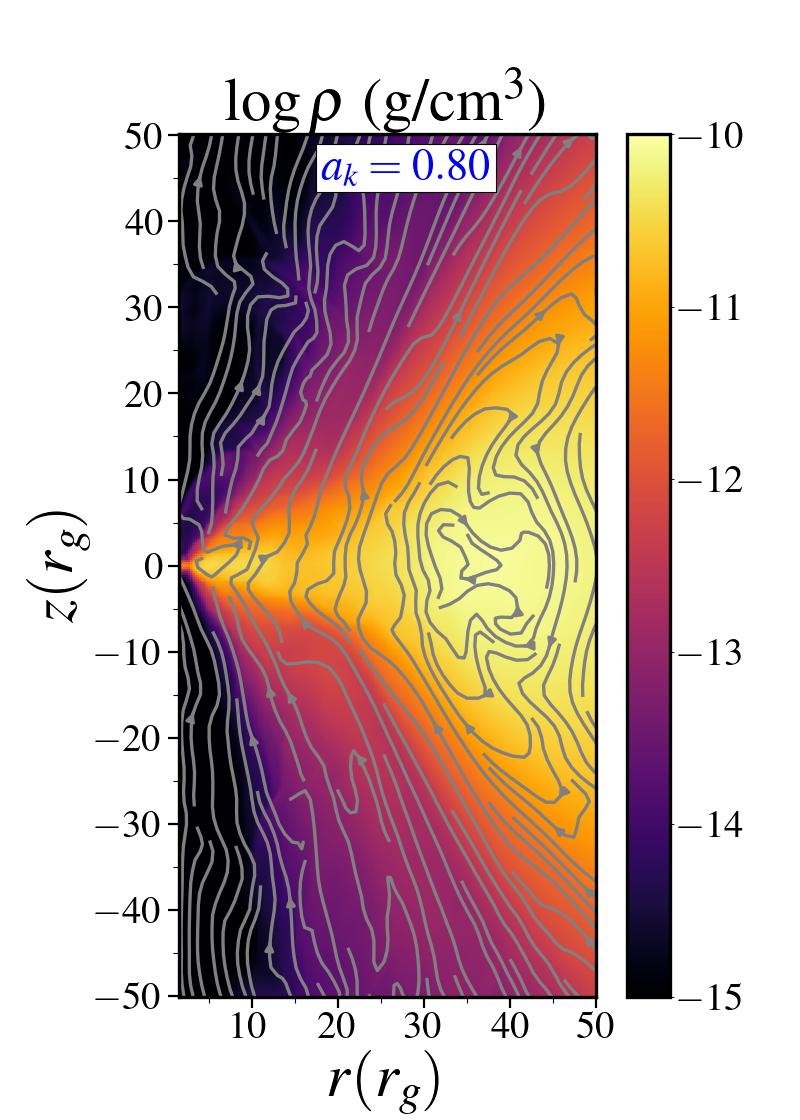} 
        \hskip -2.5mm
        \includegraphics[width=0.20\textwidth]{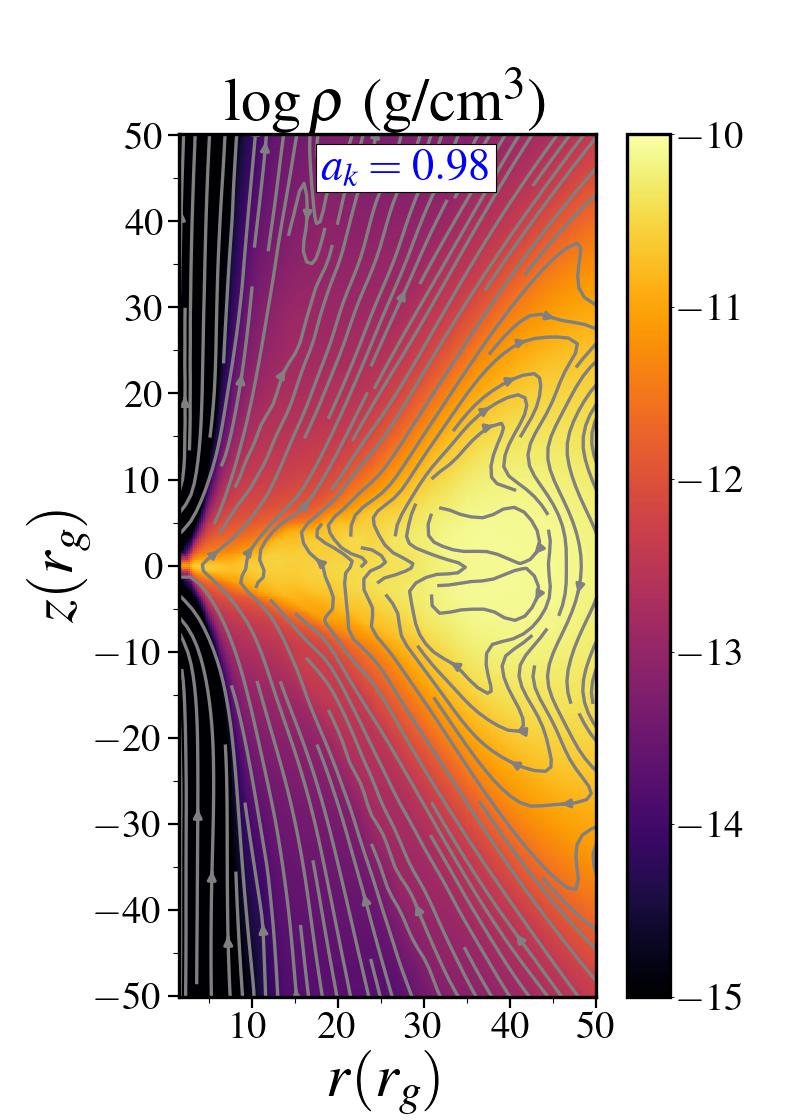} 

        \hskip -2.5mm
        \includegraphics[width=0.20\textwidth]{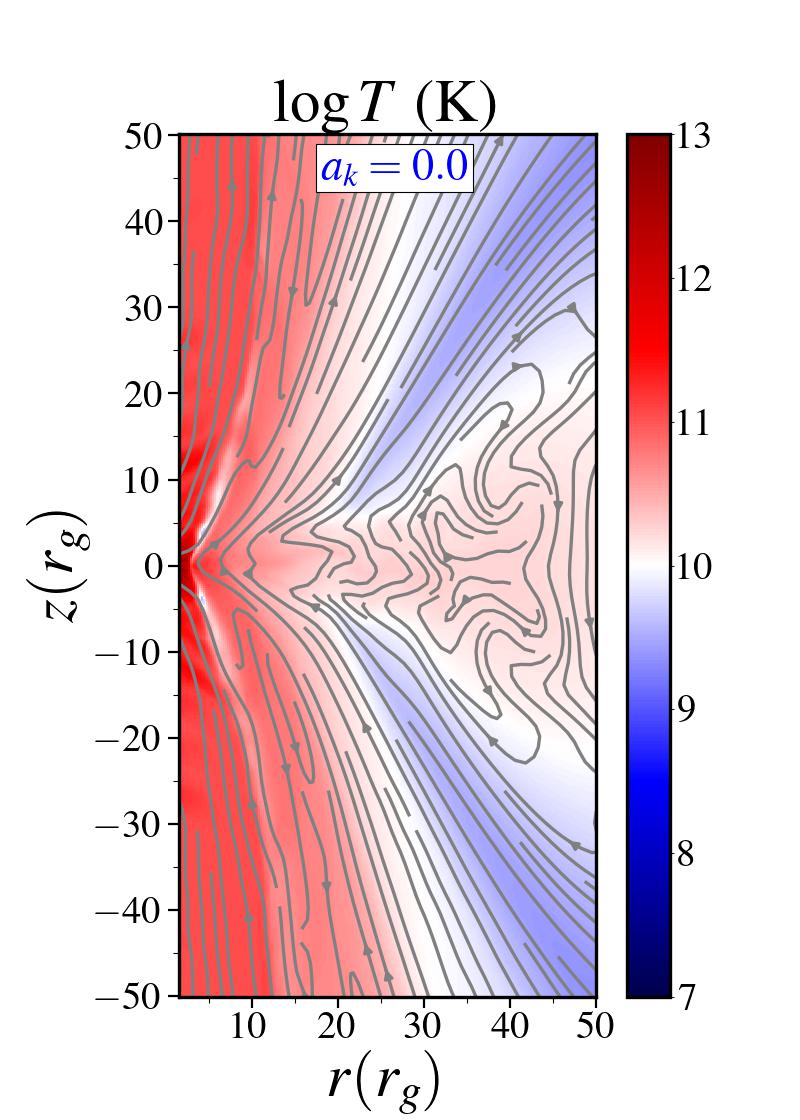} 
        \hskip -2.5mm
        \includegraphics[width=0.20\textwidth]{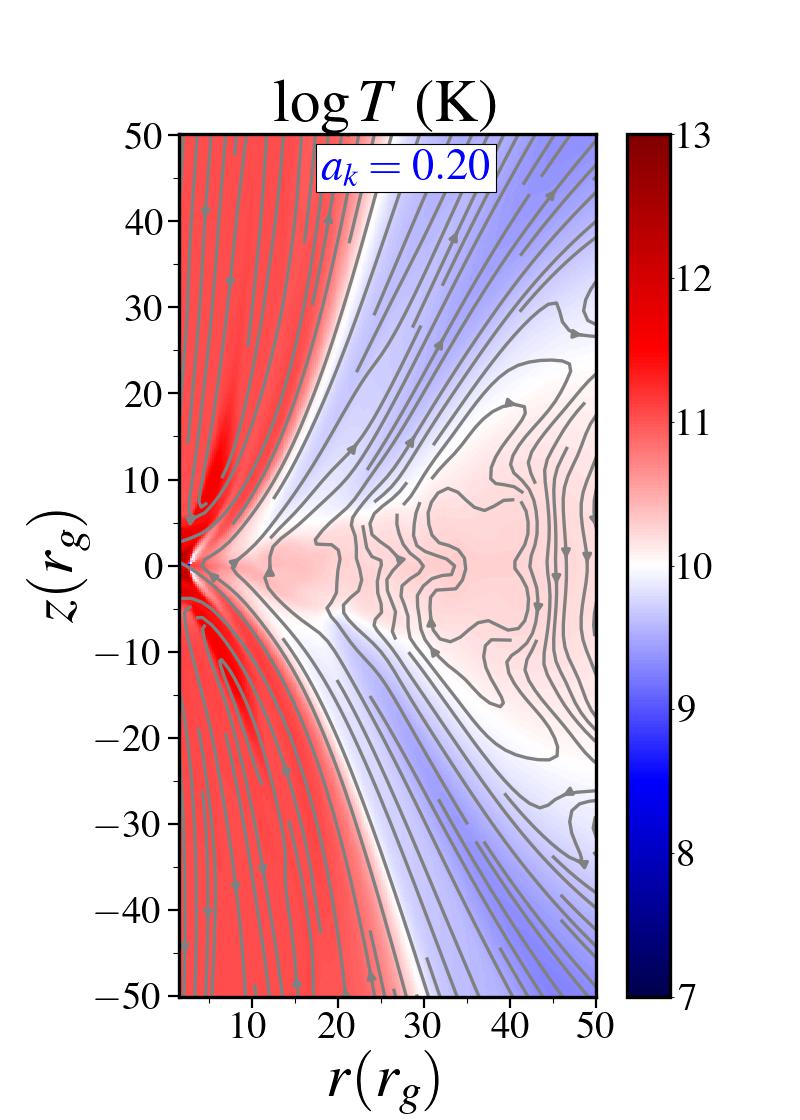} 
        \hskip -2.5mm
	\includegraphics[width=0.20\textwidth]{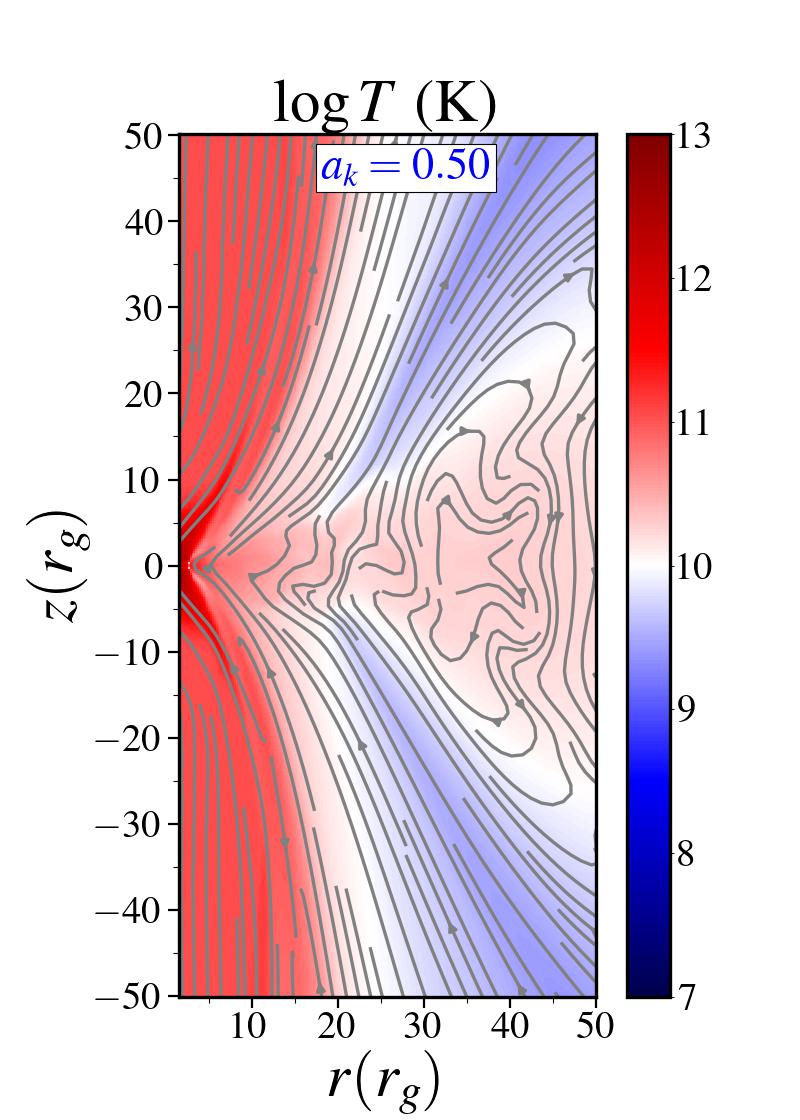} 
        \hskip -2.5mm
        \includegraphics[width=0.20\textwidth]{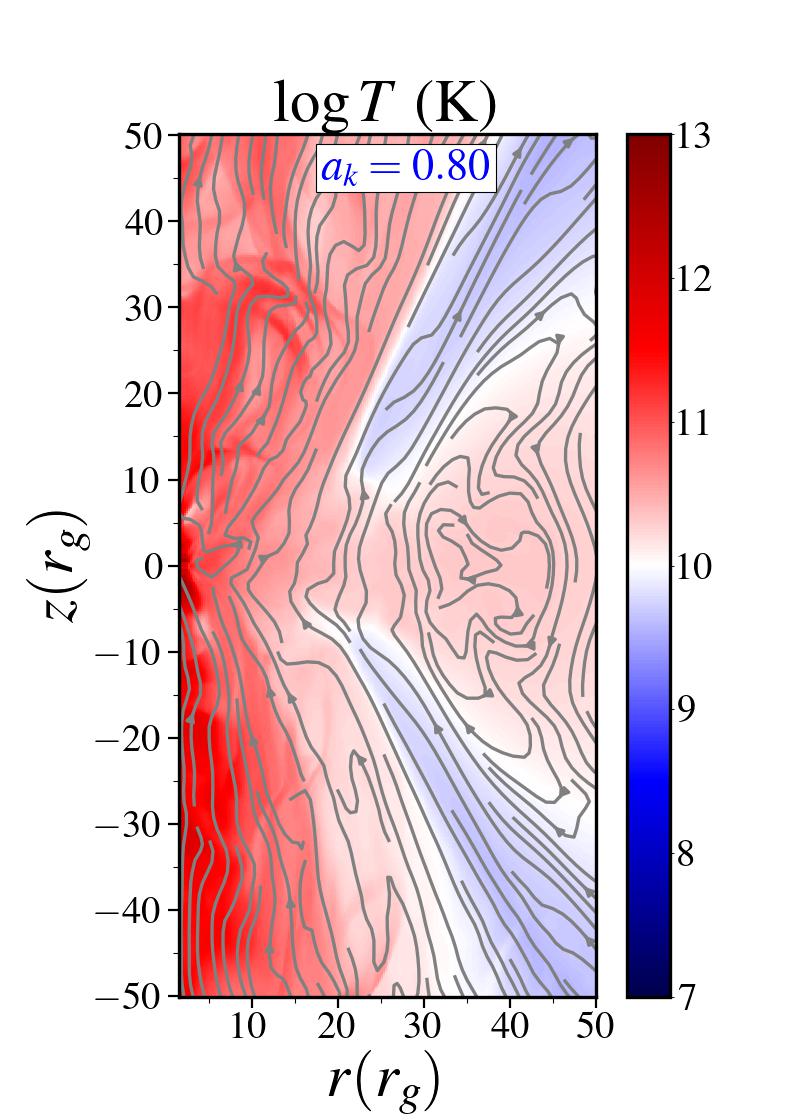} 
        \hskip -2.5mm
        \includegraphics[width=0.20\textwidth]{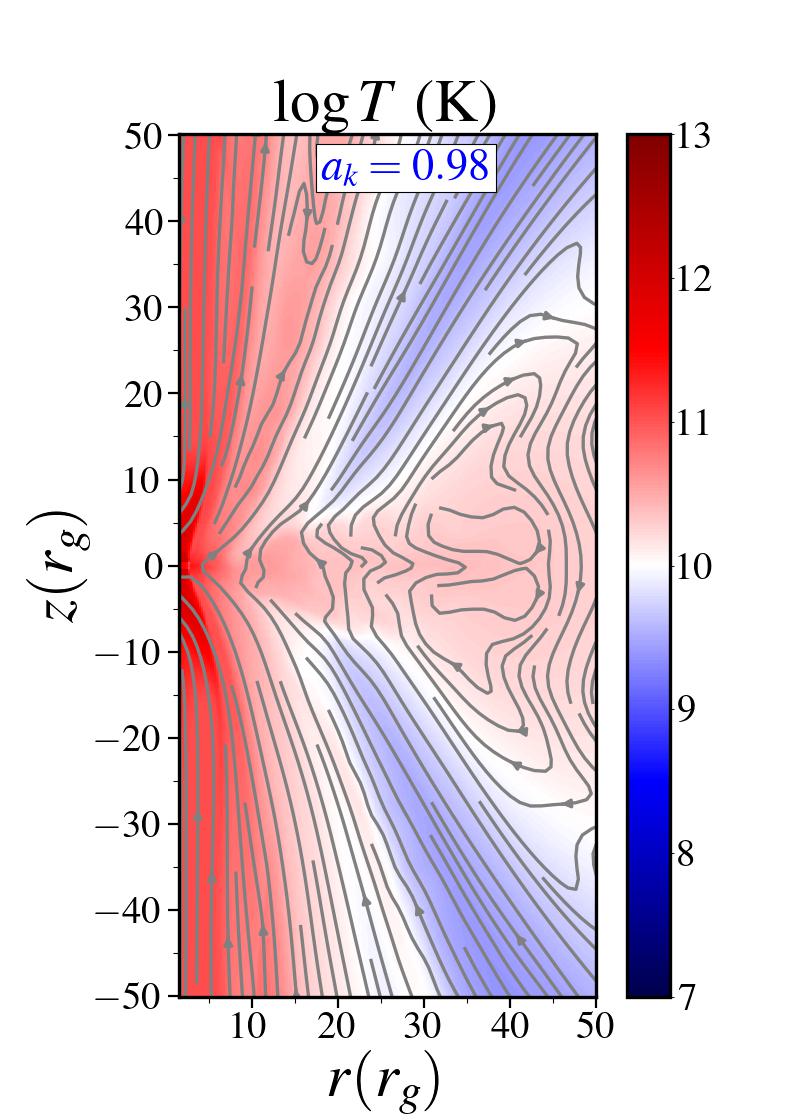} 
        
        \hskip -2.5mm       
       \includegraphics[width=0.20\textwidth]{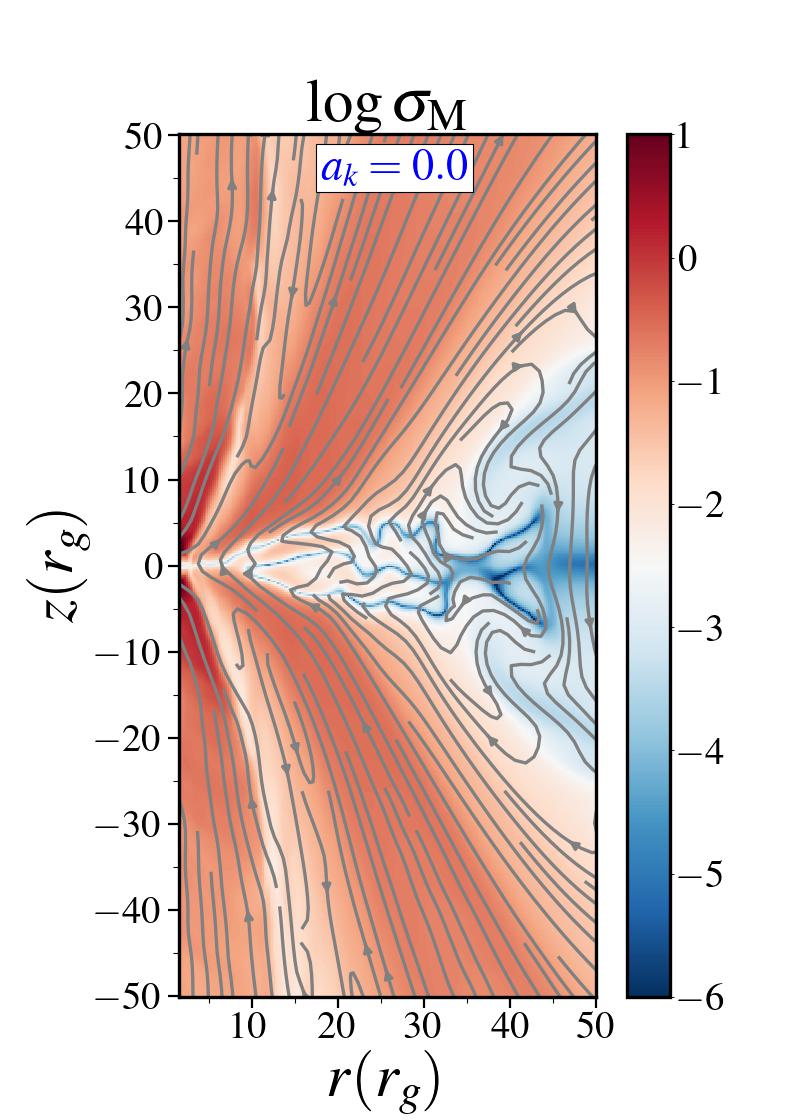} 
        \hskip -2.5mm
        \includegraphics[width=0.20\textwidth]{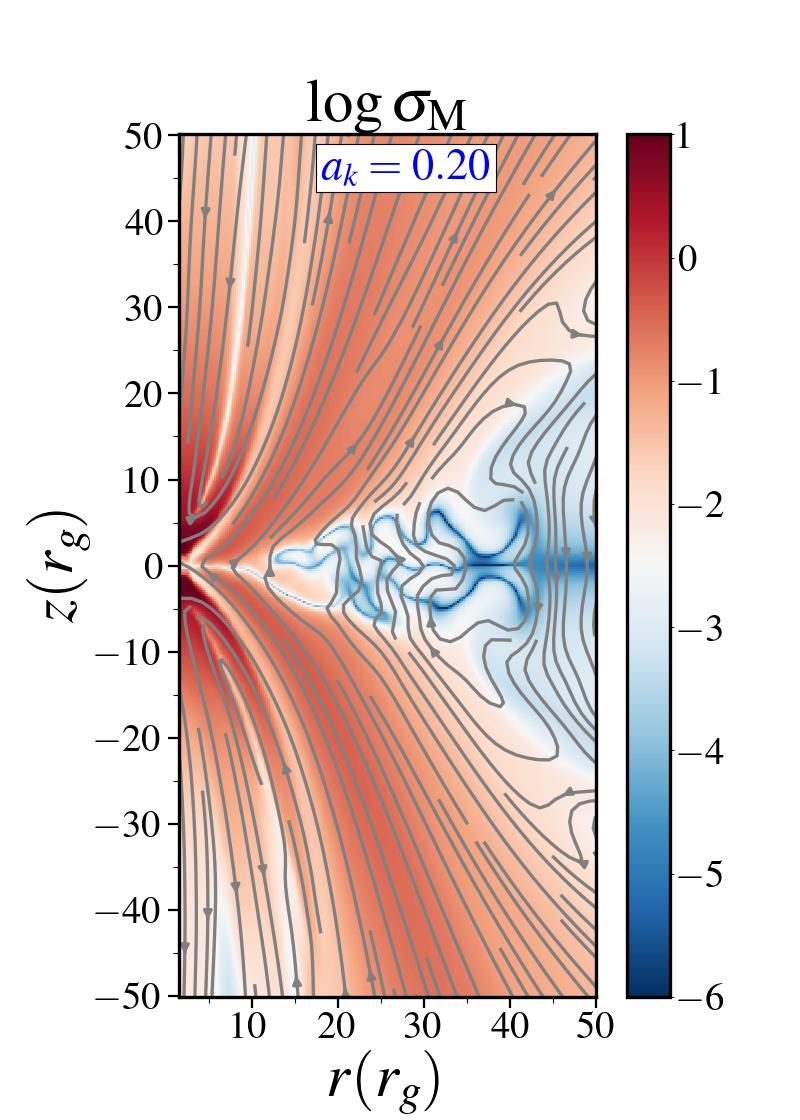} 
        \hskip -2.5mm
	\includegraphics[width=0.20\textwidth]{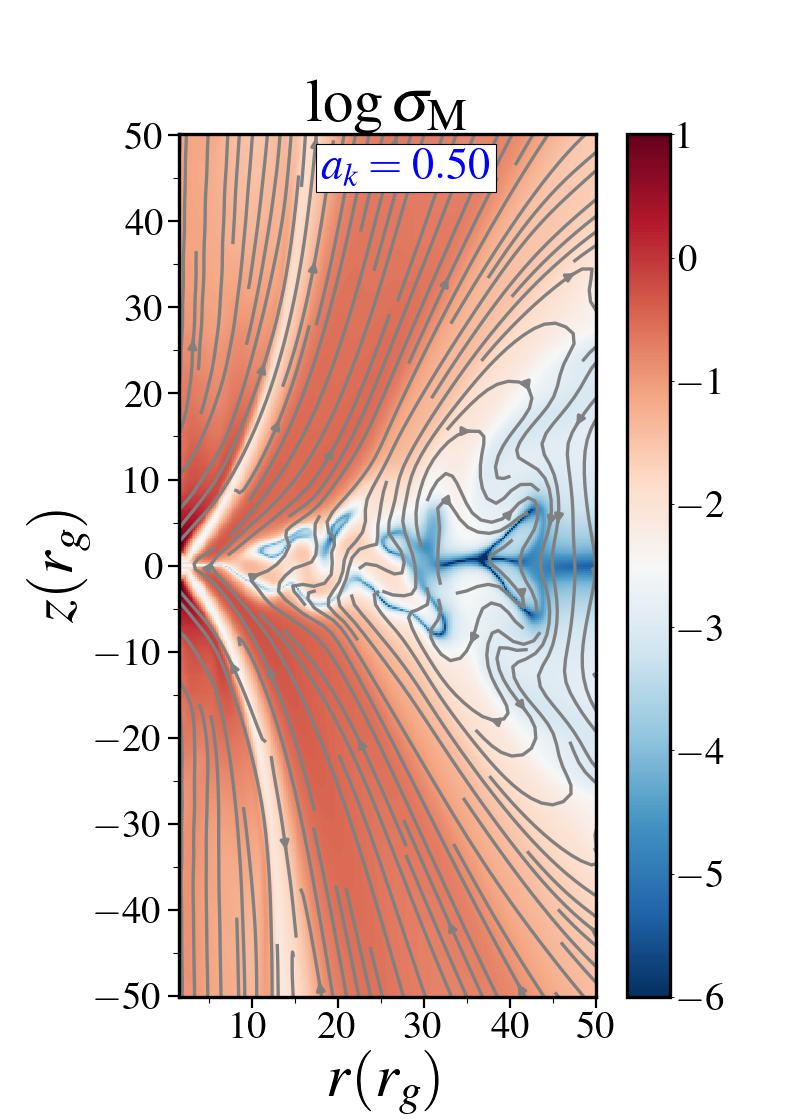} 
        \hskip -2.5mm
        \includegraphics[width=0.20\textwidth]{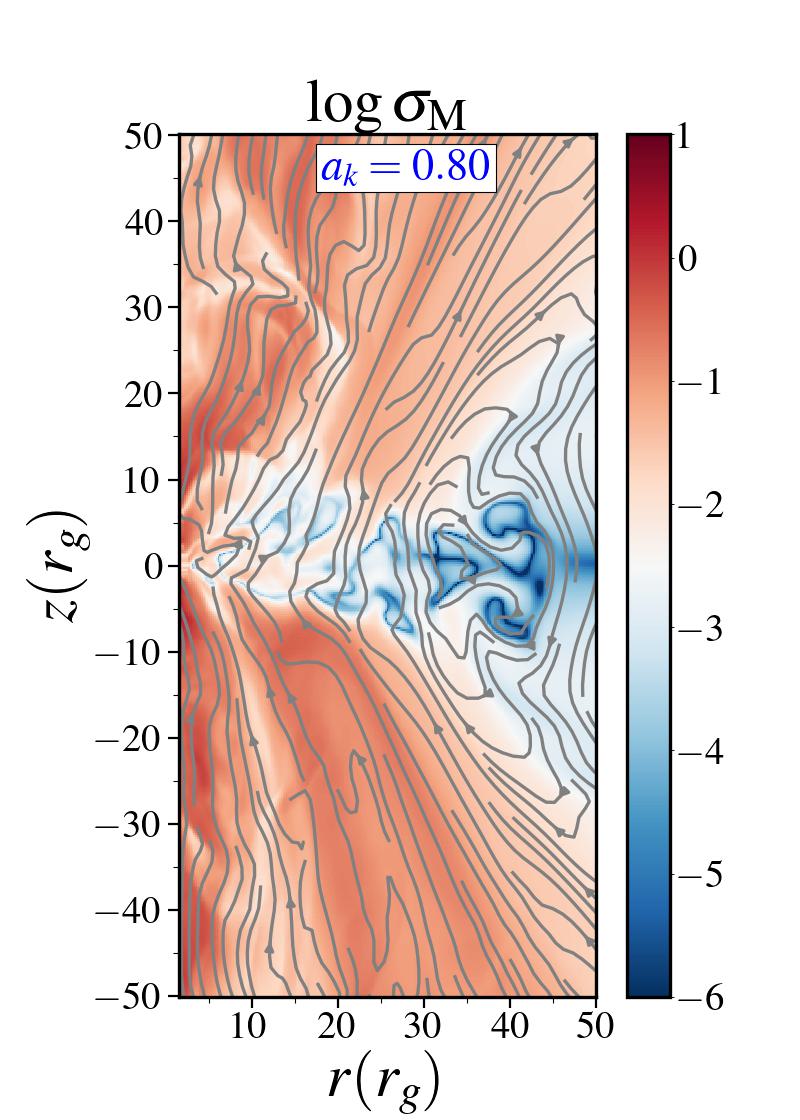} 
        \hskip -2.5mm
        \includegraphics[width=0.20\textwidth]{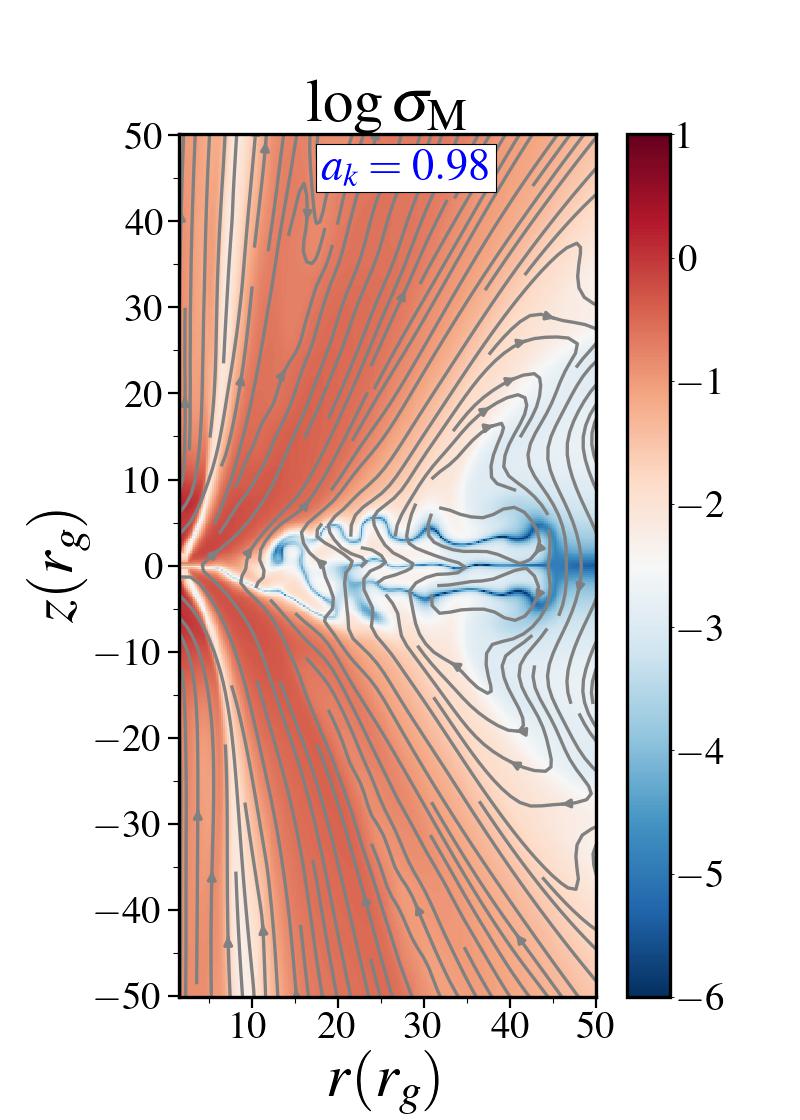} 

        \hskip -2.5mm 
        \includegraphics[width=0.20\textwidth]{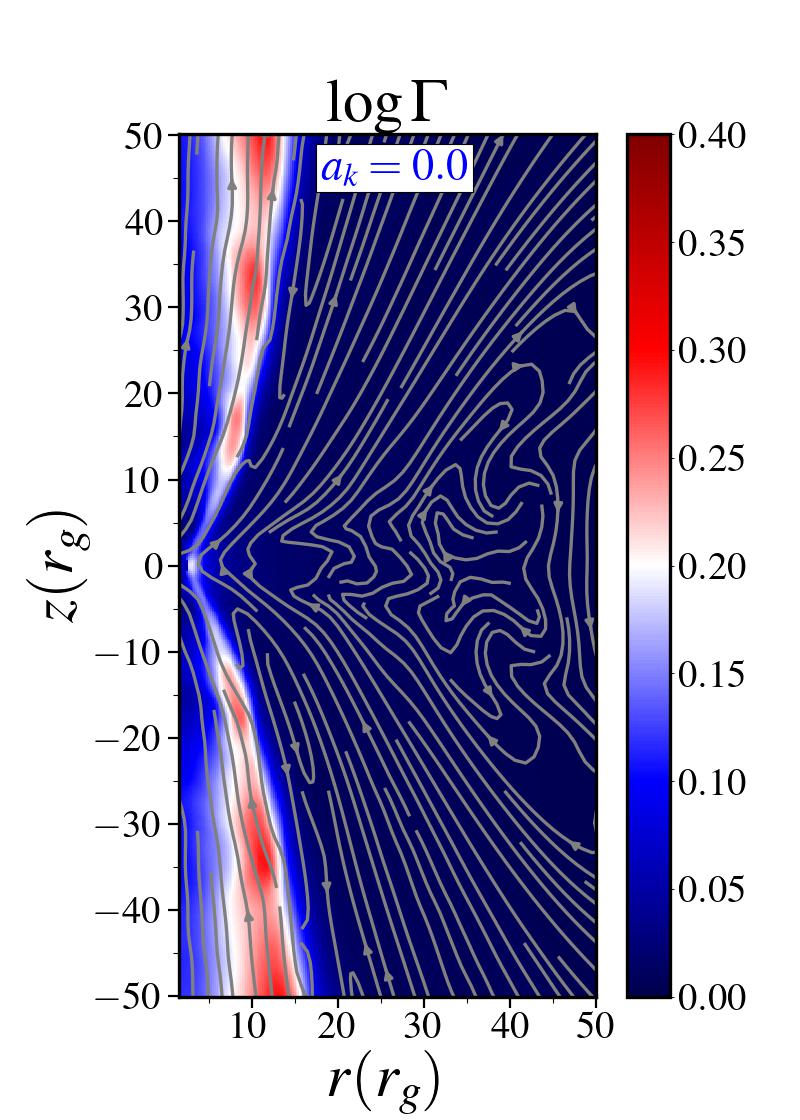} 
        \hskip -2.5mm
        \includegraphics[width=0.20\textwidth]{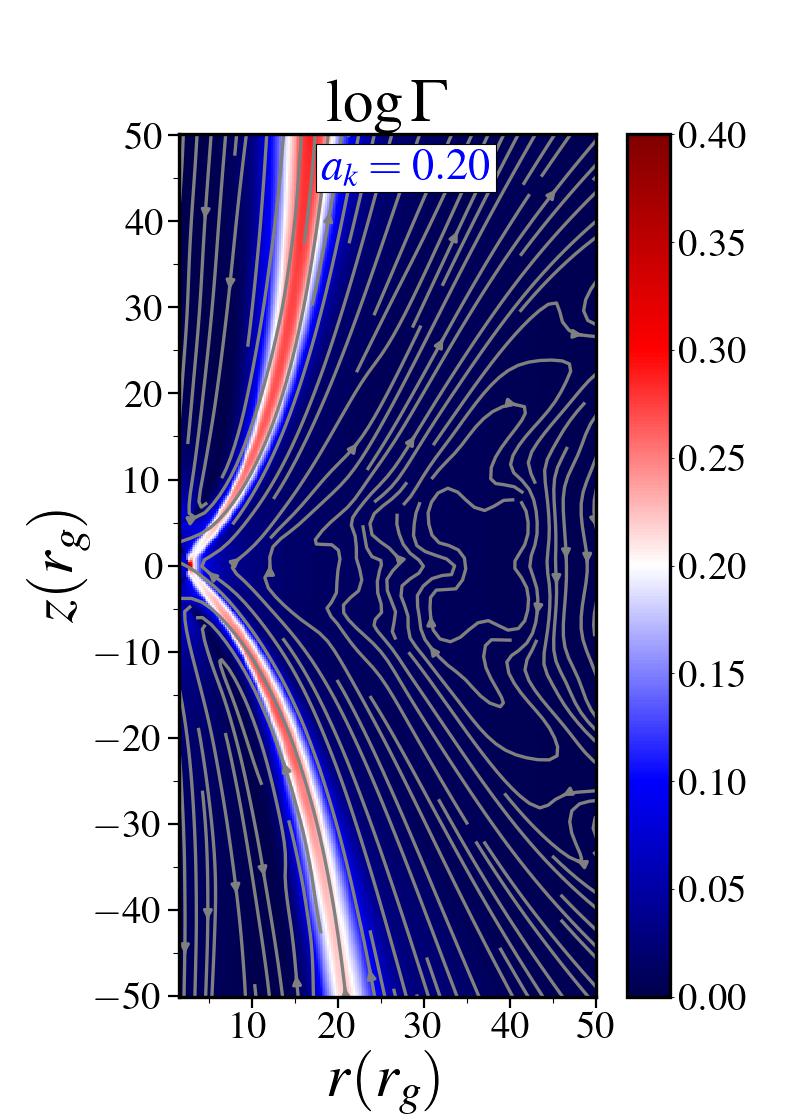} 
        \hskip -2.5mm
	\includegraphics[width=0.20\textwidth]{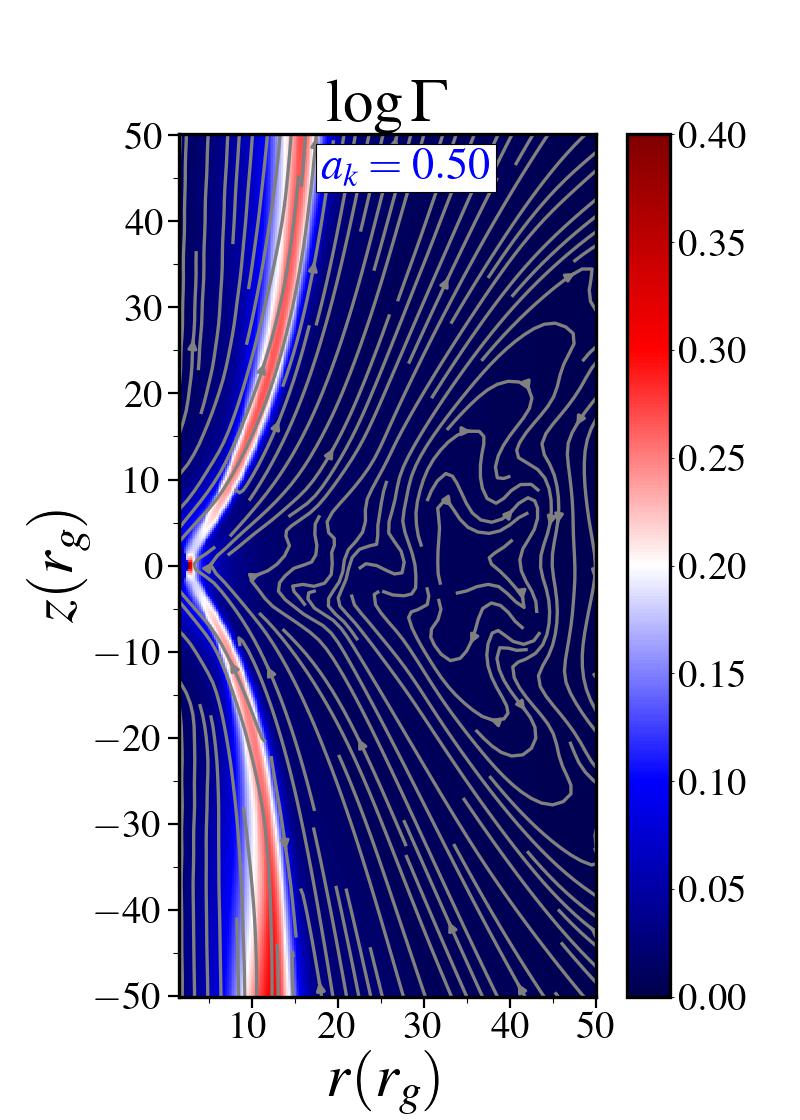} 
        \hskip -2.5mm
        \includegraphics[width=0.20\textwidth]{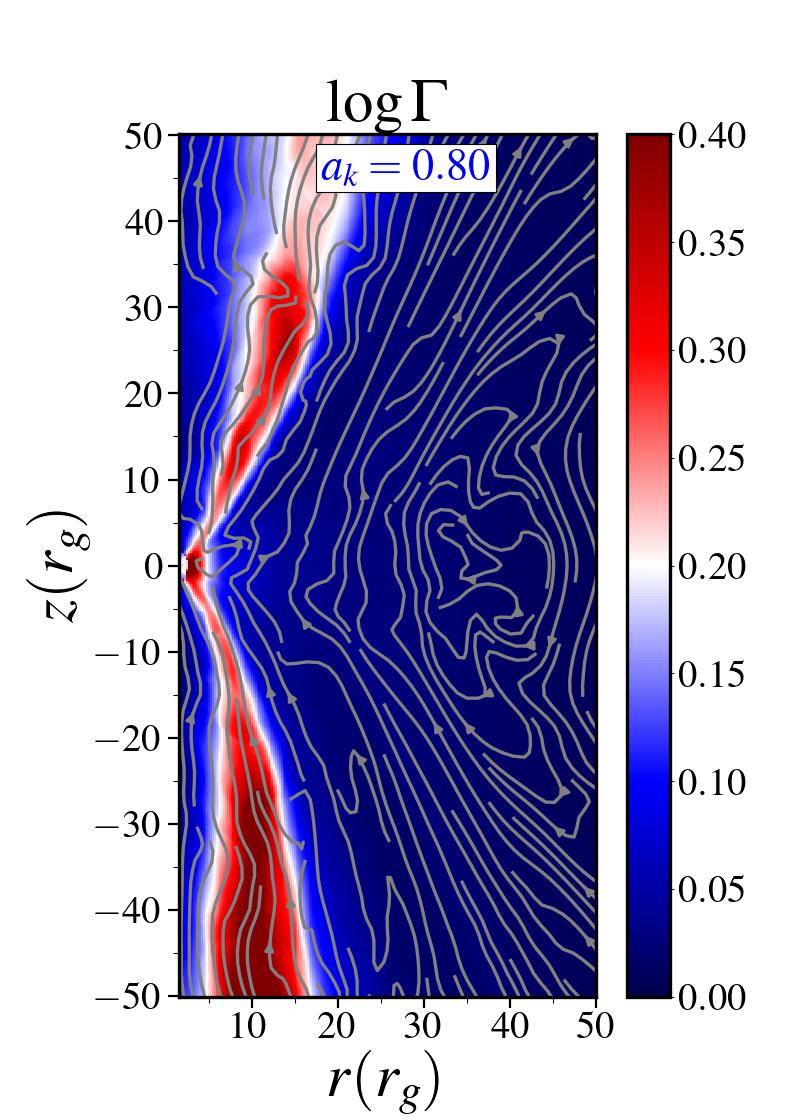} 
        \hskip -2.5mm
        \includegraphics[width=0.20\textwidth]{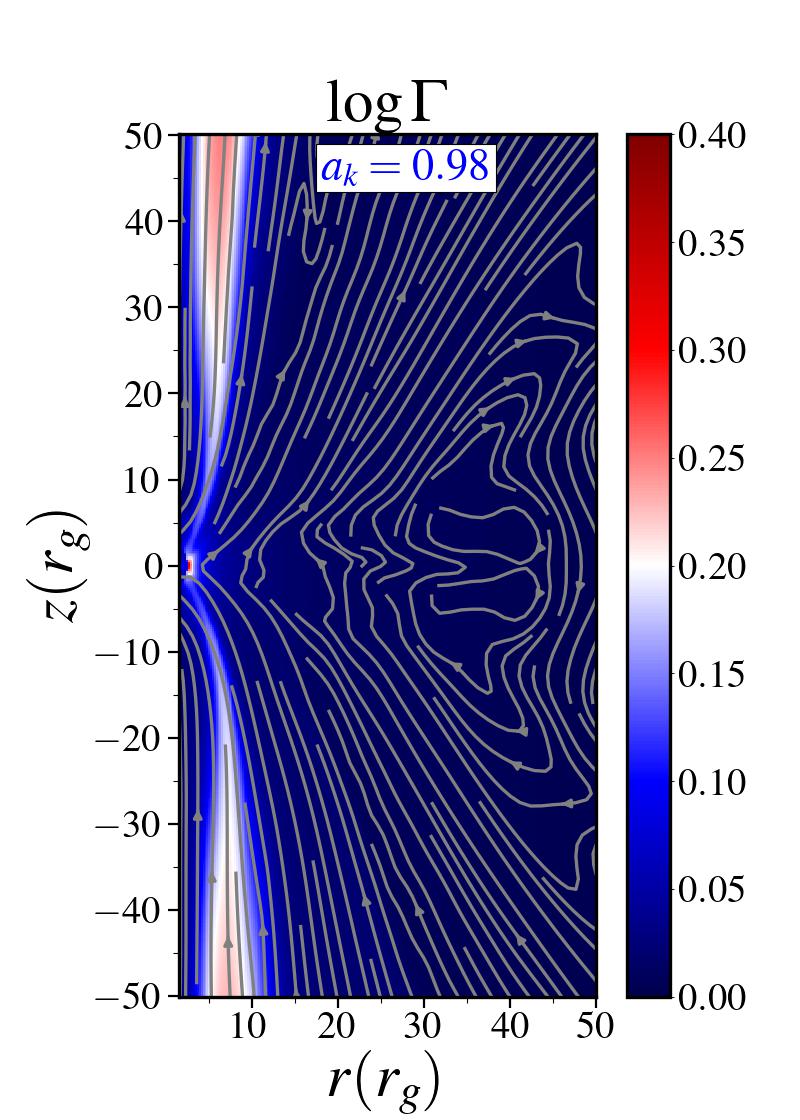} 
	\end{center}
	\caption{Comparison of azimuthal and time averaged density ($\rho$), temperature ($T$), magnetization parameter ($\sigma_{\rm M}$) and Lorentz factor ($\Gamma$) in the ($r-z$) plane, respectively. Here, we fix the black hole spin as $a_k = 0.0, 0.20, 0.50, 0.80, 0.98$ and the time average between $t = 5000 t_g$ to $6000 t_g$. See the text for details.}
	\label{Figure_3}
\end{figure*}
%%%%%%%%%%%%%%%%%%%%%%%%%%%%%%%%%%%%%%%%%%%%%%%%%%%%

%%%%%%%%%%%%%%%%%%%%%%%%%%%%%%%%%%%%%%%%%%%%%%%%%%%
%%                        Figure 4
%%%%%%%%%%%%%%%%%%%%%%%%%%%%%%%%%%%%%%%%%%%%%%%%%%%
\begin{figure*}
	\begin{center}
        \hskip -2.5mm 
        \includegraphics[width=0.20\textwidth]{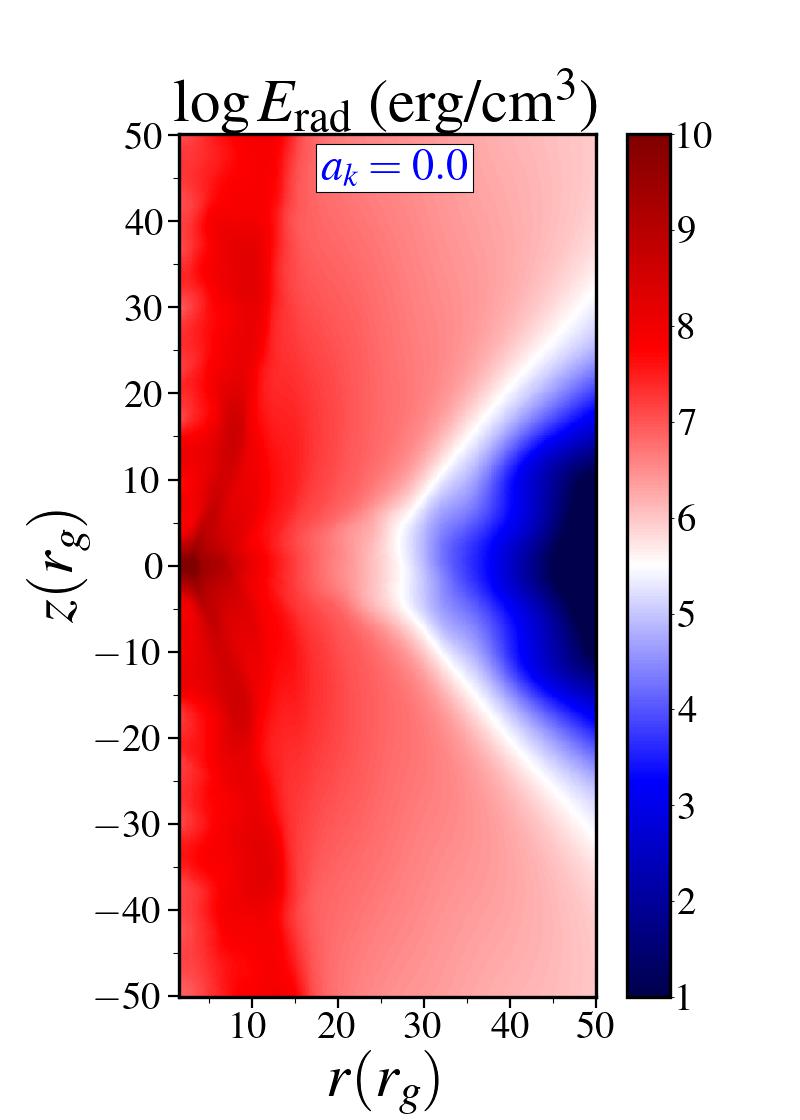} 
        \hskip -2.5mm
        \includegraphics[width=0.20\textwidth]{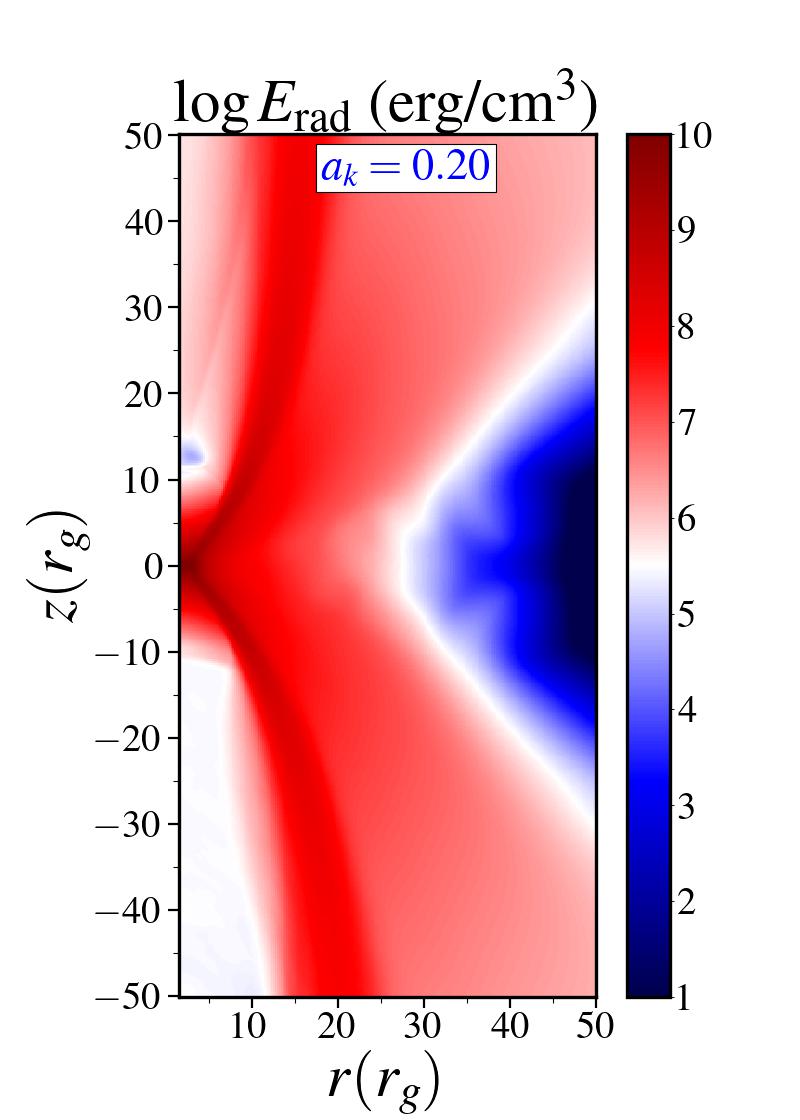} 
        \hskip -2.5mm
	\includegraphics[width=0.20\textwidth]{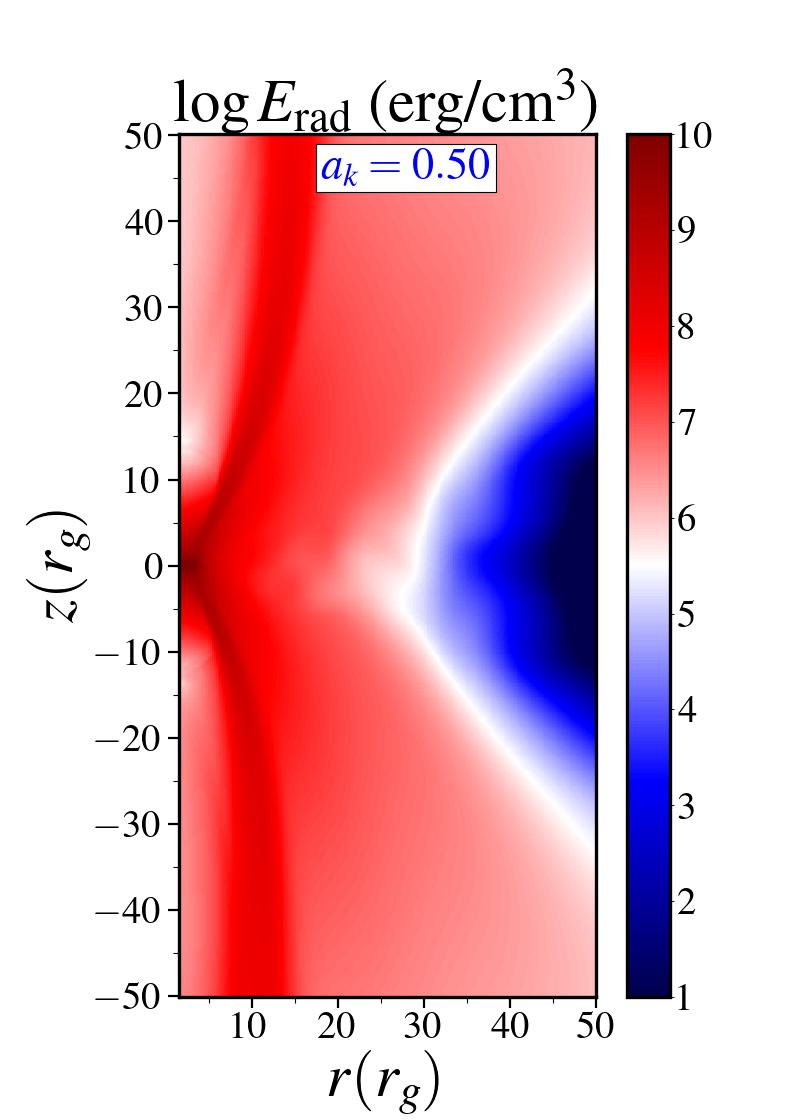} 
        \hskip -2.5mm
        \includegraphics[width=0.20\textwidth]{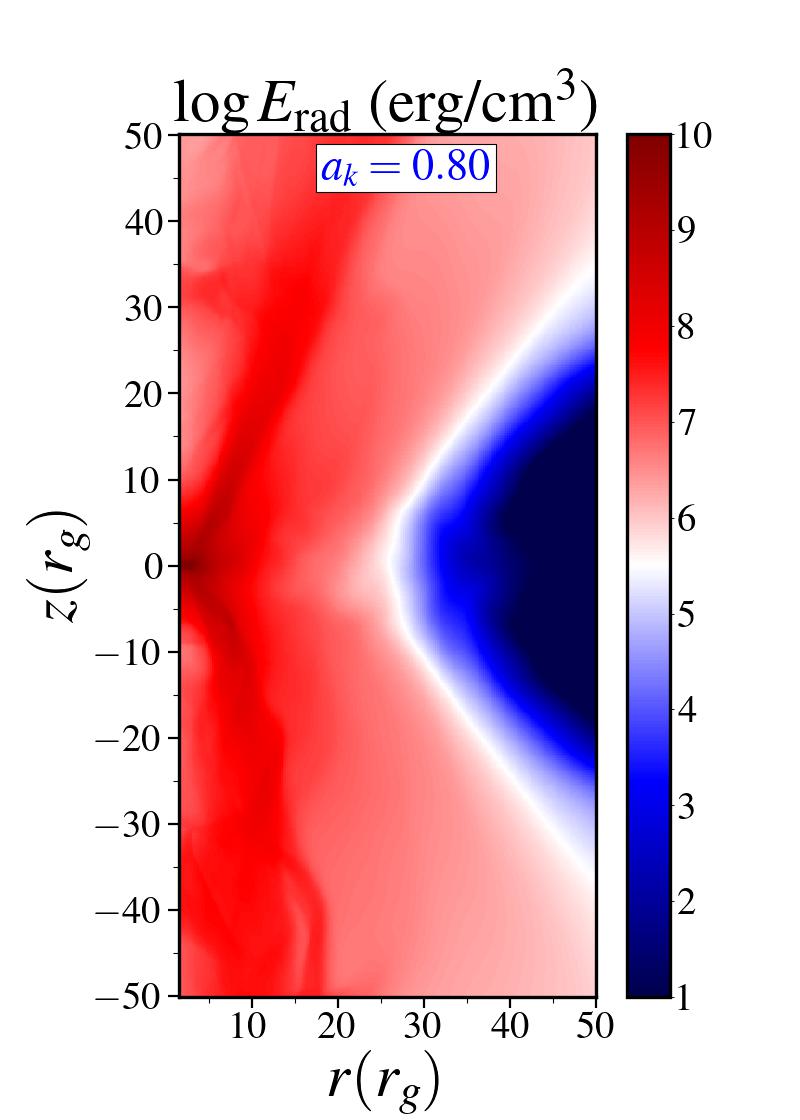} 
        \hskip -2.5mm
        \includegraphics[width=0.20\textwidth]{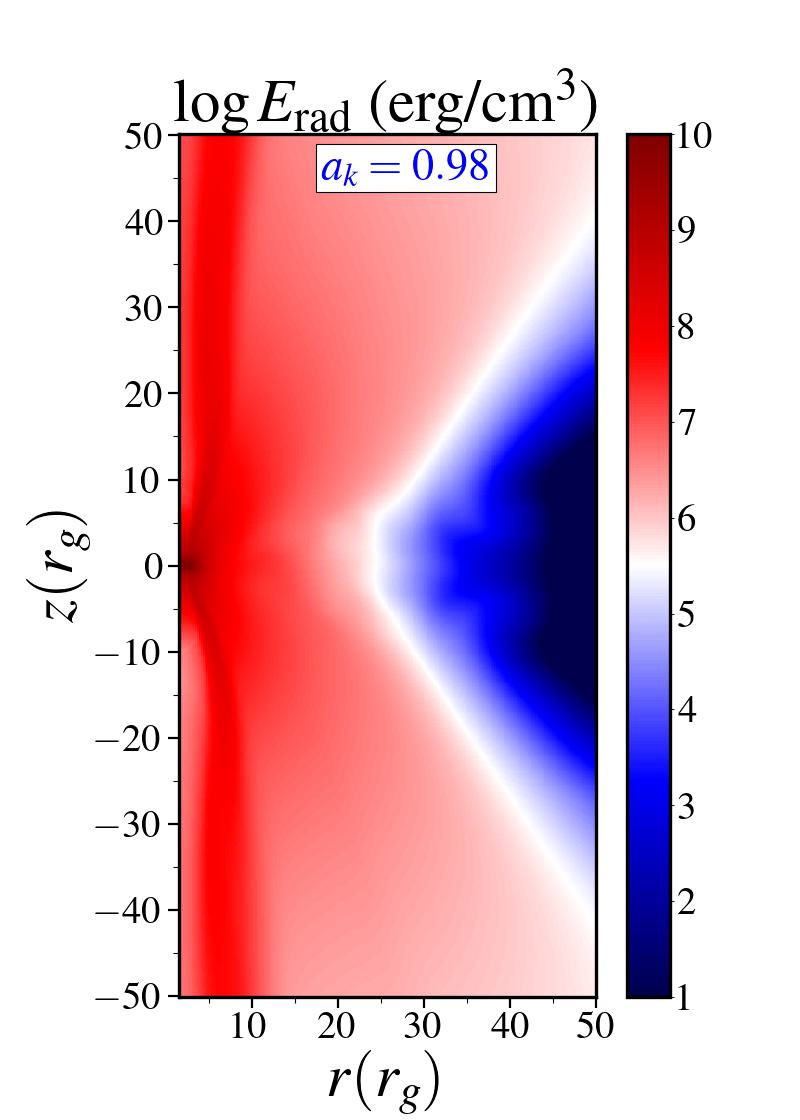} 
        \hskip -3.0mm
        \includegraphics[width=0.20\textwidth]{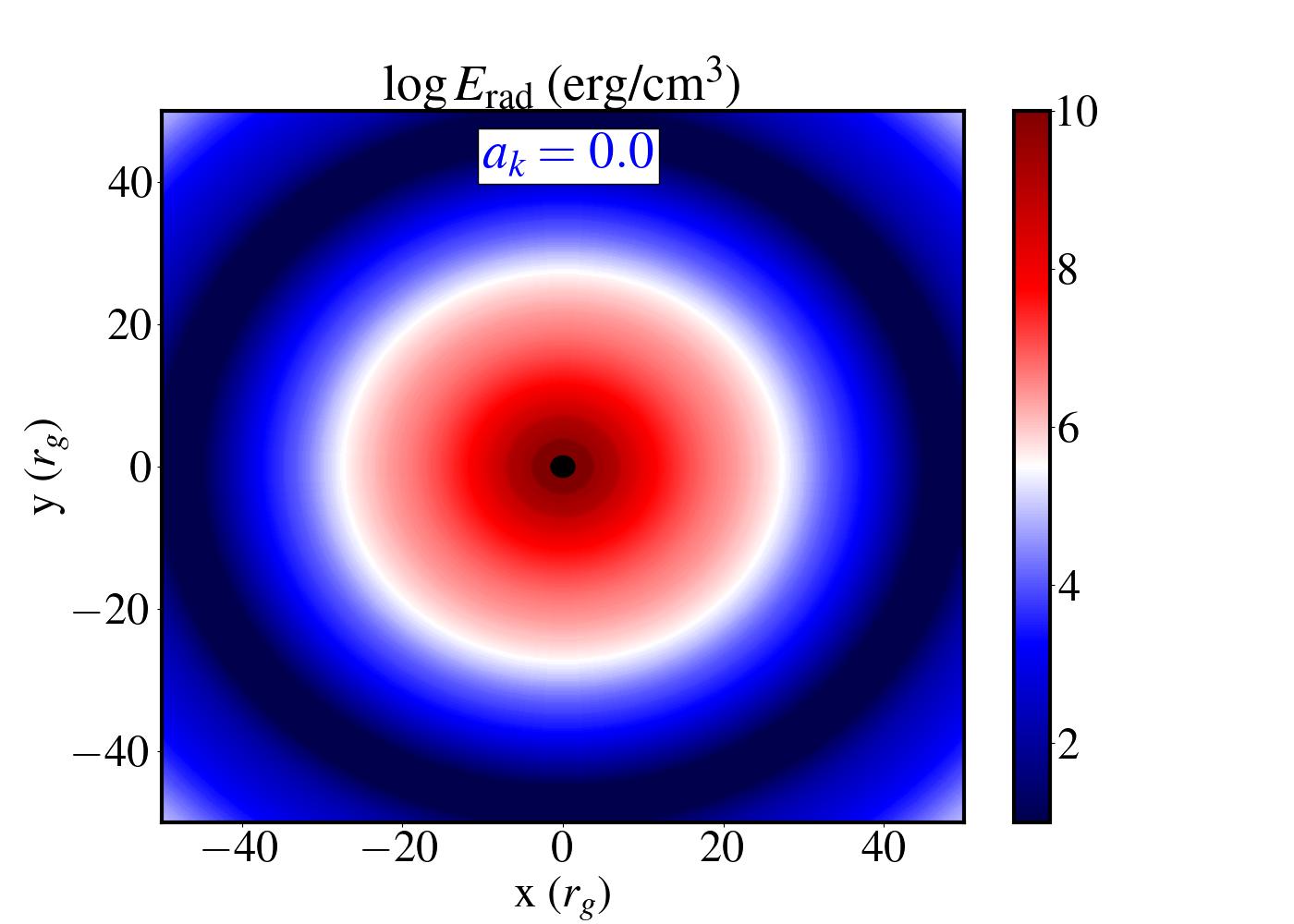} 
        \hskip -3.0mm
        \includegraphics[width=0.20\textwidth]{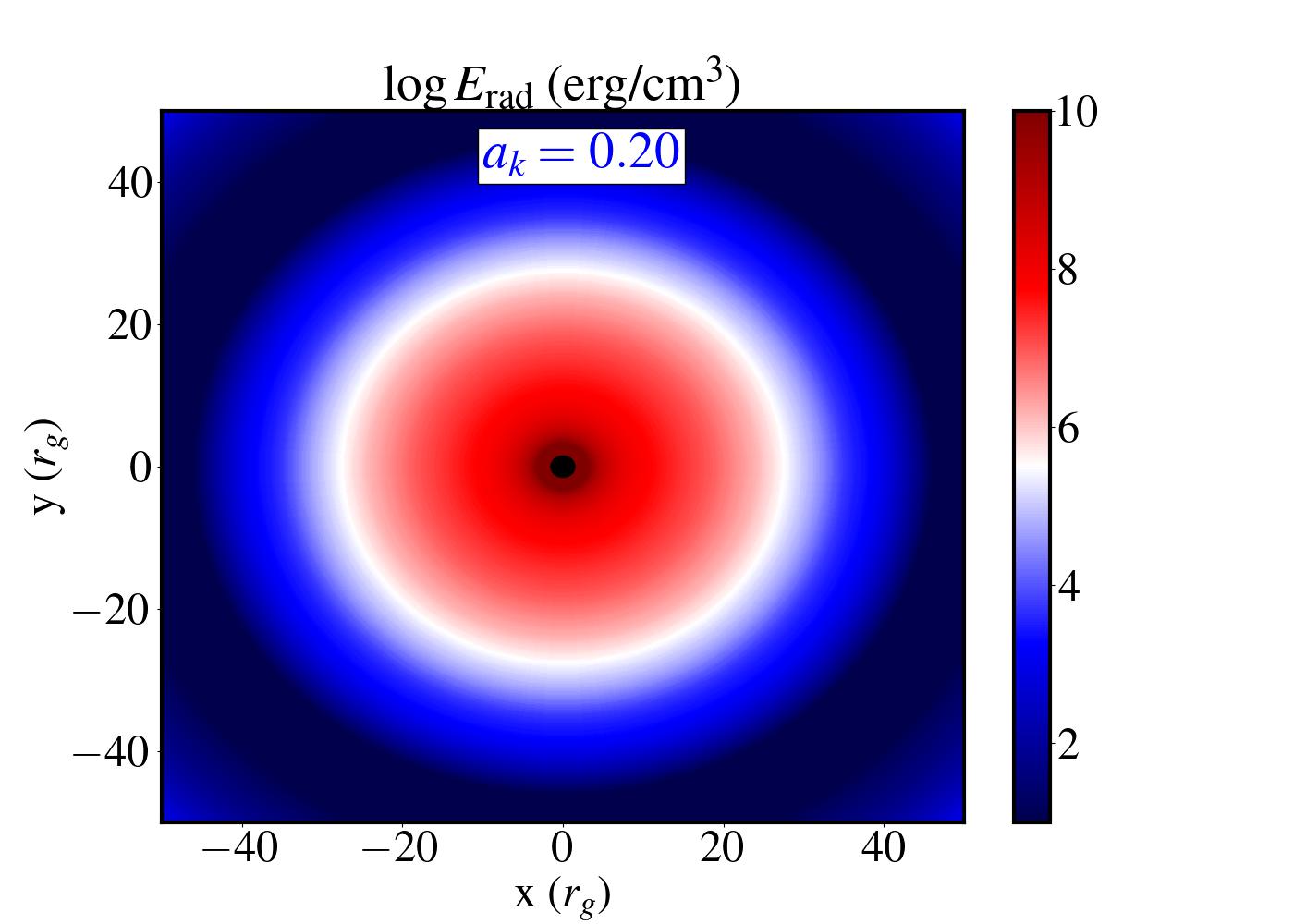} 
        \hskip -3.0mm
	\includegraphics[width=0.20\textwidth]{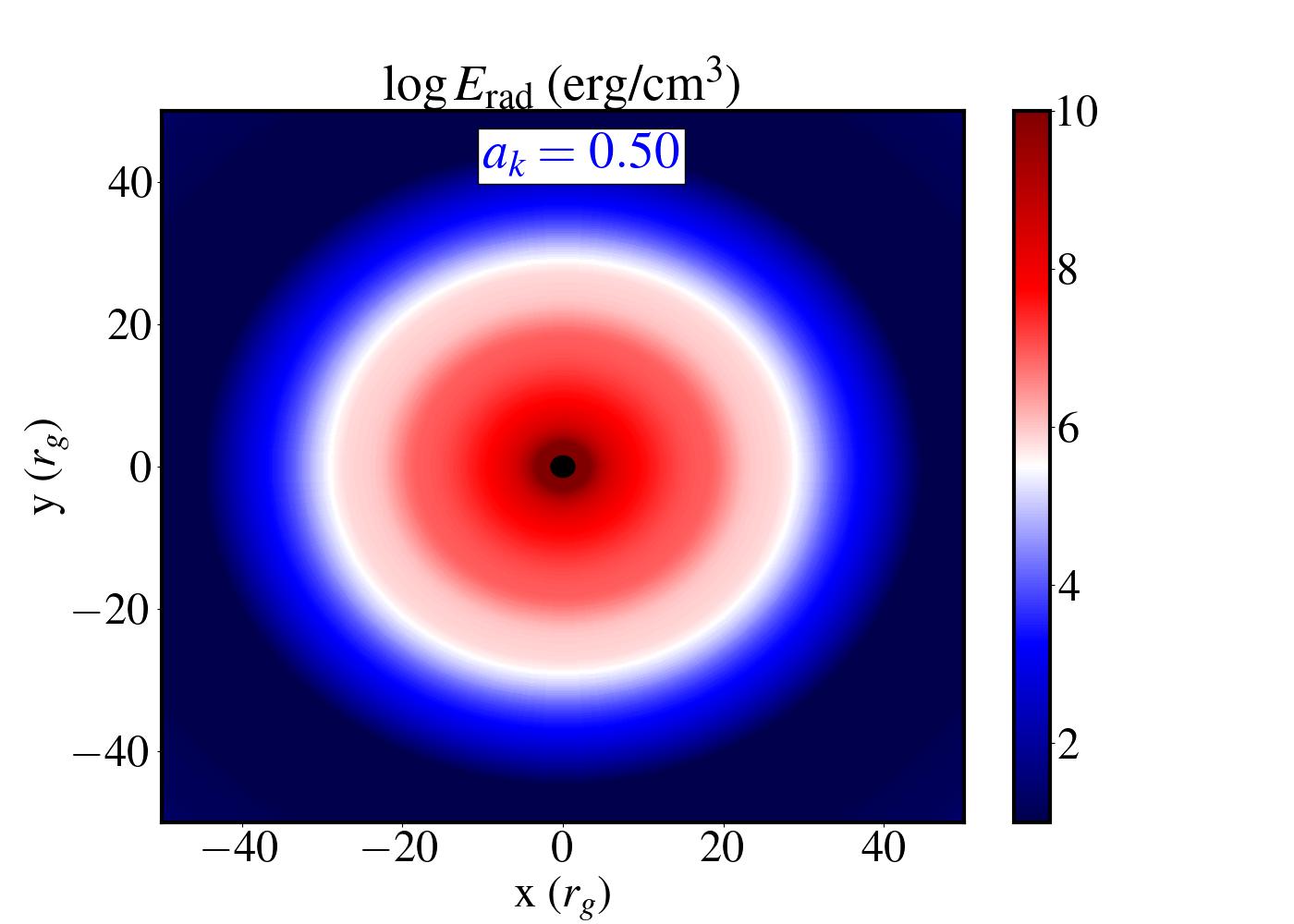} 
        \hskip -3.0mm
        \includegraphics[width=0.20\textwidth]{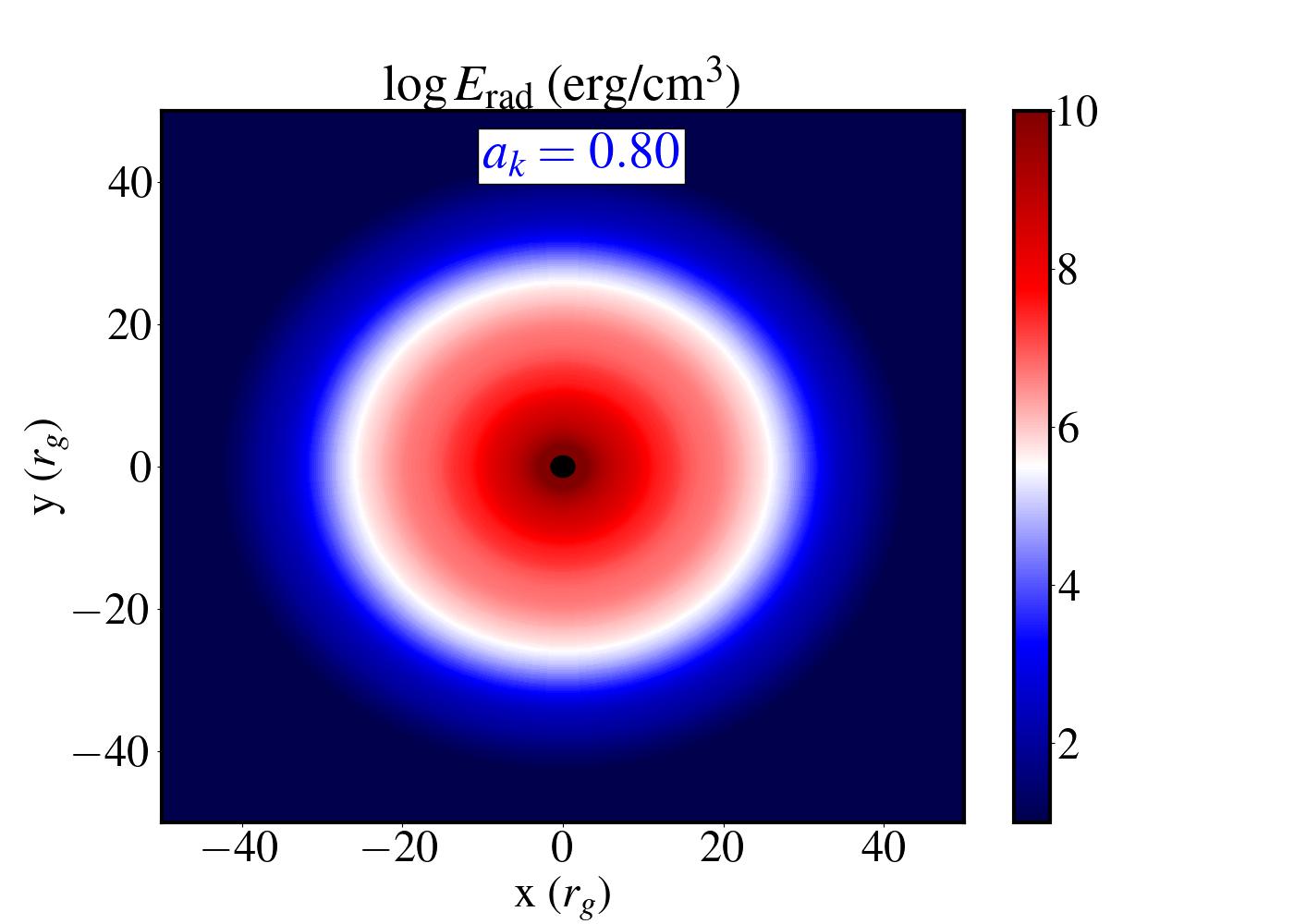} 
        \hskip -3.0mm
        \includegraphics[width=0.20\textwidth]{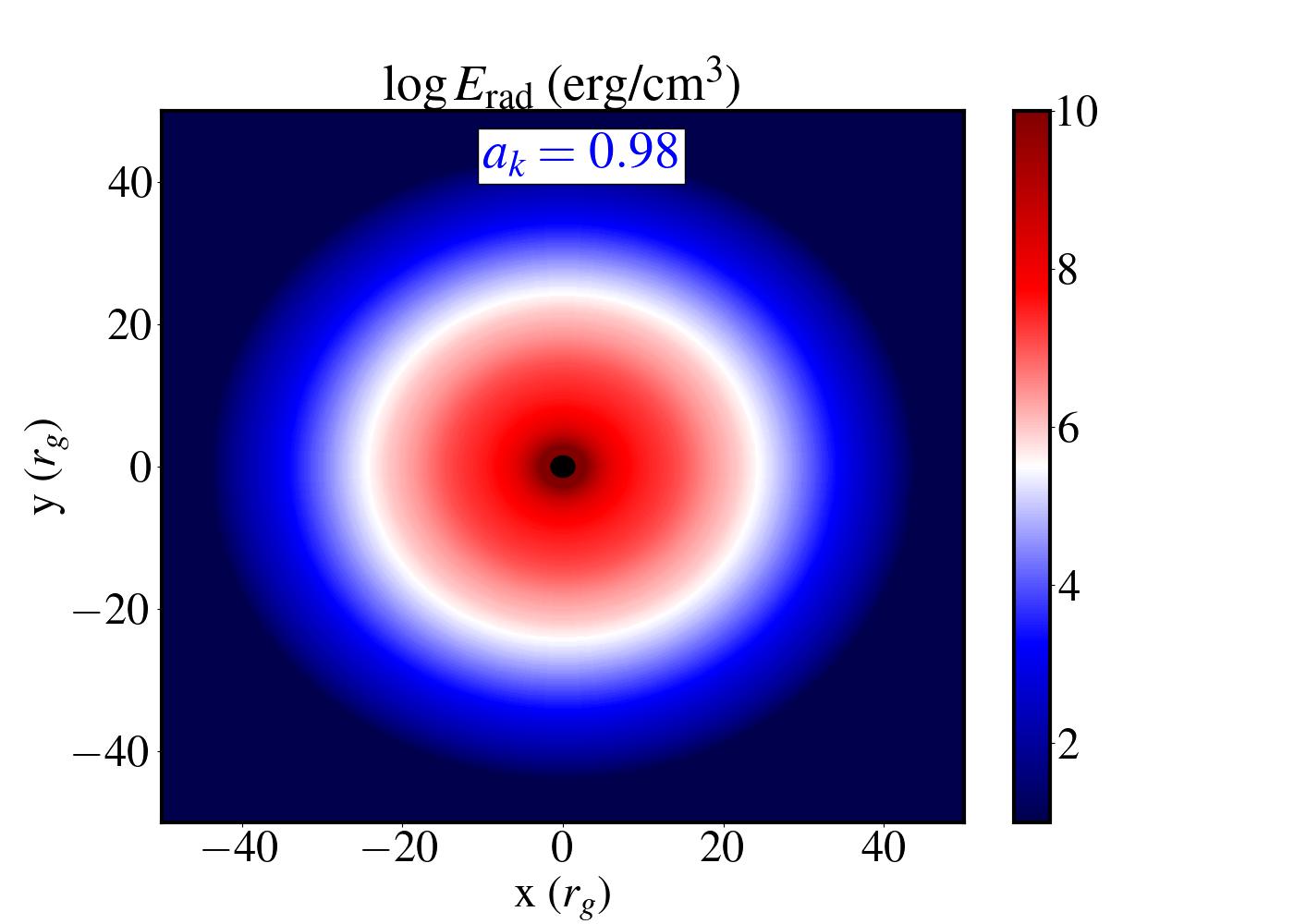} 
        \end{center}
	\caption{The distribution of azimuthal and time averaged of radiation energy density (\(E_{\rm rad}\)) is presented in the (\(r-z\)) plane using cylindrical coordinates in the upper panel and in the (\(x-y\)) equatorial plane using cartesian coordinates in the lower panel. We have fixed the black hole spin at values of \(a_k = 0.0, 0.20, 0.50, 0.80,\) and \(0.98\), with the time average between \(t = 5000 t_g\) to \(6000 t_g\). See the text for details.}
	\label{Figure_4}
\end{figure*}
%%%%%%%%%%%%%%%%%%%%%%%%%%%%%%%%%%%%%%%%%%%%%%%%%%%%

%%%%%%%%%%%%%%%%%%%%%%%%%%%%%%%%%%%%%%%%%%%%%%%%%%%
%%                        Figure 5
%%%%%%%%%%%%%%%%%%%%%%%%%%%%%%%%%%%%%%%%%%%%%%%%%%%
\begin{figure*}
	\begin{center}
        \includegraphics[width=0.32\textwidth]{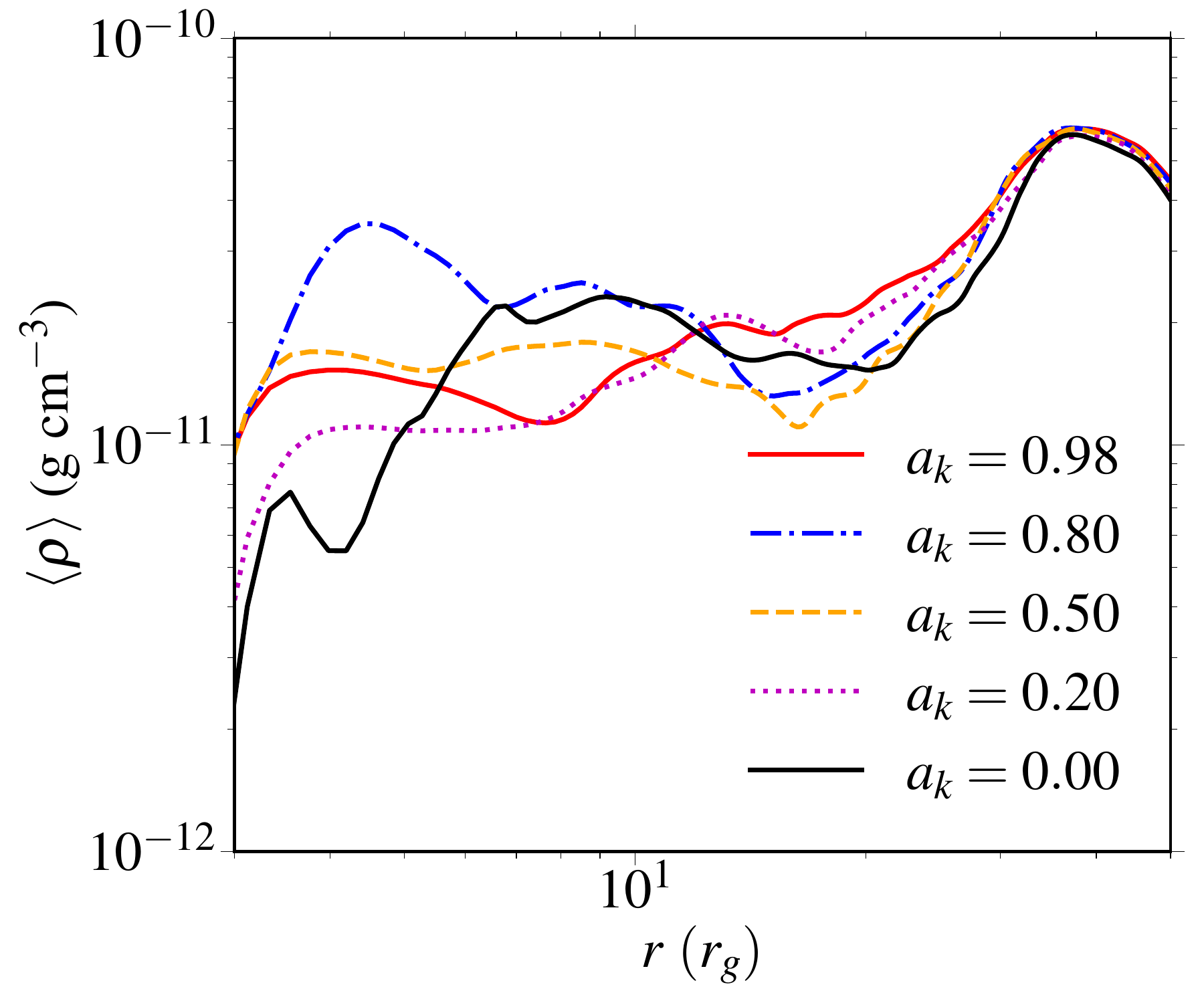} 
        \includegraphics[width=0.32\textwidth]{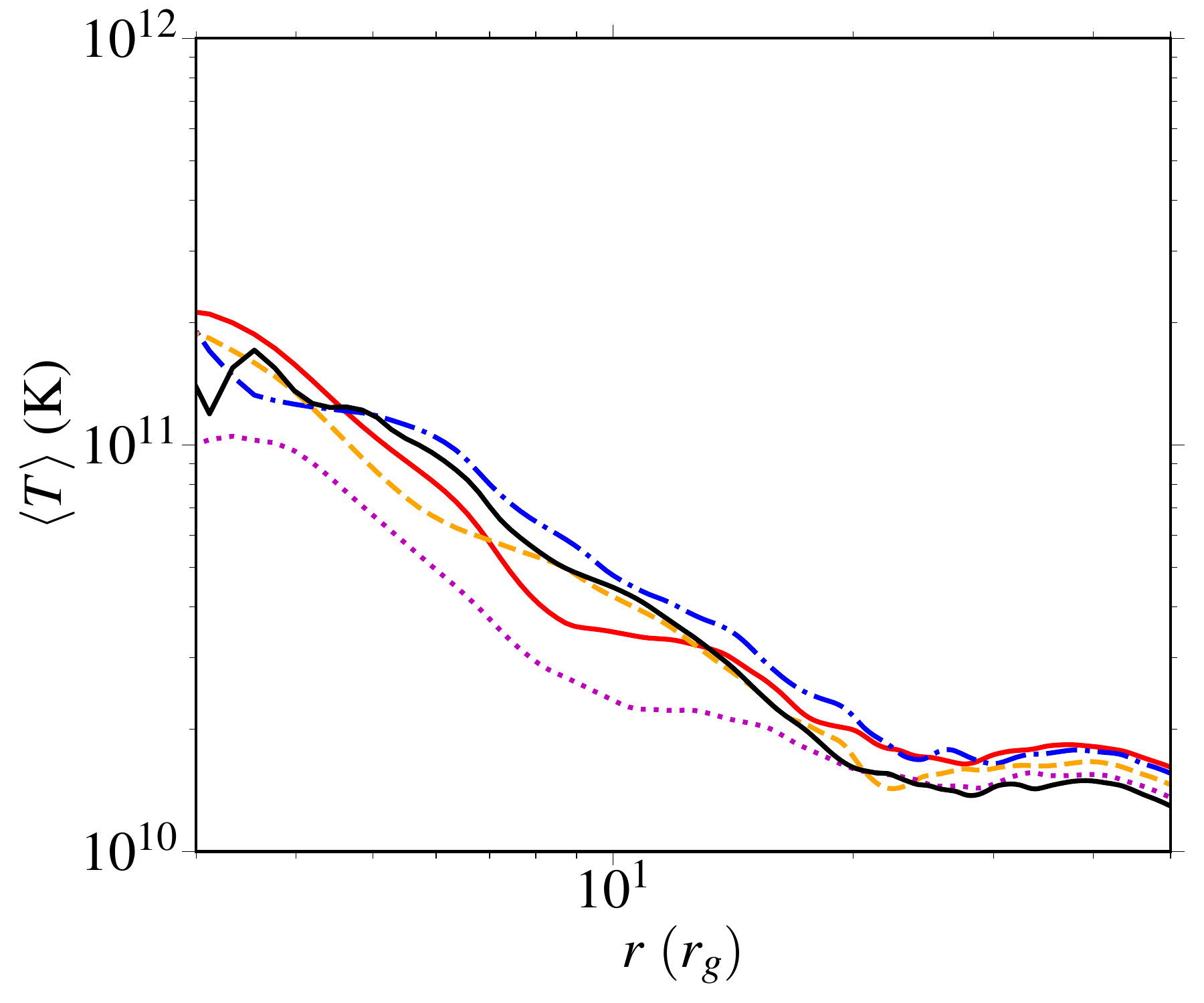} 
        \includegraphics[width=0.32\textwidth]{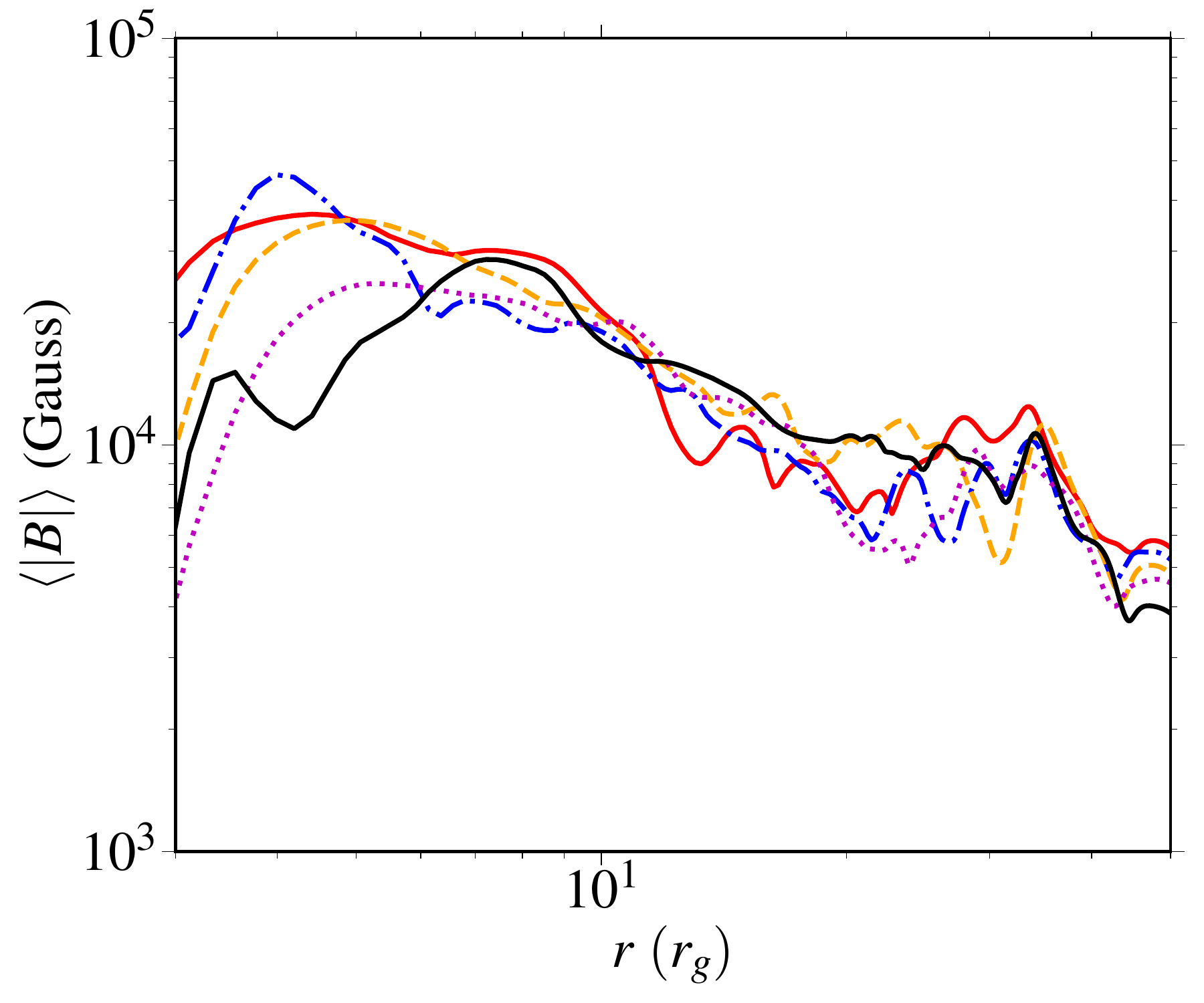} 
        \includegraphics[width=0.32\textwidth]{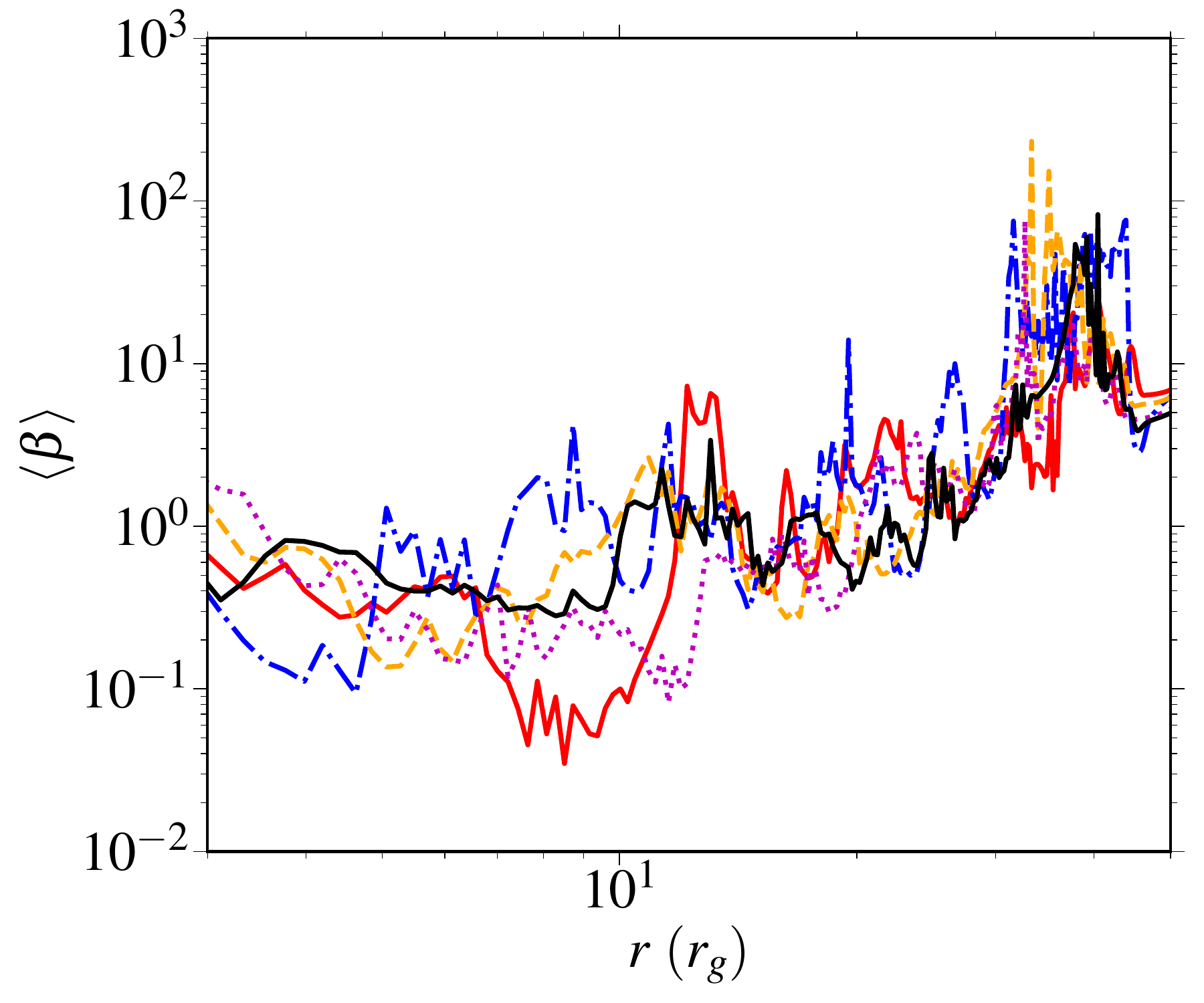} 
        \includegraphics[width=0.32\textwidth]{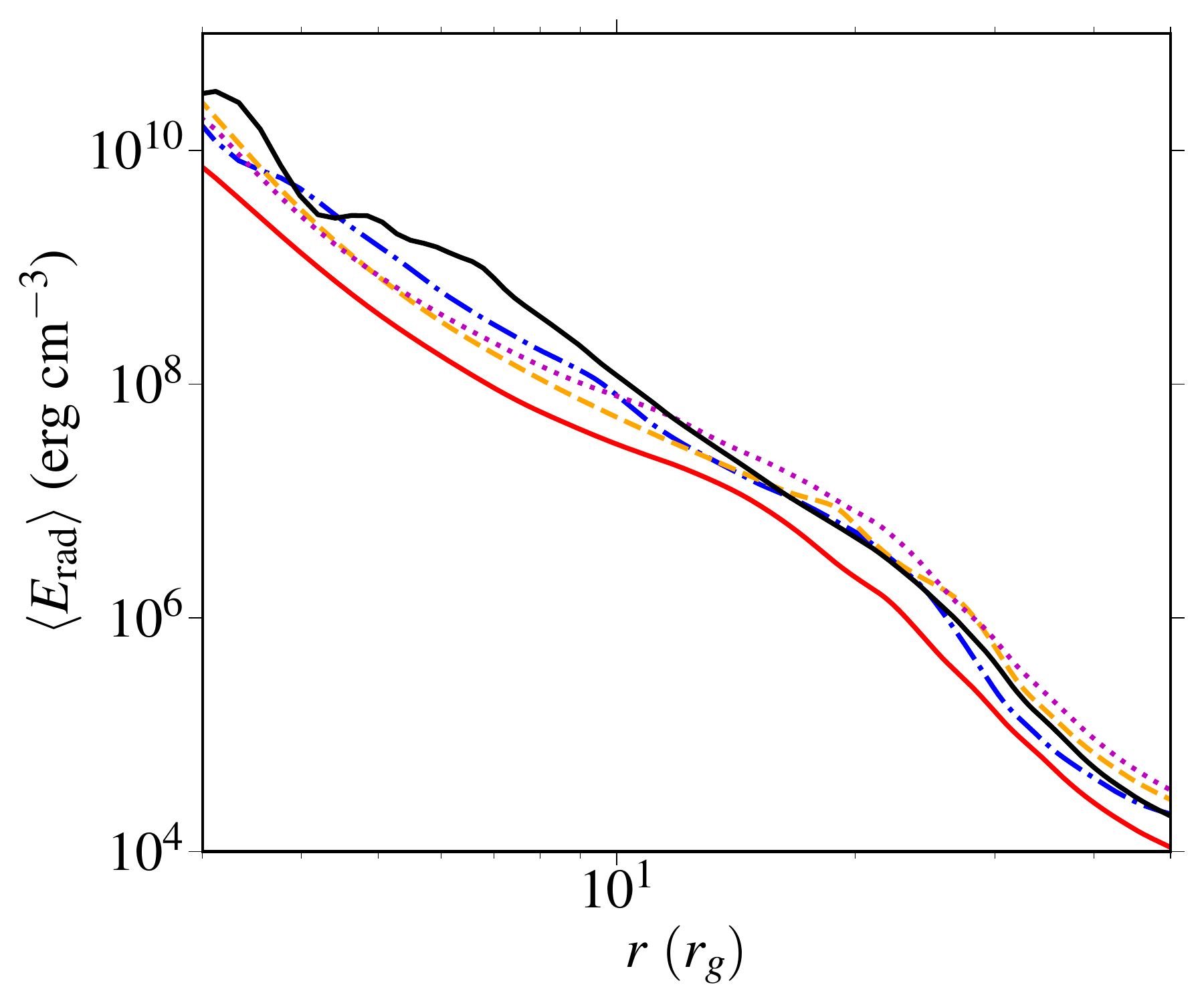} 
        \includegraphics[width=0.32\textwidth]{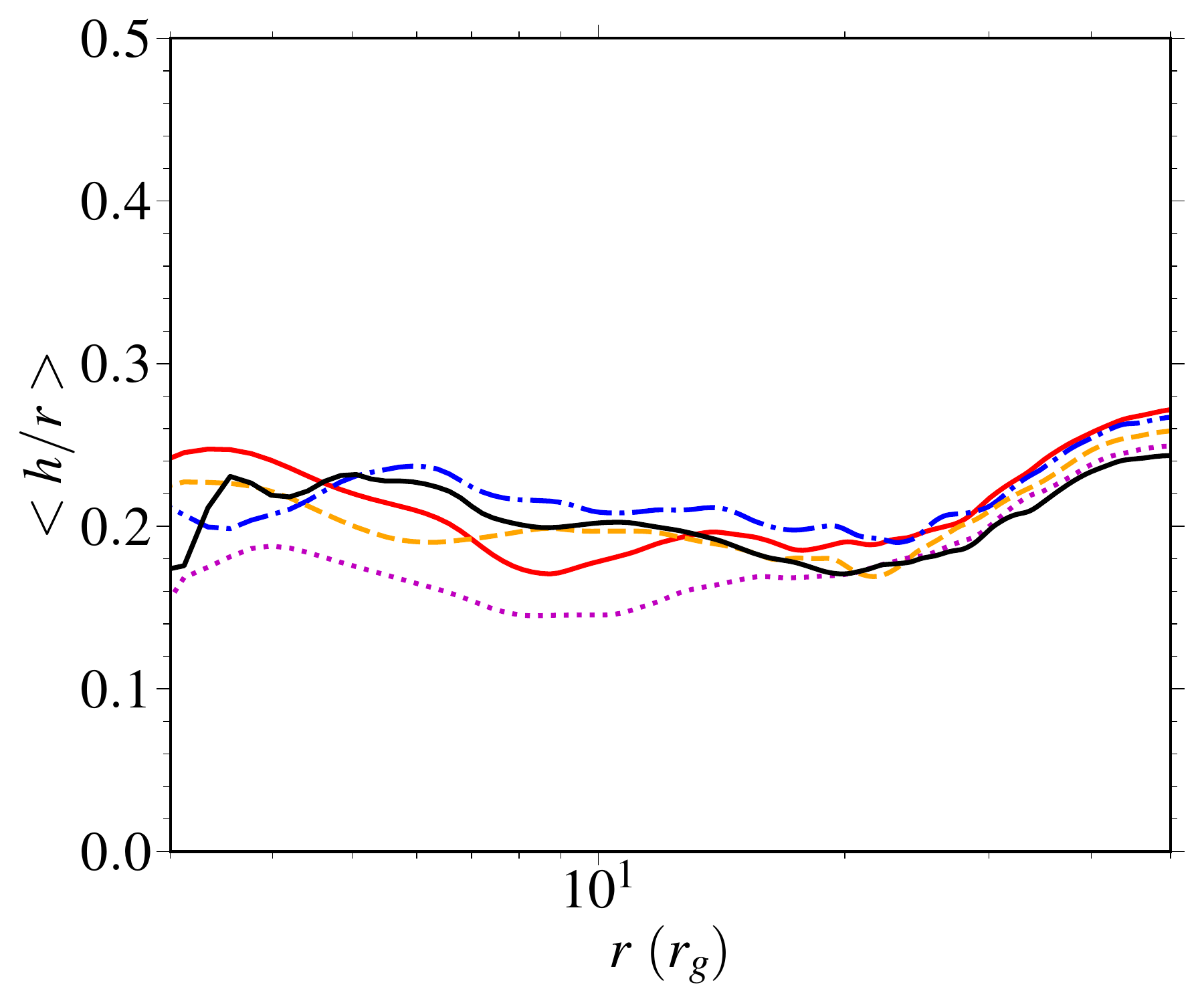} 
        \end{center}
	\caption{Radial profiles of vertically, azimuthally and density-weighted density $<\rho>$, temperature $<T>$, magnetic field $<|B|>$, plasma beta parameter $<\beta>$, radiation energy density $<E_{\rm rad}>$ and disk scale height $<h/r>$, respectively. The time average is taken from \(t = 5000 t_g\) to \(6000 t_g\).}
	\label{Figure_5}
\end{figure*}
%%%%%%%%%%%%%%%%%%%%%%%%%%%%%%%%%%%%%%%%%%%%%%%%%%%%

%%%%%%%%%%%%%%%%%%%%%%%%%%%%%%%%%%%%%%%%%%%%%%%%%%%
%%                        Figure 6
%%%%%%%%%%%%%%%%%%%%%%%%%%%%%%%%%%%%%%%%%%%%%%%%%%%
\begin{figure*}
	\begin{center}
        \includegraphics[width=0.40\textwidth]{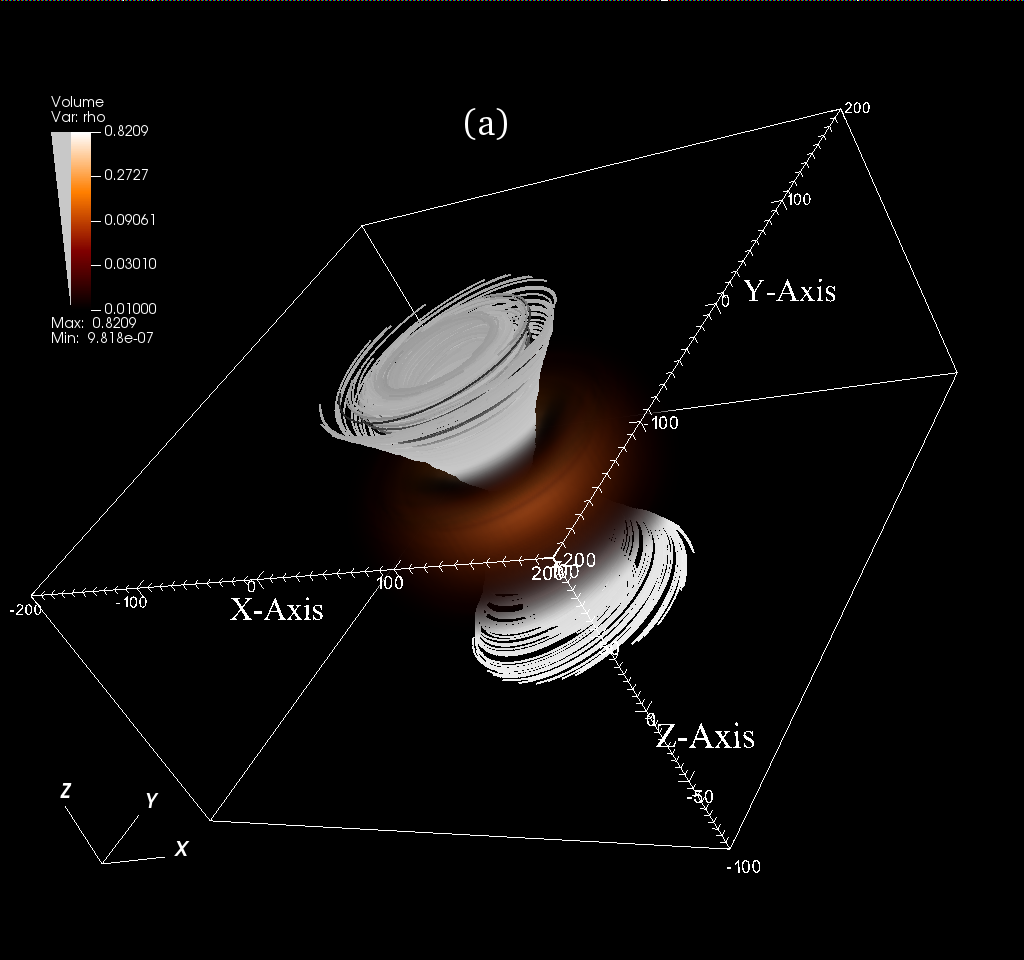} 
        \includegraphics[width=0.40\textwidth]{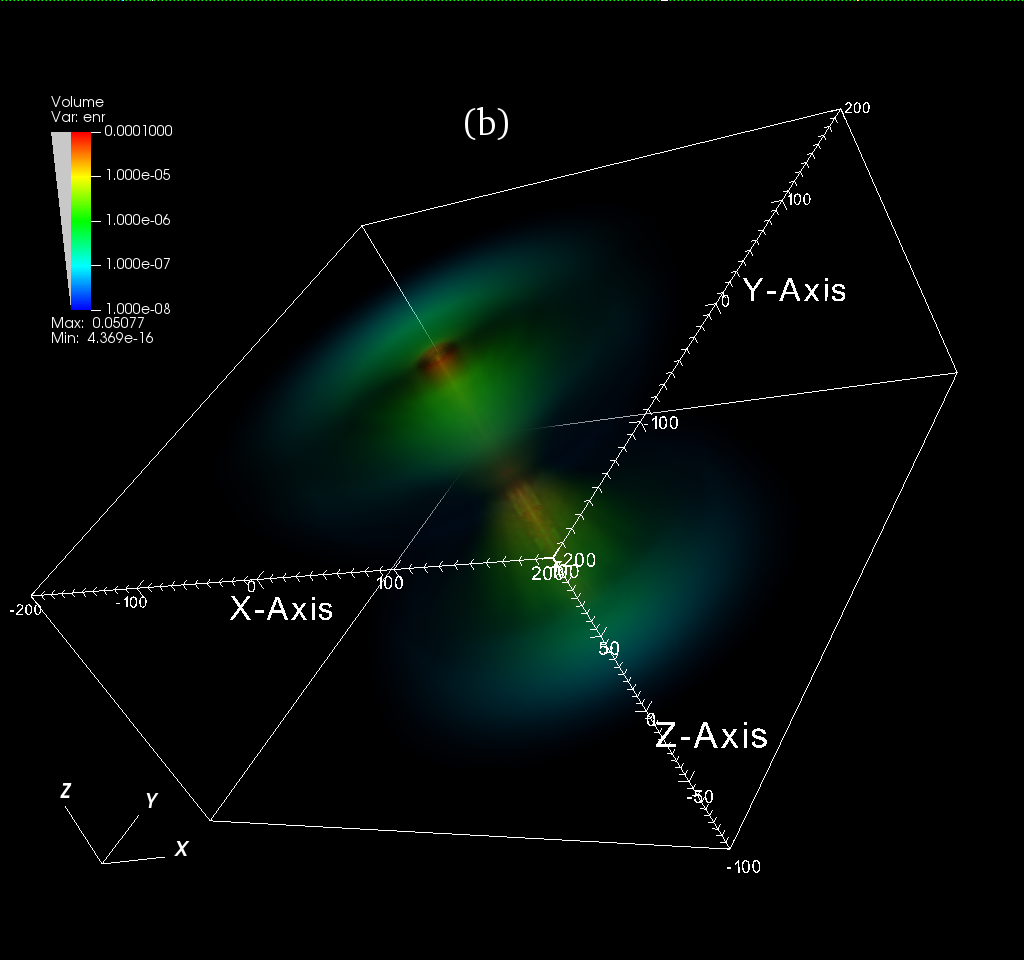} 
	\end{center}
	\caption{Volume rendering of $(a)$ density with the magnetic field lines  and $(b)$ the radiation energy density at time $t = 5000t_g$. See the text for details.}
	\label{Figure_6}
\end{figure*}
%%%%%%%%%%%%%%%%%%%%%%%%%%%%%%%%%%%%%%%%%%%%%%%%%%%%

%%%%%%%%%%%%%%%%%%%%%%%%%%%%%%%%%%%%%%%%%%%%%%%%%%%
%%                        Figure 7
%%%%%%%%%%%%%%%%%%%%%%%%%%%%%%%%%%%%%%%%%%%%%%%%%%%
\begin{figure*}
	\begin{center}
        \includegraphics[width=0.80\textwidth]{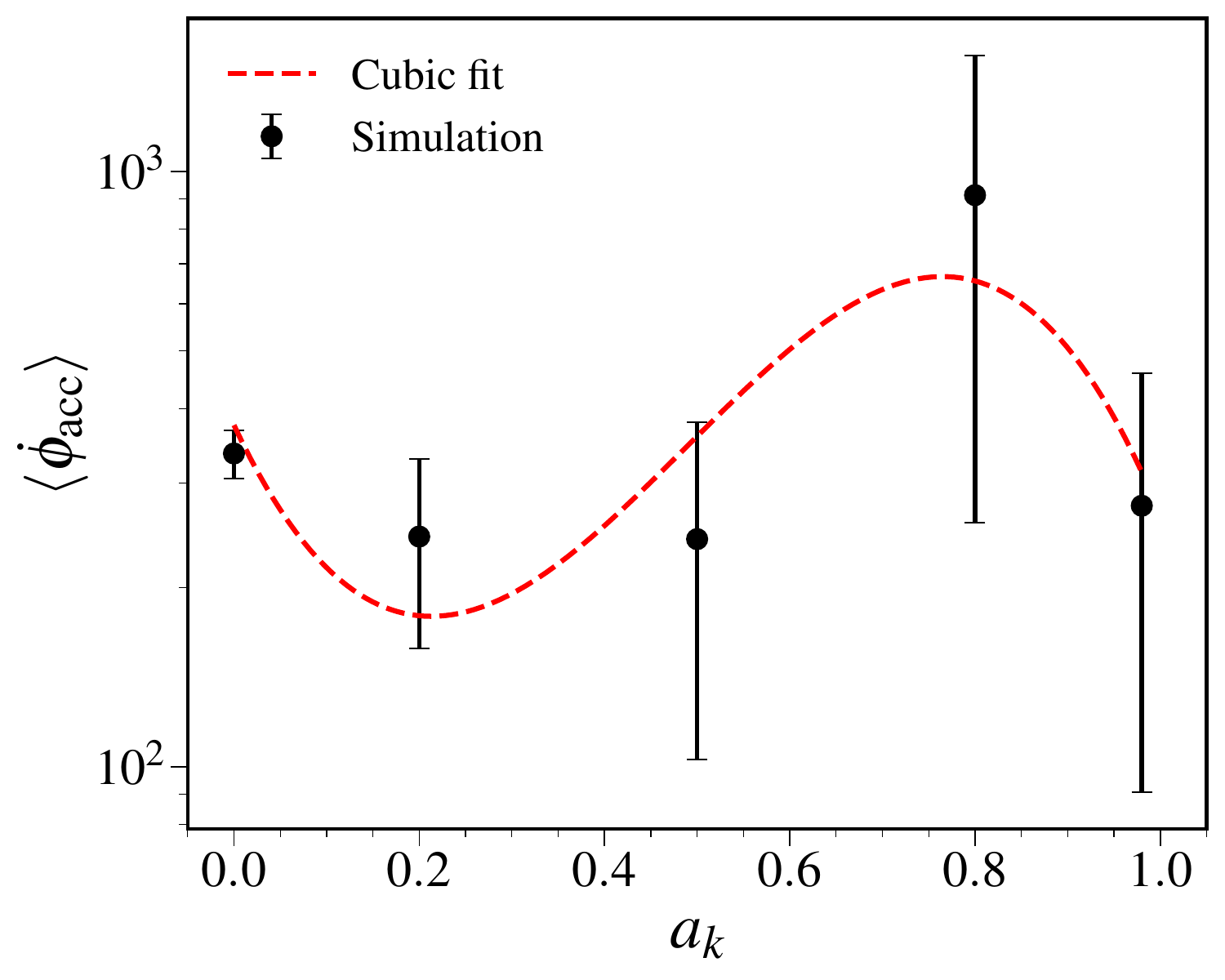} 
	\end{center}
	\caption{Time averaged saturated normalized magnetic flux ($<\dot{\phi}_{\rm acc}>$) as a function of black hole spin $(a_k)$. Black points are the simulation results with $1\sigma$ temporal error bars. The dashed red curves is a third-order polynomial fit. See the text for details.}
	\label{Figure_7}
\end{figure*}
%%%%%%%%%%%%%%%%%%%%%%%%%%%%%%%%%%%%%%%%%%%%%%%%%%%%

%%%%%%%%%%%%%%%%%%%%%%%%%%%%%%%%%%%%%%%%%%%%%%%%%%%
%%                        Figure 8
%%%%%%%%%%%%%%%%%%%%%%%%%%%%%%%%%%%%%%%%%%%%%%%%%%%
\begin{figure*}
	\begin{center}
        \hskip -2.5mm
        \includegraphics[width=0.20\textwidth]{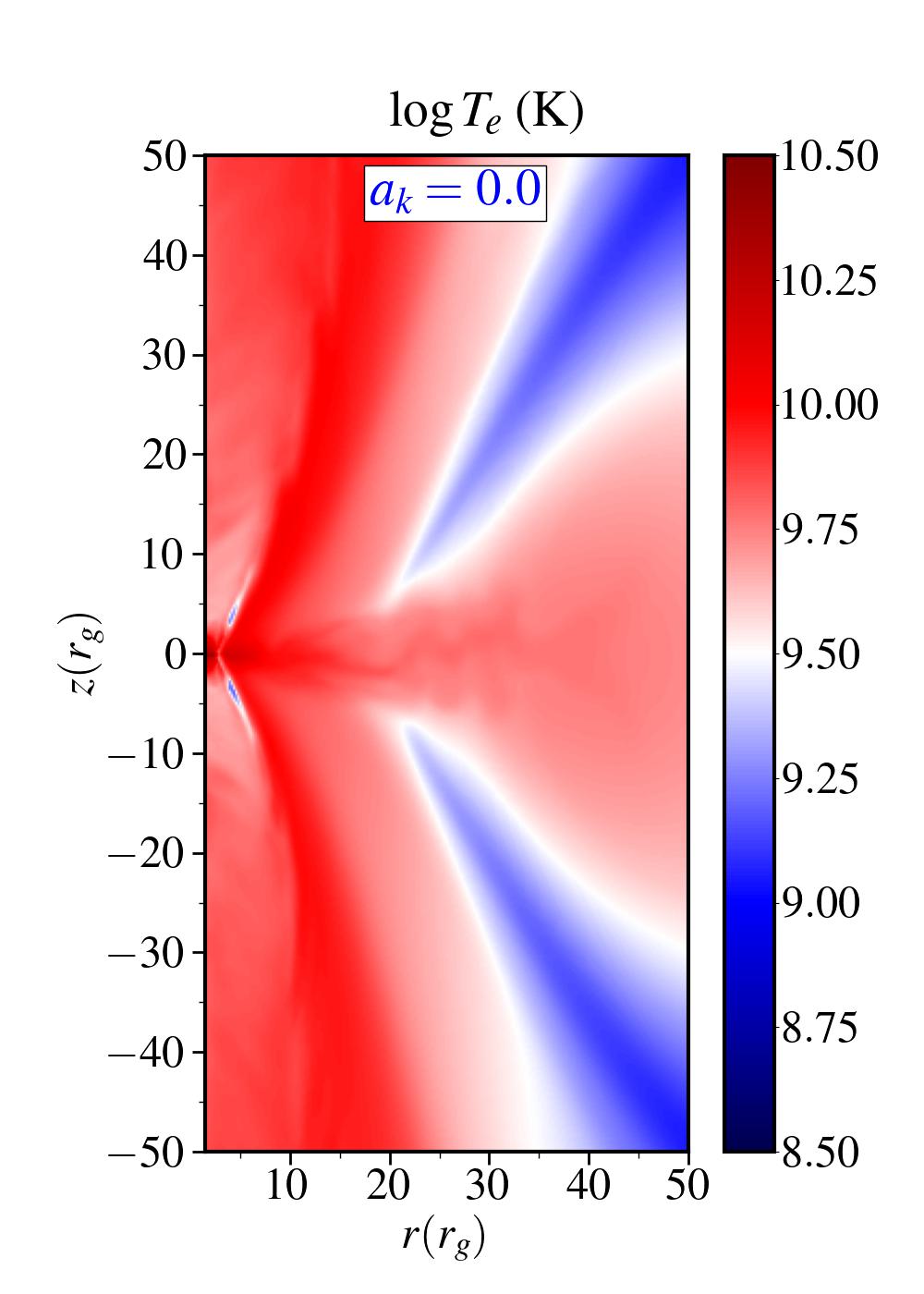} 
        \hskip -2.5mm
        \includegraphics[width=0.20\textwidth]{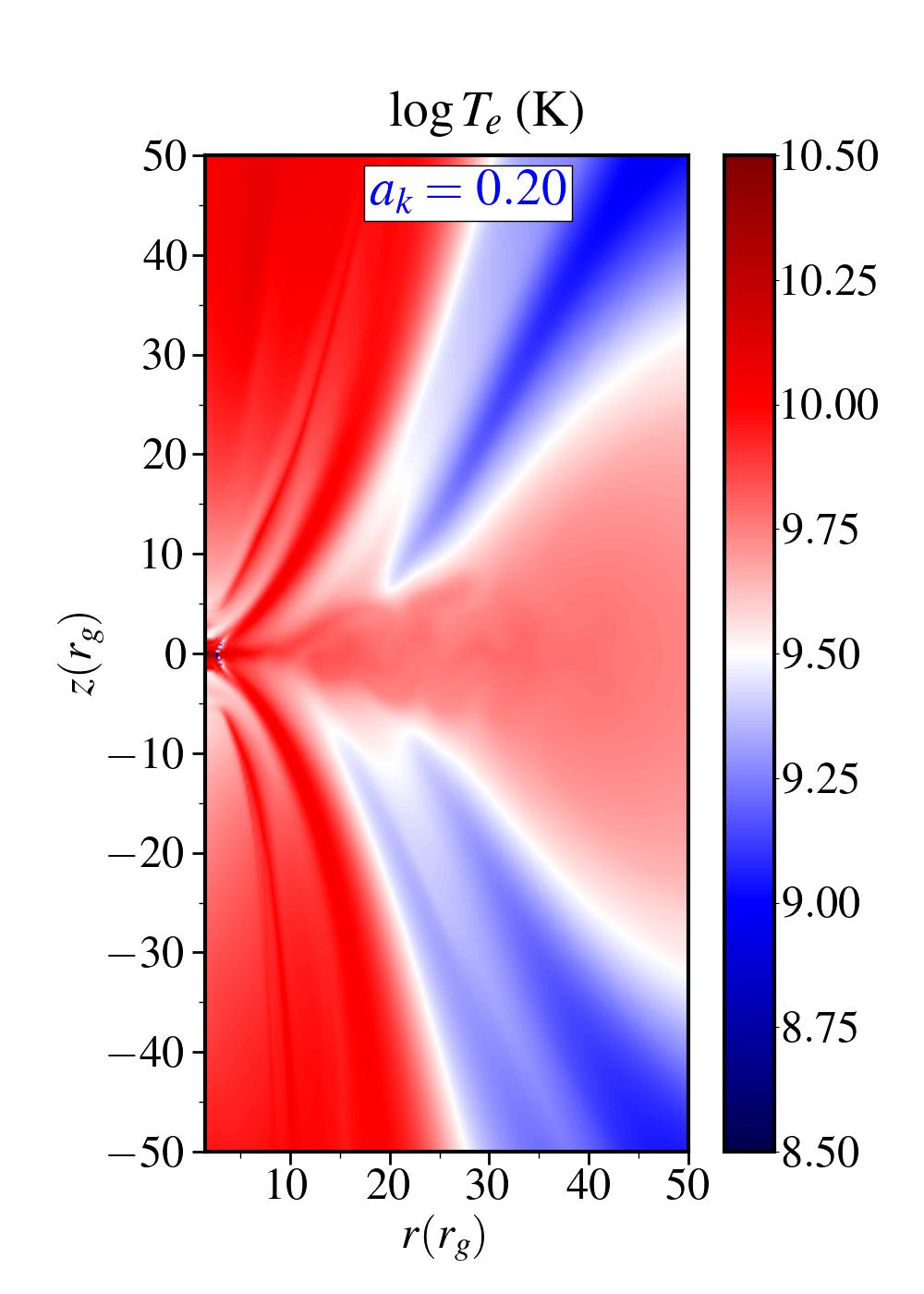} 
        \hskip -2.5mm
	\includegraphics[width=0.20\textwidth]{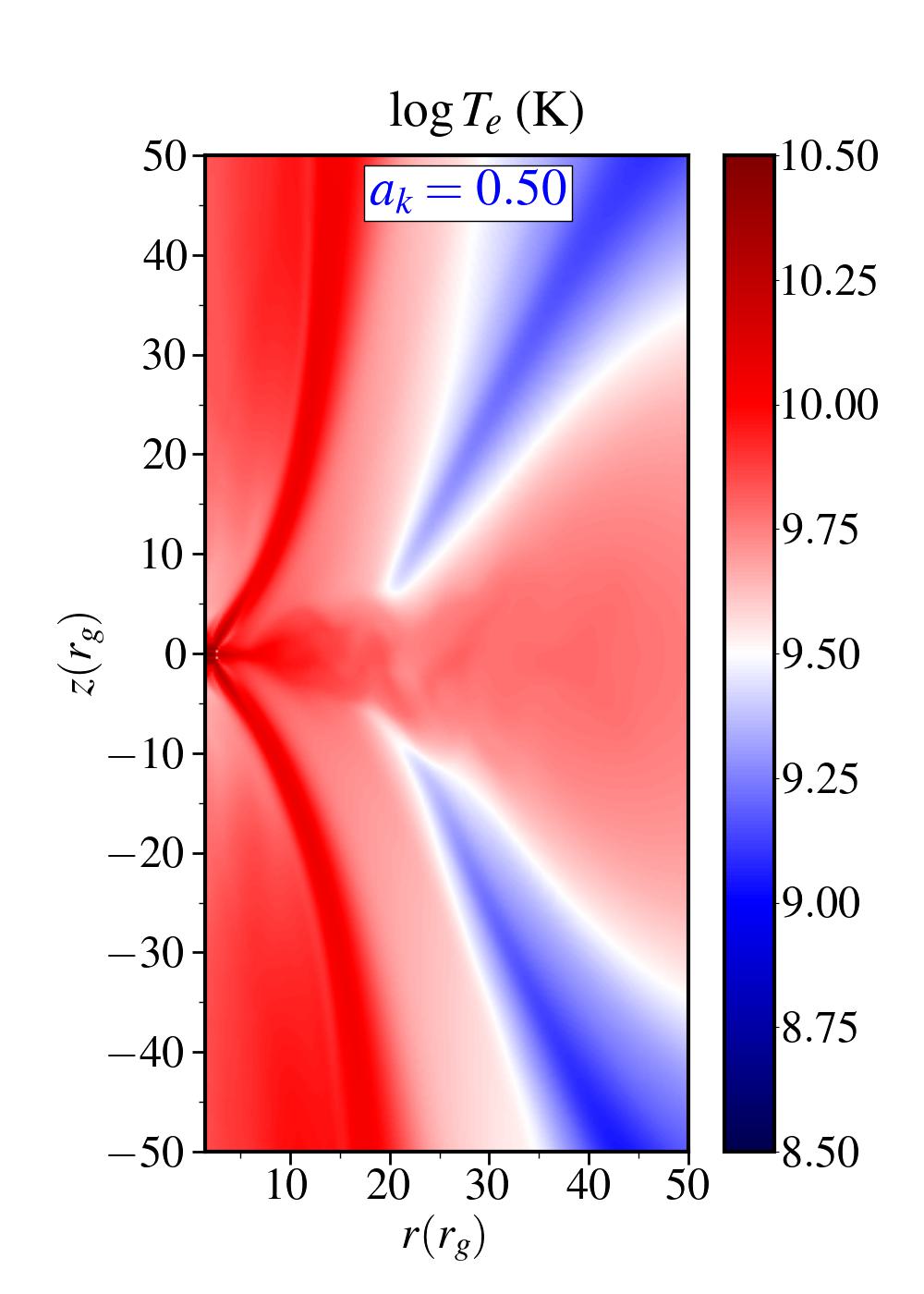} 
        \hskip -2.5mm
        \includegraphics[width=0.20\textwidth]{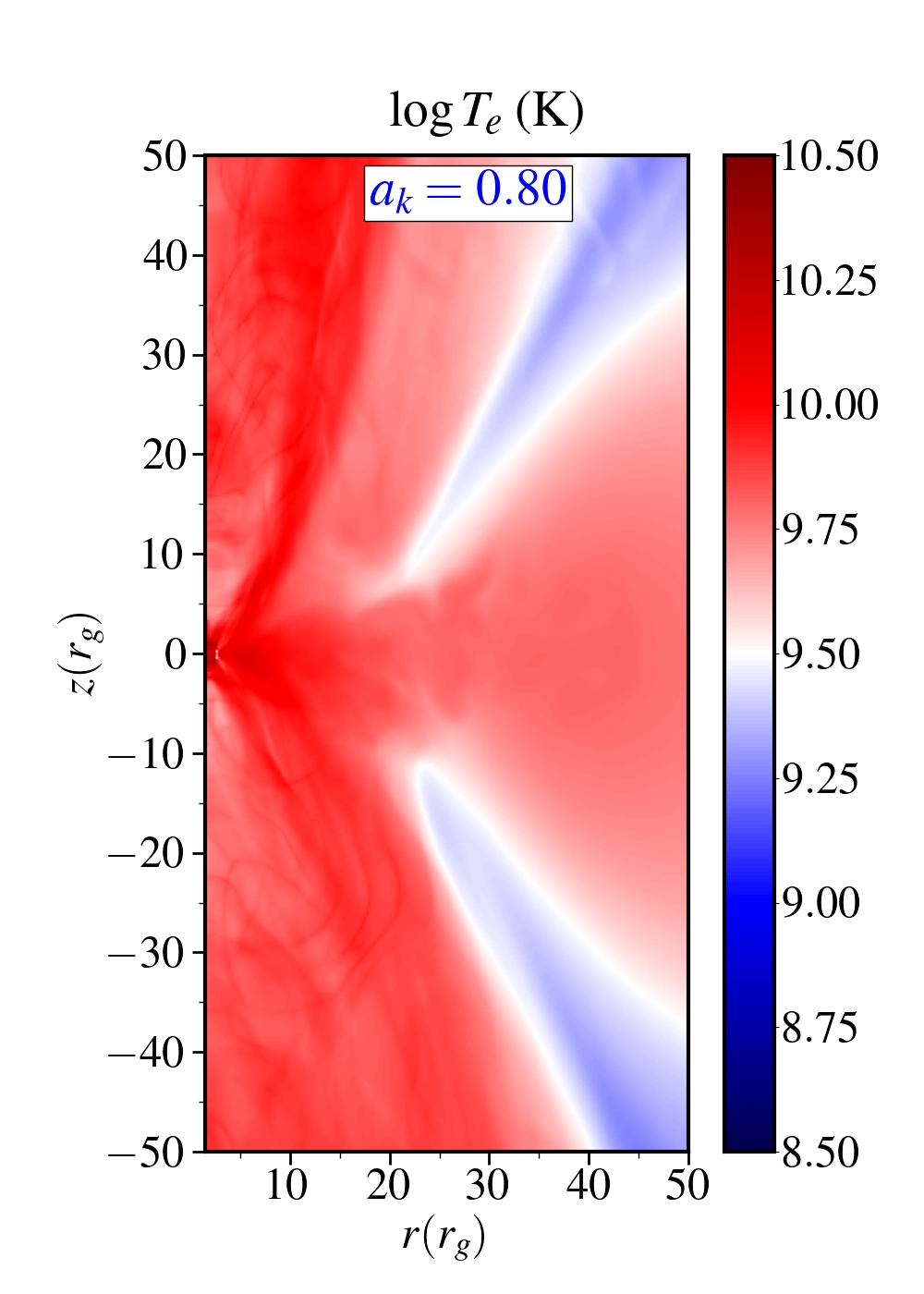} 
        \hskip -2.5mm
        \includegraphics[width=0.20\textwidth]{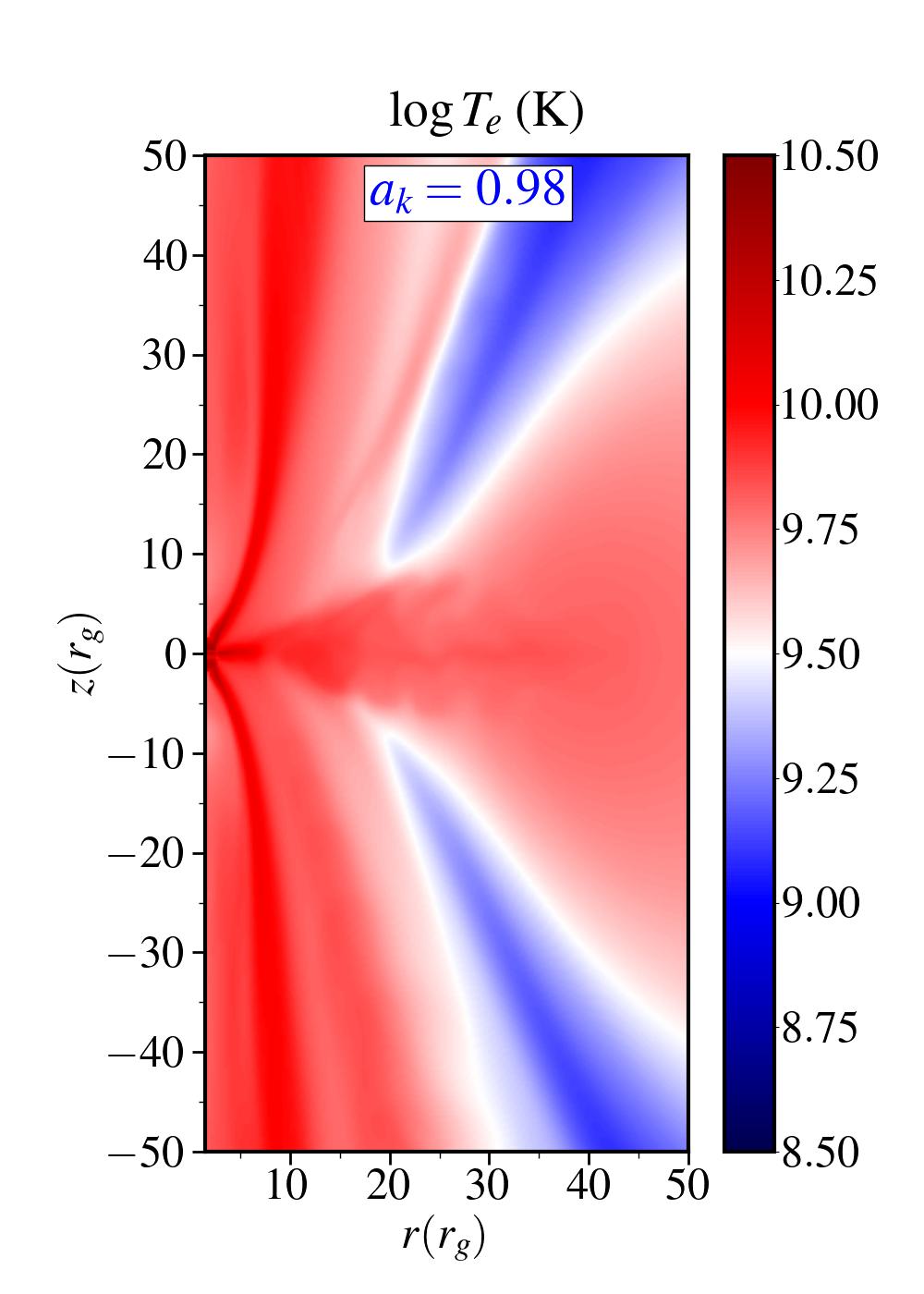} 
        
        \hskip -2.5mm 
        \includegraphics[width=0.20\textwidth]{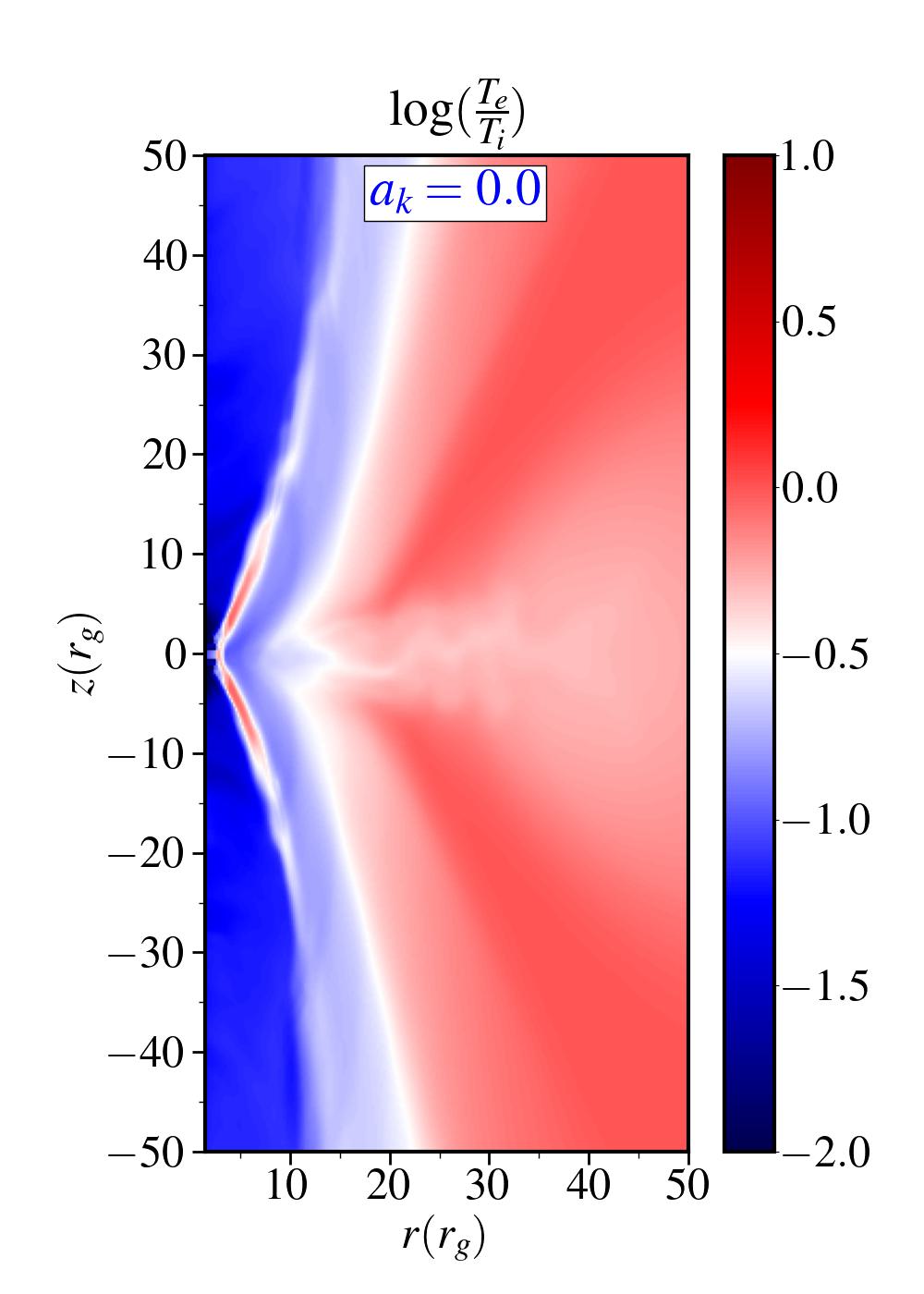} 
        \hskip -2.5mm
        \includegraphics[width=0.20\textwidth]{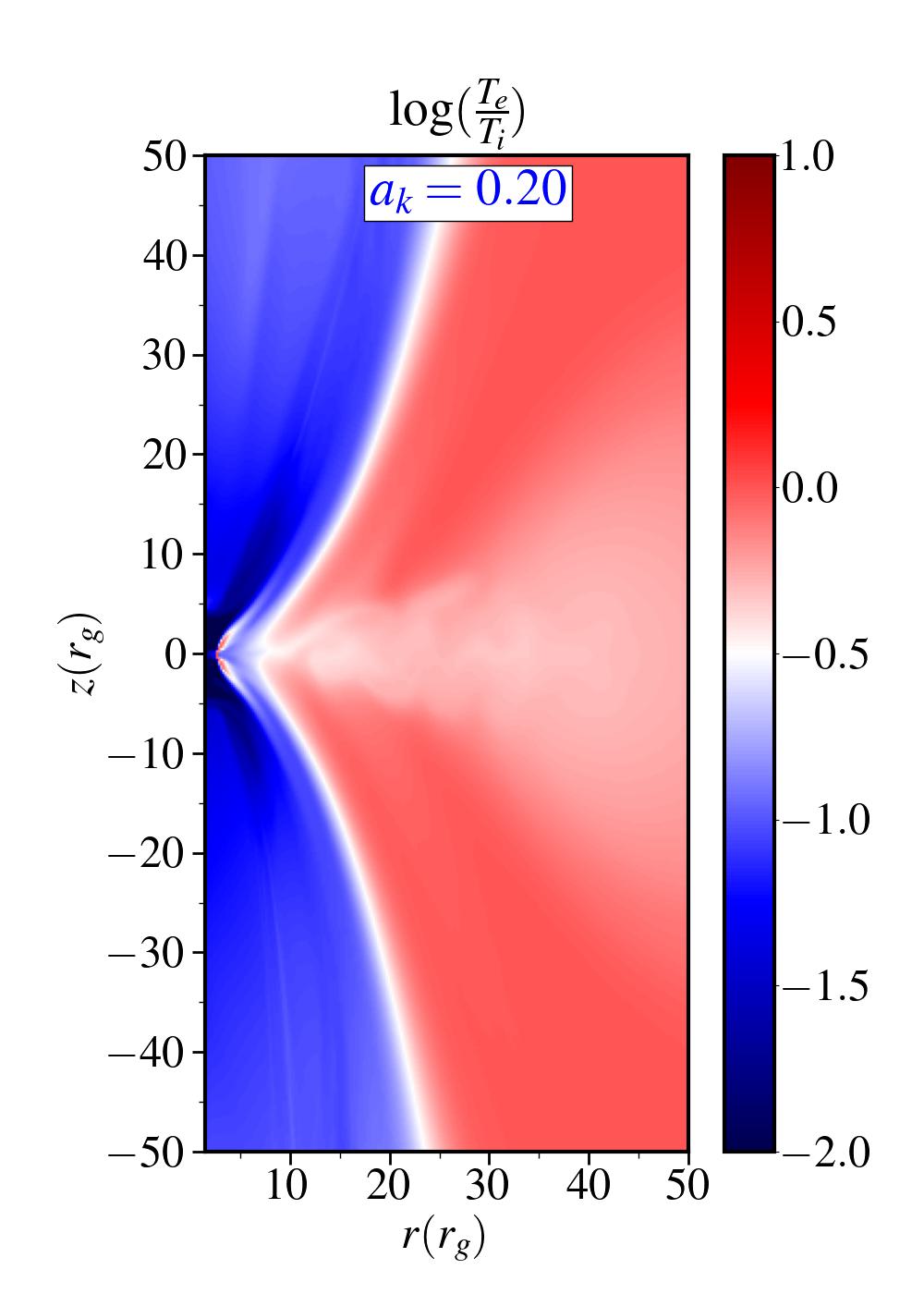} 
        \hskip -2.5mm
	\includegraphics[width=0.20\textwidth]{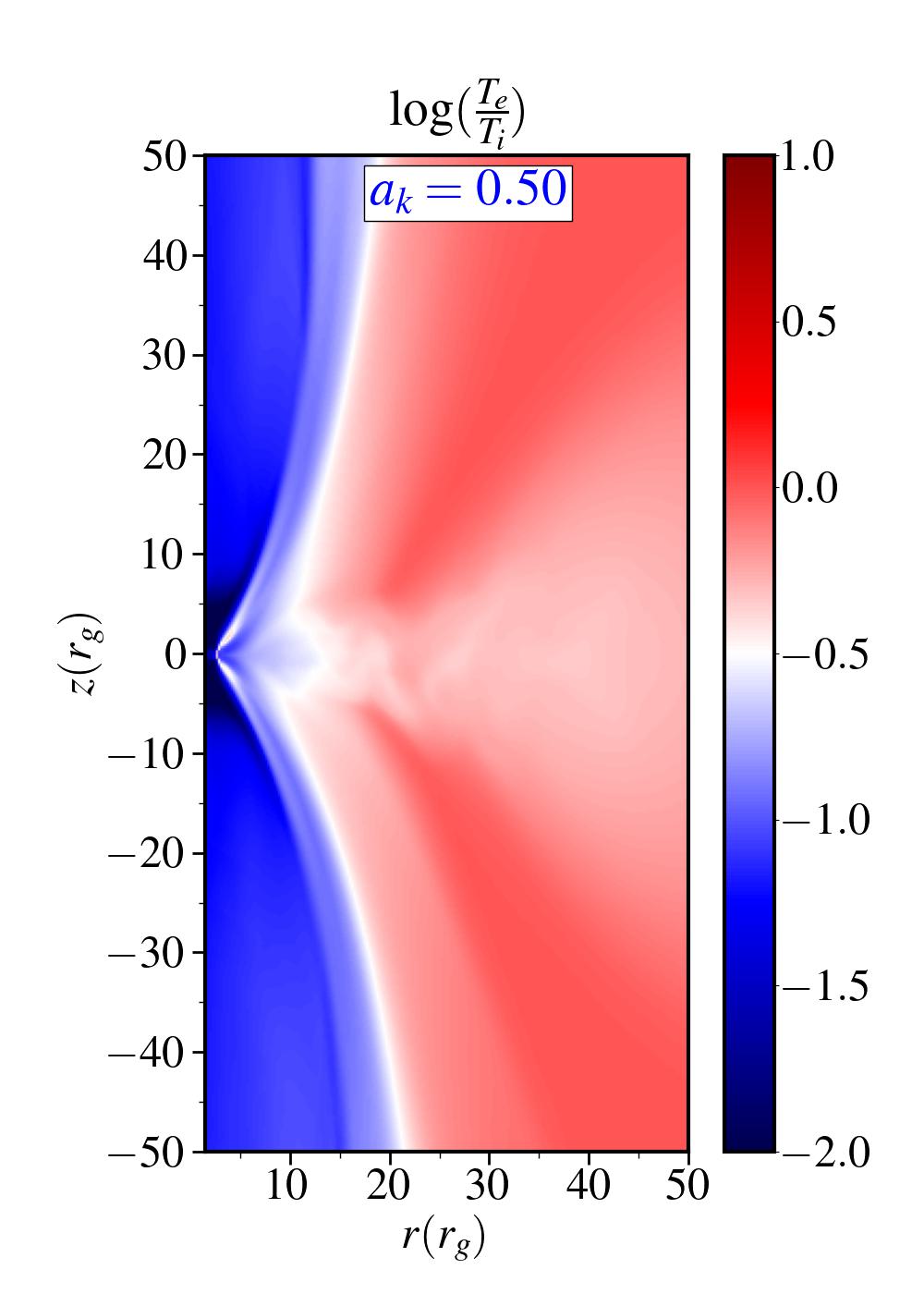} 
        \hskip -2.5mm
        \includegraphics[width=0.20\textwidth]{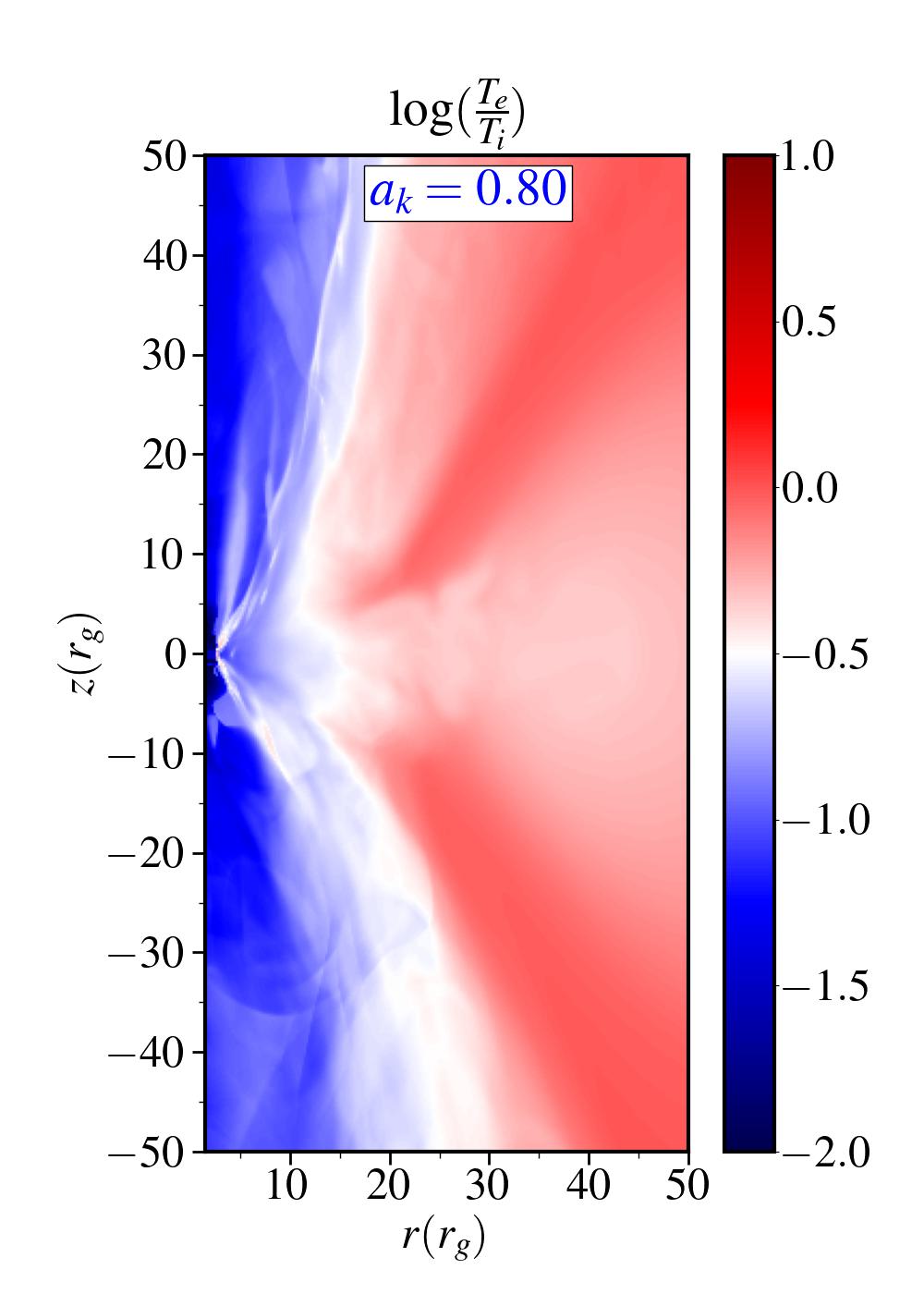} 
        \hskip -2.5mm
        \includegraphics[width=0.20\textwidth]{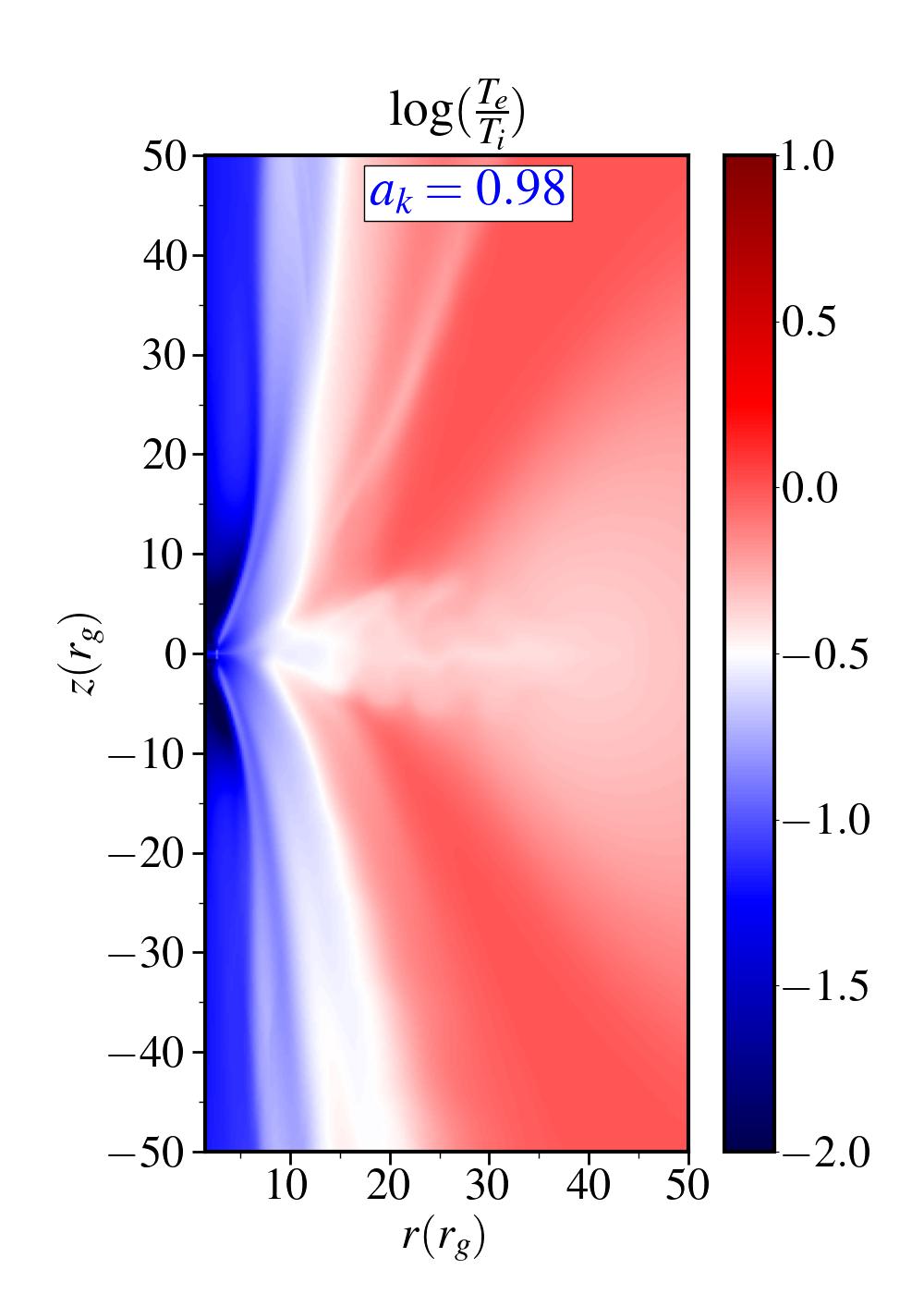} 
        \end{center}
	\caption{Distribution of azimuthal and time averaged of electron temperature ($T_e$) and ratio of the electron to ion temperature ($T_e/T_i$) in the upper row and lower row, respectively. Here, we fix the black hole spin as $a_k = 0.0, 0.20, 0.50, 0.80, 0.98$, with the time average between \(t = 5000 t_g\) to \(6000 t_g\). See the text for details.}
	\label{Figure_8}
\end{figure*}
%%%%%%%%%%%%%%%%%%%%%%%%%%%%%%%%%%%%%%%%%%%%%%%%%%%%

%%%%%%%%%%%%%%%%%%%%%%%%%%%%%%%%%%%%%%%%%%%%%%%%%%%
%%                        Figure 9
%%%%%%%%%%%%%%%%%%%%%%%%%%%%%%%%%%%%%%%%%%%%%%%%%%%
\begin{figure*}
	\begin{center}
        \includegraphics[width=0.48\textwidth]{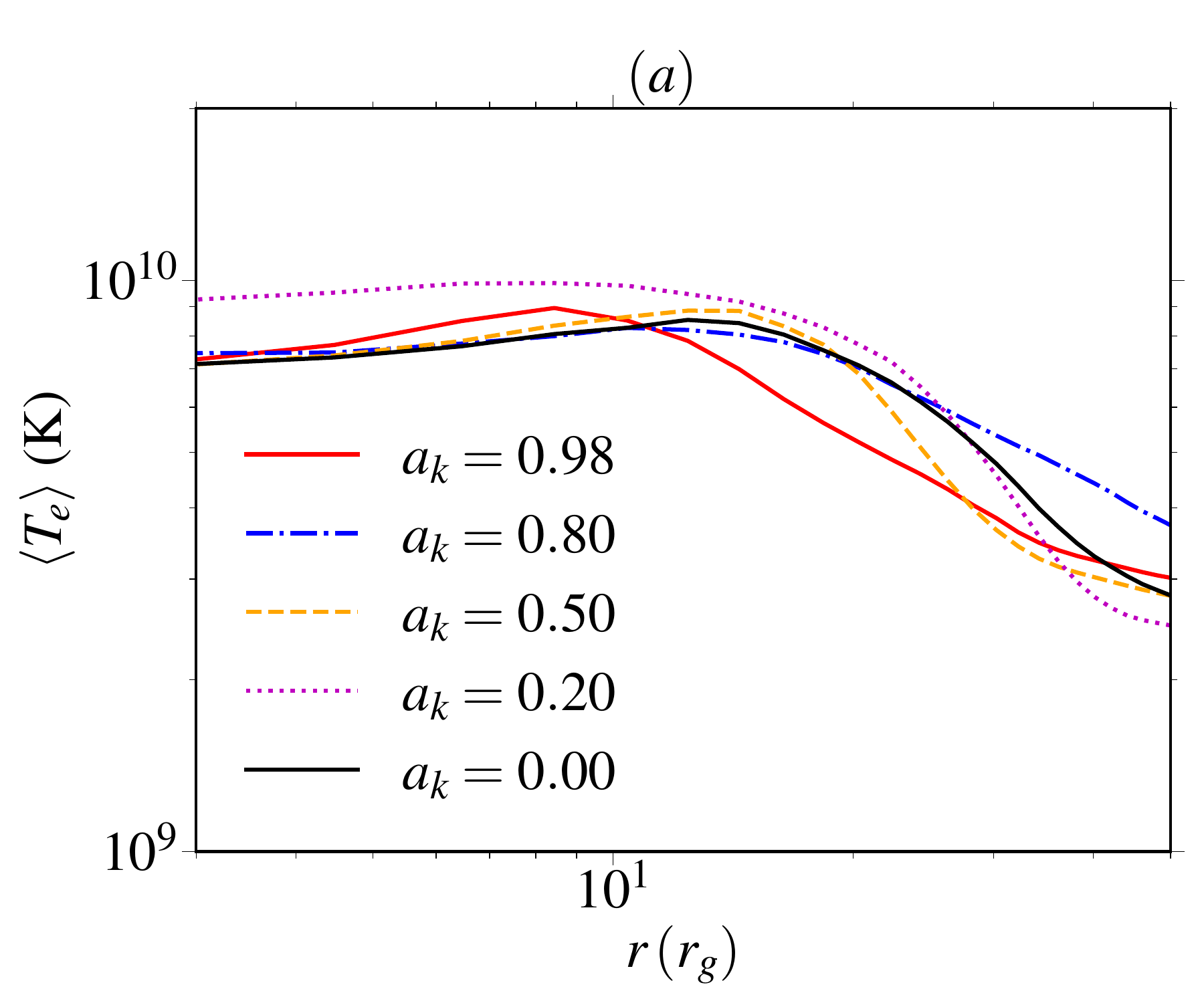} 
        \includegraphics[width=0.48\textwidth]{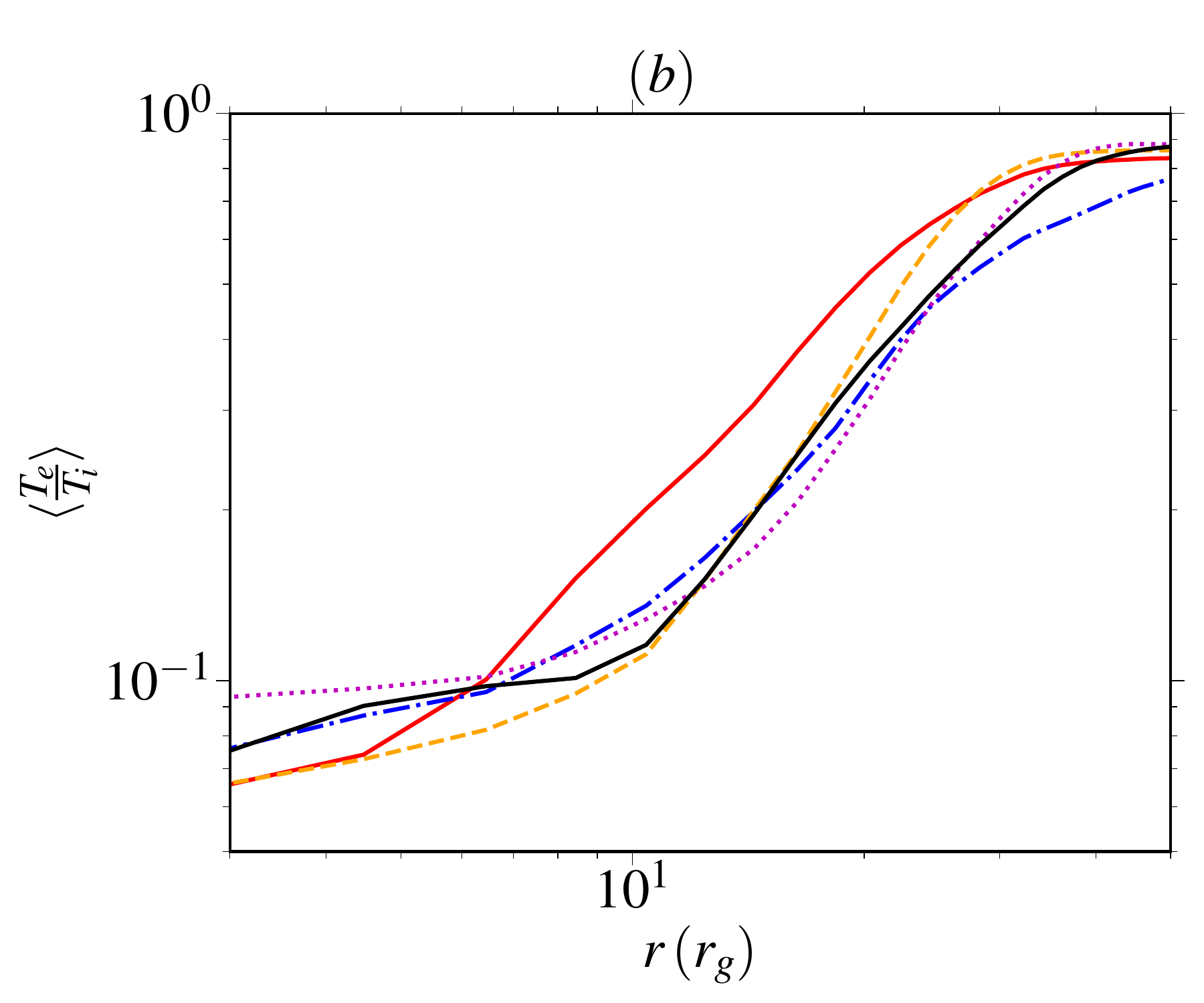} 
	\end{center}
	\caption{Radial variation of of vertically, azimuthally and density-weighted (a) electron temperature ($T_e$) and (b) ratio of the electron to ion temperature ($T_e/T_i$) for different black hole spin $a_k$ = 0.0, 0.20, 0.50, 0.80, 0.98. The time average is taken from \(t = 5000 t_g\) to \(6000 t_g\). See the text for details.}
	\label{Figure_9}
\end{figure*}
%%%%%%%%%%%%%%%%%%%%%%%%%%%%%%%%%%%%%%%%%%%%%%%%%%%%

%%%%%%%%%%%%%%%%%%%%%%%%%%%%%%%%%%%%%%%%%%%%%%%%%%%
%%                        Figure 10
%%%%%%%%%%%%%%%%%%%%%%%%%%%%%%%%%%%%%%%%%%%%%%%%%%%
\begin{figure*}
	\begin{center}
        \includegraphics[width=0.80\textwidth]{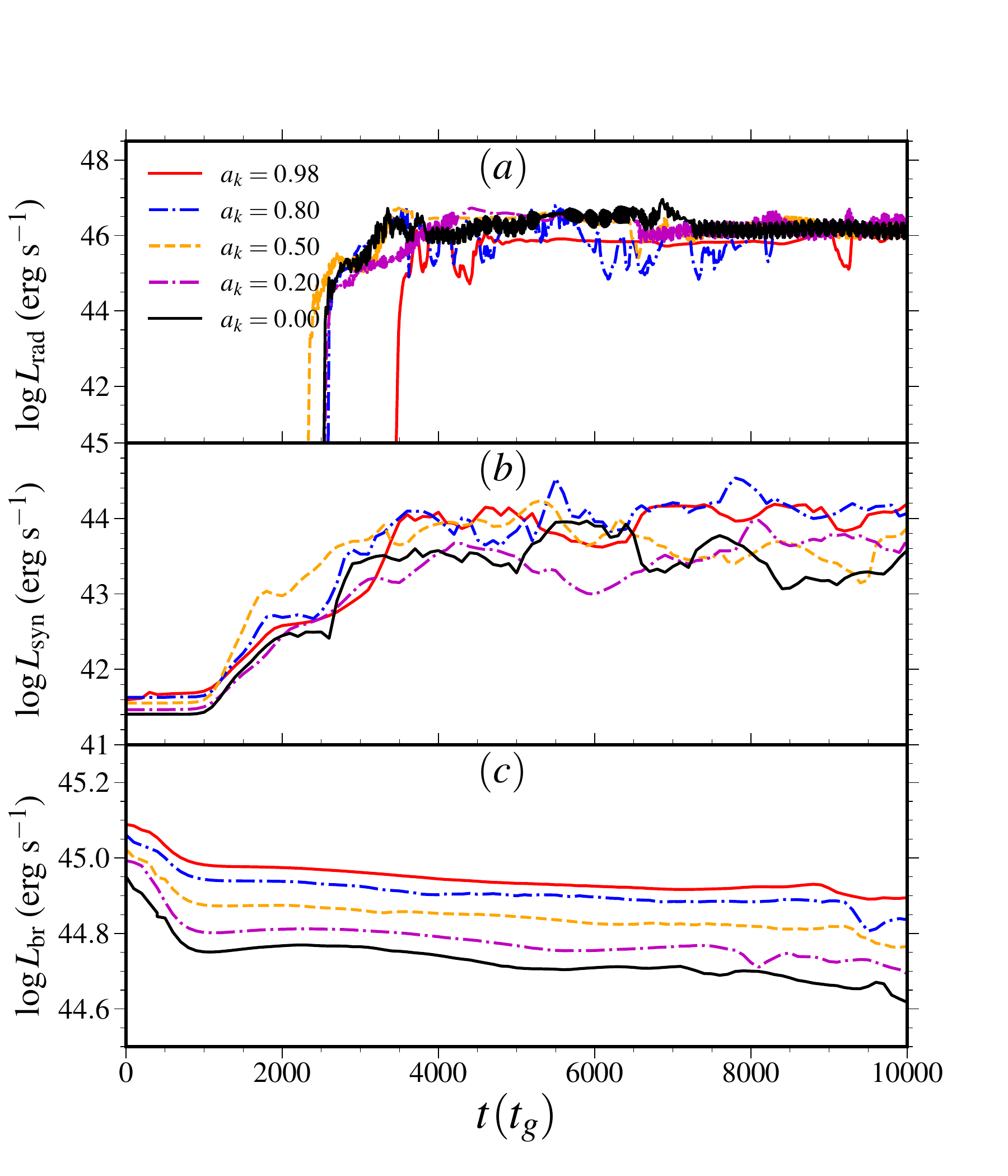} 
	\end{center}
	\caption{Variation of (a) radiative luminosity ($L_{\rm rad}$), (b) synchrotron ($L_{\rm syn}$), and (c) bremsstrahlung ($L_{\rm br}$) with time for different black hole spin $a_k$ = 0.0, 0.20, 0.50, 0.80, 0.98. See the text for details.}
	\label{Figure_10}
\end{figure*}
%%%%%%%%%%%%%%%%%%%%%%%%%%%%%%%%%%%%%%%%%%%%%%%%%%%%

%%%%%%%%%%%%%%%%%%%%%%%%%%%%%%%%%%%%%%%%%%%%%%%%%%%
%%                        Figure 11
%%%%%%%%%%%%%%%%%%%%%%%%%%%%%%%%%%%%%%%%%%%%%%%%%%%
\begin{figure*}
	\begin{center}
        \includegraphics[width=0.80\textwidth]{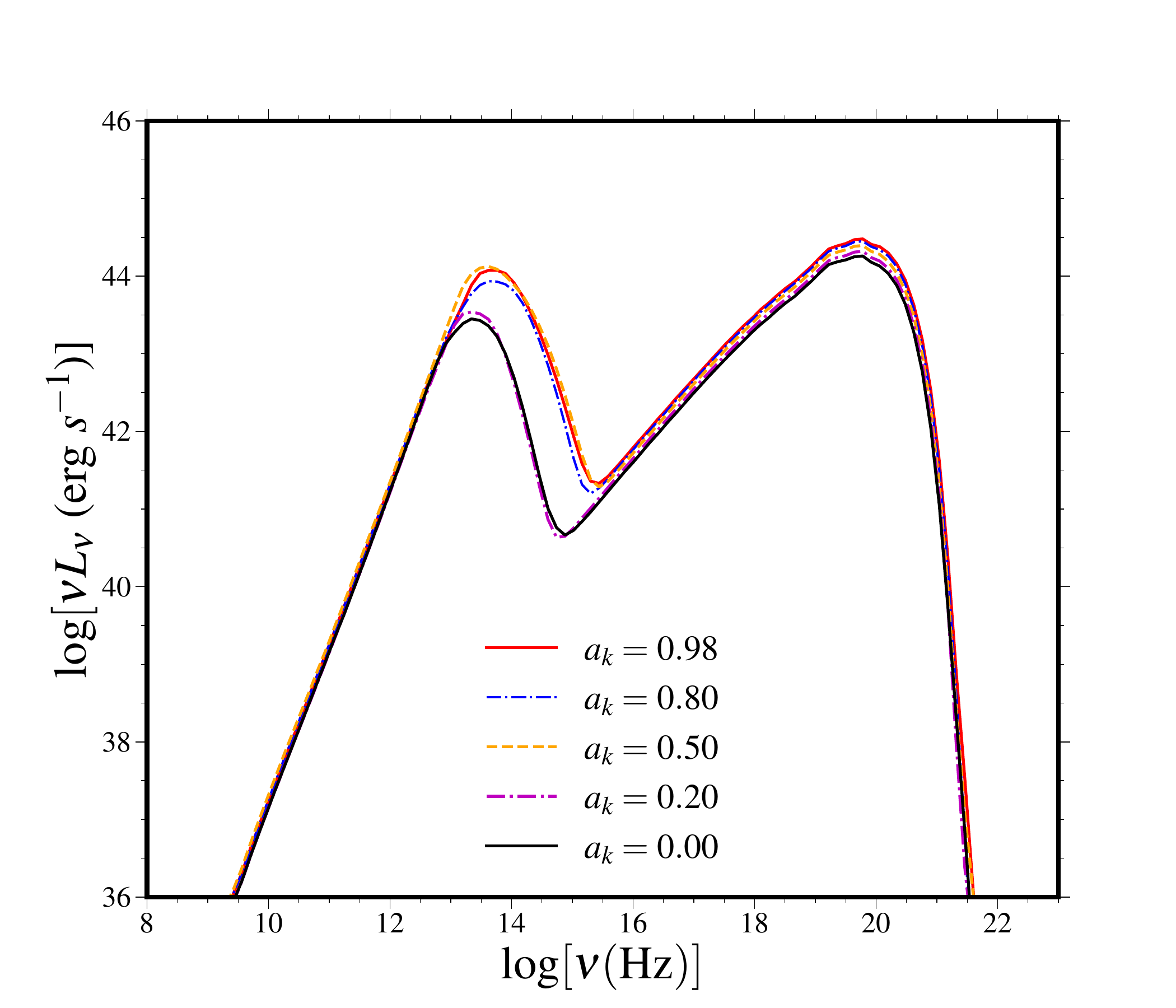} 
	\end{center}
	\caption{Comparison of spectral energy distribution (SED) by varying black hole spin ($a_k$), with the time average between \(t = 5000 t_g\) to \(6000 t_g\). See the text for details.}
	\label{Figure_11}
\end{figure*}
%%%%%%%%%%%%%%%%%%%%%%%%%%%%%%%%%%%%%%%%%%%%%%%%%%%%

%%%%%%%%%%%%%%%%%%%%%%%%%%%%%%%%%%%%%%%%%%%%%%%%%%
\section{Simulation results}
\label{results}
%%%%%%%%%%%%%%%%%%%%%%%%%%%%%%%%%%%%%%%%%%%%%%%%%%

The initial equilibrium torus surrounding the black hole is established by fixing the flow angular momentum ($\lambda$) and the spin ($a_k$) of the black hole. We present the density distribution in a 2D slice of the ($r-z$) plane for the initial equilibrium torus, as illustrated in Figure \ref{Figure_1}a. The corresponding volume rendering of density in 3D space is shown in Figure \ref{Figure_1}b. The gray lines represent the magnetic field lines, which are embedded within the torus \citep{Aktar-etal24a, Aktar-etal24b, Aktar-etal25}. We compare the temporal evolution of the mass accretion rate ($\dot{M}_{\rm acc}$) in Eddington units ($\dot{M}_{\rm Edd}$) for 3D models, as depicted in Figure \ref{Figure_2}a. This comparison includes different black hole spin values, ranging from non-spinning to highly spinning, specifically $a_k = 0.0, 0.20, 0.50, 0.80,$ and $0.98$. The definition of the Eddington mass accretion rate is given in Table~\ref{Table-1}. We find that, for all models, the mass accretion rate saturates at $\dot{M}_{\rm acc} \gtrsim 10^{-2} \dot{M}_{\rm Edd}$. At such accretion rates, the flow is expected to be radiatively efficient, with radiation pressure and radiative cooling significantly influencing the thermodynamic state and dynamical structure of the accretion flow. We also observe that the average thermal disk height lies in the range $0.15 - 0.8$ in the quasi-stationary state for all the spin model. It also indicate that the flow is radiatively efficient. All model parameters are summarized in Table \ref{Table-2}. To investigate the magnetic state of the flow, we examine the temporal evolution of normalized magnetic flux (\(\dot{\phi}_{\rm acc}\)) accumulated at the event horizon, as shown in Figure \ref{Figure_2}b. We observe that all the models belong to the MAD states, regardless of the black hole spin. The dashed black lines indicate the threshold value for entering the MAD state (\(\dot{\phi}_{\rm acc} = 50\)). We find that the accretion flow enters the MAD state at \(t \gtrsim 2500 t_g\) for models with spin values \(a_k = 0.0, 0.20, 0.50, 0.80\). However, the model with spin \(a_k = 0.98\) enters the MAD state at around \(t \gtrsim 3500 t_g\), as shown in Figure \ref{Figure_2}b. We observe the flux eruption events after the magnetic flux has saturated around a specific value for all the models \citep{Tchekhovskoy-etal11, Chatterjee-Narayan-22, Aktar-etal25}. We now analyze the MAD state by examining the spatial average of plasma beta (\(\beta_{\rm ave}\)) throughout the entire computational domain, to determine the magnetic state, following \citet{Aktar-etal25}, as depicted in Figure \ref{Figure_2}c. The dashed black lines indicate the threshold where gas pressure equals magnetic pressure (\(\beta_{\rm ave} = 1\)). We observe that the flow enters the MAD state when \(\beta_{\rm ave} \lesssim 1\). Importantly, we find that the flow enters the MAD state simultaneously with the results of our examination of the normalized magnetic flux (\(\dot{\phi}_{\rm acc}\)), as shown in Figure \ref{Figure_2}b. We present the mass accretion rate, normalized magnetic flux, and average plasma beta parameters, utilizing a moving time average to minimize unnecessary noisy variations. 

In this section, we compare the azimuthal and time averaged of 2D distribution of flow variables for all models at time averaged between \( t = 5000 t_g \) to \( 6000 t_g \), after the flow enters the MAD state, as illustrated in Figure \ref{Figure_3}. We examine the inflow equilibrium for the accretion flow of our simulation models as described in appendix \ref{appendix-A} \citep{Penna-etal-10, Narayan-etal12, Zhang-etal-24, Dihingia-etal-25}. It is found that all the models reach an inflow equilibrium state roughly below radial distance $ \lesssim 50 r_g$ within the time \( t = (5000 - 6000 ) t_g\). The first row of Figure \ref{Figure_3} shows the distribution of gas density (\(\rho\)) for various black hole spins: \(a_k = 0.0, 0.02, 0.50, 0.80, 0.98\). We find that the temperature near the jet region is extremely high (\(T \gtrsim 10^{13}\) K), as depicted in the second row of Figure \ref{Figure_3}. In fact, the temperature in the jet region is \(10^3\) times higher than in the disk region for all cases. The third row of Figure \ref{Figure_3} displays the corresponding magnetization parameter (\(\sigma_{\rm M}\)). We observe that the magnetization is very high (\(\sigma_{\rm M} \gtrsim 10\)) near the horizon and in the jet region \citep{Aktar-etal24b, Aktar-etal25}. We also investigate the distribution of the Lorentz factor ($\Gamma$) in the fourth row of Figure \ref{Figure_3}. Our findings indicate that the Lorentz factor in the jet spine is $\Gamma \gtrsim 2.5$ for all spin models. In contrast, the Lorentz factor for the jet sheath is around $\Gamma \gtrsim 1.5$, while in the disk region it is $\Gamma \sim 1.0$ \citep{Fromm-etal-22}. Moreover, we examine the distribution of radiation energy density (\(E_{\rm rad}\)) as shown in Figure \ref{Figure_4}. For comparison, we present the radiation energy density in the (\(r-z\)) plane using cylindrical coordinates in the upper panel and in the (\(x-y\)) equatorial plane using Cartesian coordinates in the lower panel. We observe that the radiation energy density is significantly higher (\(\gtrsim 10^{10}\)) near the horizon and in the jet region compared to the torus region, as indicated in the upper panel of the (\(r-z\)) plane. Similarly, in the lower panel depicting the (\(x-y\)) plane, we also find that the radiation energy density is considerably higher (\(\gtrsim 10^{10}\)) near the horizon (\(r \lesssim 20 r_g\)). In general, the MAD state occurs through intermittent magnetic interchange events, leading to episodes of rapid inflow (see Figure \ref{Figure_2}a) \citep{Narayan-etal-03, Igumenshchev-08, McKinney-etal-12}. Near the black hole, the timescale for this inflow can become comparable to the timescale for photon escape \citep{Begelman-79, Ohsunga-etal-05, Sadowski-etal-13}, resulting in partial photon trapping and an increased radiation energy density in the inner region. Notably, these trends are consistent across all spin models \citep{Aktar-etal24b}. Overall, it appears that the spin of the black hole has a minimal effect on the overall dynamics of the accretion process within the flow \citep{Jiang-etal-23, Aktar-etal24b}. To gain a more rigorous understanding, we analyze and compare the average radial profiles of several quantities: density ($\rho$), temperature ($T$), magnetic field strength ($|B| \equiv \sqrt{B^2}$), plasma beta parameter ($\beta$), radiation energy density ($E_{\rm rad}$), and thermal scale height ($h/r$), as shown in Figure \ref{Figure_5}. We compute the vertically, azimuthally, and time-averaged values of each quantity $Q \in (\rho, T, |B|, \beta, h/r)$, weighted by density, using the following equation \citep{Narayan-etal-22}:
\begin{align}
\label{average_eqn}
<Q>(r) = \frac{\int \int \int Q \rho ~dz~ d\phi~ dt}{\int \int \int \rho~ dz~d\phi~dt}.
\end{align}
For all spin models, the time averaging is conducted between $t = (5000-6000)t_g$, where $a_k = 0.98, 0.80, 0.50, 0.20, 0.0$. Here, we calculate and compare all quantities below the inflow equilibrium radius, specifically for $r \lesssim 50 r_g$. We observe similar trends in the density profiles across all spin models. Notably, the temperature in the horizon region is nearly more than ten times greater than that in the outer region ($r = 50 r_g$). Similarly, the magnetic field strength in the horizon region is over ten times higher compared to the outer disk in MAD models. The radial variation of plasma beta parameters indicates that, in the inner region of the disk for MAD models, magnetic pressure predominates or is comparable to gas pressure, as demonstrated in Figure \ref{Figure_2}c. Moreover, the radiation energy density ($E_{\rm rad}$) is significantly higher (greater than $10^6$) near to the horizon region of the disk compared to the outer part, as indicated in Figure \ref{Figure_4}. The thermal scale height varies between $0.15$ and $0.25$ for all spin models, suggesting that the flow is radiatively efficient \citep{Zhang-etal-25}. We do not observe any correlation between the radial profiles of the various flow variables and the black hole spin.

Now we present the volume rendering of density (\(\rho\)) at time \(t = 5000 t_g\), when the flow enters into the MAD state, as illustrated in Figure \ref{Figure_6}a. In this scenario, we fix the black hole's spin at \(a_k = 0.98\). The gray lines in the figure represent the magnetic field lines. To examine the magnetic field structure in the jet region, we focus on the magnetic field lines located within \(r < 20 r_g\) \citep{Aktar-etal25}. We observe that the magnetic field lines twist along the rotation axis of the black hole in the jet region. This twisting of magnetic field is a critical feature of the Blandford–Znajek mechanism, which is responsible for extracting rotational energy from spinning black holes and launching relativistic jets \citep{Blandford-Znajek77}. Additionally, we display the volume rendering of radiation energy density in Figure \ref{Figure_6}b. We observe that the radiation energy density is significantly high along the rotation axis in the jet for the MAD state compared to the disk region, as also depicted in Figure \ref{Figure_4}.

We further investigate the relationship between the saturated MAD parameter during the MAD state and black hole spin, as illustrated in Figure \ref{Figure_7}, based on the work of \citet{Narayan-etal-22, Zhang-etal-24}. We calculate the time-averaged saturated MAD parameter over the interval \( t = 5000-10000 t_g \) for all models. The simulation data is represented by black dots, with \(1\sigma\) temporal error bars included. We fit the simulation MAD parameters using a third-order polynomial, as shown by the dashed red lines in Figure \ref{Figure_7}. The relationship between the black hole spin parameter \( a_k \) and the MAD parameters is roughly described by third-order polynomials, similar to findings from other GRMHD codes \citep{Narayan-etal-22, Zhang-etal-24}. However, additional simulation data points, ranging from retrograde to prograde flow, are needed to confirm the results.

%%%%%%%%%%%%%%%%%%%%%%%%%%%%%%%%%%%%%%%%%%%%%%%%%%
\subsection{Two temperature model for 3D flow}
\label{two-temp}
%%%%%%%%%%%%%%%%%%%%%%%%%%%%%%%%%%%%%%%%%%%%%%%%%%

In this work, we consider a single-temperature model, where we assume that the electron temperature ($T_e$) is equal to the ion temperature ($T_i$) for our 3D-Rad-RMHD simulation model in PLUTO. Additionally, we only account for bremsstrahlung emission between ions and electrons as the radiation source. However, synchrotron and Compton emissions cannot be overlooked when evaluating luminosity and spectrum. Motivated by this, we adopt a self-consistent two-temperature model to evaluate both synchrotron and bremsstrahlung luminosity and their corresponding spectra, following the same approach as outlined by \citet{Okuda-etal23} and \citet{Aktar-etal24b}.

To estimate the synchrotron and bremsstrahlung emissions, we solve the radiation energy equilibrium equation. In this equation, the energy transfer rate from ions to electrons due to Coulomb interactions, denoted as $(Q^{ie})$, is equal to the sum of the synchrotron cooling rate ($Q_{\rm syn}$) and the bremsstrahlung cooling rate ($Q^{ie}_{\rm br}$) in electrons. Thus, we have:
\begin{align}\label{tot_cooling_rate}
Q^{ie} = Q_{\rm syn} + Q_{\rm br}.
\end{align}
Here, the Coulomb interaction rate $(Q^{ie})$ is given by \citet{Stepney-Guilbert-83} as follows:
\begin{align} \label{Syn_rate}
\begin{split}
Q^{ie} & = 5.61 \times 10^{-32} \frac{n_e n_i (T_i -T_e)}{K_2(1/\Theta_e) K_1(1/\Theta_i)} \\
       & \times \left[ \frac{2(\Theta_e+\Theta_i)^2+1}{(\Theta_e+\Theta_i)} K_1 \left(\frac{\Theta_e+\Theta_i}{\Theta_e \Theta_i}\right) + 2K_0 \left(\frac{\Theta_e+\Theta_i}{\Theta_e \Theta_i}\right) \right] \\
       & \text{erg~cm}^{-3}~\text{s}^{-1}.
\end{split}
\end{align}
In this equation, $n_e$ and $n_i$ represent the number densities of electrons and ions, respectively. The terms $K_0$, $K_1$, and $K_2$ are the modified Bessel functions, while $\Theta_e = \frac{k_{\rm B}T_e}{m_e c^2}$ and $\Theta_i = \frac{k_{\rm B}T_i}{m_p c^2}$ are the dimensionless electron and ion temperatures, respectively. Here, $k_{\rm B}$ is the Boltzmann constant, $m_e$ is the electron mass, $m_p$ is the proton mass, and $c$ is the speed of light.

To calculate the bremsstrahlung cooling rate, we adopt the prescriptions from \citet{Stepney-Guilbert-83, Esin-etal96}. The free-free bremsstrahlung cooling rate for an ionized plasma, which consists of electrons and ions, is expressed as follows:
\begin{align} \label{Brem_rate}
Q_{\rm br} = Q^{ie}_{\rm br} + Q^{ee}_{\rm br},
\end{align}
where 
\begin{align}
Q^{ie}_{\rm br} = 1.48 \times 10^{-22} n_e^2 F^{ie} (\Theta_e)~~~~ {\rm erg~ cm^{-3}~ s^{-1}}.
\end{align}
In this equation, \( F^{ie}(\Theta_e) \) is defined as:
\begin{align}
F^{ie}(\Theta_e)= 
\begin{cases}
    4 \sqrt{\frac{2 \Theta_e}{\pi^3}} (1 + 1.781\Theta_e^{1.34}), & \text{if } \Theta_e < 1, \\
    \frac{9 \Theta_e}{2 \pi}[\ln(1.12 \Theta_e + 0.48) + 1.5],  & \text{if } \Theta_e > 1.
\end{cases}
\end{align}

We also have the expression for \( Q^{ee}_{\rm br} \):
\begin{align}
\begin{split}
Q^{ee}_{\rm br} & =
\begin{cases}
    2.56 \times 10^{22} n_e^2 \Theta_e^{1.5}(1 + 1.10\Theta_e + \Theta_e^2 - 1.25\Theta_e^{2.5}), & \text{if } \Theta_e < 1, \\
    3.40 \times 10^{-22} n_e^2 \Theta_e [\ln(1.123 \Theta_e) + 1.28],  & \text{if } \Theta_e > 1.
\end{cases}\\
& {\rm erg~cm^{-3}~s^{-1}}.
\end{split}
\end{align}

In the presence of a magnetic field, the hot electrons emit radiation through a thermal synchrotron process. The synchrotron cooling rate can be expressed as follows \citep{Narayan-Yi95, Esin-etal96}:
\begin{align}
  \begin{aligned}
    Q_{\rm syn} = & \frac{2\pi k_{\rm B} T_i \nu_c^3}{3Hc^2} + 6.76 \times 10^{-28} \frac{n_e}{K_2\left(\frac{1}{\Theta_e}\right) a_1^{1/6}} \\
                 & \times \left[ \frac{1}{a_4^{11/2}} \Gamma\left(\frac{11}{2}, a_4\nu_c^{1/3}\right) + \frac{a_2}{a_4^{19/4}} \Gamma\left(\frac{19}{4}, a_4\nu_c^{1/3}\right) \right. \\
                 & + \left. \frac{a_3}{a_4^4} \left(a_4^3 \nu_c + 3a_4^2 \nu_c^{2/3} + 6a_4 \nu_c^{1/3} + 6\right) \exp(-a_4\nu_c^{1/3}) \right] \\
                 & \text{erg~cm}^{-3}~\text{s}^{-1} ,
\end{aligned}
\end{align}
where \( H \) is the scale height of the disk, and the coefficients \( a_{1-4} \) are defined as:
\begin{align}
a_1 = \frac{2}{3\nu_0 \Theta_e^2}, \quad a_2 = \frac{0.4}{a_1^{1/4}}, \quad a_3 = \frac{0.5316}{a_1^{1/2}}, \quad a_4 = 1.8899 a_1^{1/3}.
\end{align}
In these equations, \( \nu_0 = \frac{eB}{2 \pi m_e c} \) and \( \nu_c = \frac{3}{2} \nu_0 \Theta_e^2 x_{\rm M} \) are the characteristic synchrotron frequencies, where \( e \) is the electron charge, and \( B \) is the strength of the magnetic field. The value of \( x_{\rm M} \) can be obtained from the following equation:
\begin{align}
  \begin{aligned}
\exp{(1.8899 x_{\rm M}^{1/3})} = & 2.49 \times 10^{-10} \frac{4 \pi n_e r}{B} \frac{1}{\Theta_e^3 K_2 \left(\frac{1}{\Theta_e}\right)} \\
                               & \times \left( \frac{1}{x_{\rm M}^{7/6}} + \frac{0.40}{x_{\rm M}^{17/12}} + \frac{0.5316}{x_{\rm M}^{5/3}} \right).
  \end{aligned}
\end{align}
Here, the Gamma function is defined as \( \Gamma(a,x) = \int_x^\infty t^{a-1} e^{-t}~dt \).

To incorporate the two-temperature model based on single-temperature simulation results, we employ the following approach of \citet{Okuda-etal23}. We consider:
\begin{align}\label{tot_temp}
T_e + T_i = 2 T,
\end{align}
where we assume $P_{\rm gas} = nk_B T = n_i k_B T_i + n_e k_B T_e$ with \( n = n_e + n_i \) and \( n_e = n_i \), with \( n \) and \( T \) representing the number density and temperature of the single-temperature model. We numerically solve equations (\ref{tot_cooling_rate}) and (\ref{tot_temp}) to evaluate the electron temperature \( (T_e) \) and ion temperature \( (T_i) \) for our model.

In this section, we compare the azimuthal and time averaged of the distribution of electron temperature ($T_e$) as shown in Figure \ref{Figure_8}. We plot the electron temperature distribution for different black hole spins: $a_k = 0.0, 0.20, 0.50, 0.80, 0.98$, for the time average between $t = 5000 t_g$ to $6000t_g$, which is depicted in the upper panel of Figure \ref{Figure_8}. It is observed that the electron temperature is significantly higher in the jet region and near the horizon, reaching about $T_e \sim 10^{11}$ K, compared to the disk and torus regions, where $T_e \sim 10^{9}$ K across all spin models. Additionally, we analyze the ratio of electron to ion temperature ($T_e/T_i$) in the lower panel of Figure \ref{Figure_8}. The ratio of electron temperature to ion temperature, \(T_e/T_i\), generally indicates how thermal energy is distributed between electrons and ions. We present the distribution of \(T_e/T_i\) for various spin values \(a_k = 0.0, 0.20, 0.50, 0.80, 0.98\) for the time average between $t = 5000 t_g$ to $6000t_g$. It is observed that the \(T_e/T_i\) ratio is approximately \(0.01\) in the jet region for all spin cases. In contrast, the temperature ratio increases in other regions. In the presence of Coulomb interactions, the electron-ion collision rate is decreased in low-density regions, such as the jet region \citep{Dihingia-etal23}. To gain a clearer understanding, we further analyze the radial variation of electron temperature \((T_e)\) and the temperature ratio \((T_e/T_i)\), as shown in Figures \ref{Figure_9}a and \ref{Figure_9}b. These figures display the vertically, azimuthally, and density-weighted electron temperature, as well as the temperature ratio, averaged over the time interval \(t = 5000 - 6000 t_g\) within the inflow equilibrium radius. Our results indicate that the electron temperature increases as we approach the horizon from the outer region, as illustrated in Figure \ref{Figure_9}a. In contrast, Figure \ref{Figure_9}b reveals that the temperature ratio decreases towards the horizon from the outer region across all models. In the high-density region of the outer disk, where the region is optically thick, it is indicated that $T_e/T_i \sim 1$. Conversely, in the low density horizon region, which is optically thin, $T_e/T_i < 0.1$.

\subsubsection{Calculation of luminosity and Spectrum}
\label{spectral_ana}
%============================================================================

In this section, we calculate the luminosity and spectral energy distribution (SED) for our model. In order to generate spectrum, we follow \citet{ Manmoto-etal-97} prescription to calculate radiative cooling processes. Here, we consider three radiative cooling processes, namely bremsstrahlung emission, synchrotron emission, and Comptonization of soft photons. Assuming a locally plane-parallel gas configuration at each radius ($r$) for the accretion flow, we calculate the radiation energy flux $F_\nu$ following \citet{Manmoto-etal-97} as
\begin{align} \label{spec_flux}
F_\nu = \frac{2 \pi}{\sqrt{3}} B_\nu \left[1 - \exp(-2 \sqrt{3} \tau_\nu^*)\right],
\end{align}
where $\tau_\nu^* = (\sqrt{\pi}/2) \kappa(0)H$ is the optical depth for absorption of the accretion in the vertical direction. Here $\kappa(0)$ is the the absorption coefficient on the equatorial plane.  The absorption coefficient $\kappa_{\nu}$ is defined as $\kappa_\nu = \chi_{\nu}/(4 \pi B_\nu)$. Here $\chi_\nu = \chi_{\nu,\rm br}+\chi_{\nu,\rm syn}$ is the total emissivity. 

Now, the bremsstrahlung emissivity $(\chi_{\nu, {\rm br}})$ can be written as \citep{Manmoto-etal-97}
\begin{align} \label{spec_brem_emisi}
\chi_{\nu, {\rm br}} = Q_{\rm br} \overline{G} \exp(\frac{h\nu}{k_{\rm B}T_e}),
\end{align}
where $Q_{\rm br}$ is the bremsstrahlung cooling rate as depicted in equation (\ref{Brem_rate}). Here $\overline{G}$ is the Gaunt factor, which is given by \citep{Rybicki-Lightman-79}
\begin{align}
   \overline{G} = 
\begin{cases}
    \frac{h}{k_{\rm B}T_e} \left(\frac{3}{\pi} \frac{k_{\rm B} T_e}{h \nu}\right)^{1/2}, & \text{if } \frac{k_{\rm B}T_e}{h \nu} < 1\\
    \frac{h}{k_{\rm B}T_e} \frac{\sqrt{3}}{\pi} \ln{\left(\frac{4}{\xi}\frac{k_{\rm B} T_e}{h \nu}\right)}, & \text{if } \frac{k_{\rm B}T_e}{h \nu} > 1.
\end{cases}
\end{align}

We also consider the synchrotron emissivity $\chi_{\nu,\rm syn}$ by a relativistic Maxwellian distribution of electrons following \citet{Narayan-Yi95} as
\begin{align} \label{spec_sync_emisi}
\chi_{\nu,\rm syn} = 4.43 \times 10^{-30} \frac{4 \pi n_e \nu}{K_2 (1/\Theta_e)} I(x),
\end{align}
where $x = \frac{4 \pi m_e c \nu}{3 e B \Theta_e^2}$ and 
\begin{align}
I(x) = \frac{4.0505}{x^{1/6}} \left(1+\frac{0.40}{x^{1/4}}+ \frac{0.5316}{x^{1/2}}\right) \exp(-1.8899x^{1/3}).
\end{align}

Further, we consider the effect of Compton scattering process in our model. In this work, we adopt a more accurate energy enhancement factor due to the Compton scattering following \citet{Coppi-Blandford-90}. 

Now, we evaluate the bremsstrahlung and synchrotron luminosity based on the frequency-dependent radiation flux following the equation (\ref{spec_flux}). The frequency-dependent bremsstrahlung ($L_{\rm br}$) and synchrotron ($L_{\rm syn}$) luminosity can be obtained throughout the disk surface as  
\begin{align}\label{brem_lumi}
L_{\rm br} = \int_S \int_\nu {F_{\nu},}_{\rm br}~d\nu~dS,
\end{align}
and
\begin{align}\label{sync_lumi}
L_{\rm syn} = \int_S \int_\nu {F_{\nu},}_{\rm syn}~d\nu~ dS,
\end{align}
where the surface integration is carried out all over the disk surface. $F_{\nu,\rm syn}$ and $F_{\nu,\rm br}$ can be calculated using synchrotron and bremsstrahlung emissivities using equation (\ref{spec_brem_emisi}) and (\ref{spec_sync_emisi}), respectively. On the other hand, the radiative luminosity can be calculated at the outer $z$-boundary and the outer $r$-boundary surfaces as follows 
\begin{align}\label{tot_lumi}
L_{\rm rad} = \int_S \bm{F}_{\rm rad}~ \bm{dS},
\end{align}
where $\bm{F}_{\rm rad}$ is the radiation energy ﬂux at the boundary surfaces. In general, the radiative luminosity (\( L_{\rm rad} \)) is roughly equivalent to the bremsstrahlung luminosity (\( L_{\rm br} \)) when free-free emission is predominantly occurring within the disk \citep{Okuda-etal23, Aktar-etal24b}. However, this relationship may differ if the free-free emission is also significant outside the disk.

In this section, we begin by calculating the total radiative luminosity (\(L_{\rm rad}\)), which can be determined from the radiative flux across all computational surfaces, as shown in equation (\ref{tot_lumi}). Total radiative luminosity is a crucial characteristic of radiation-dominated accretion flows. Examining \(L_{\rm rad}\) enables us to ensure that our calculations are accurate, provided that the relevant opacity is properly specified in the Rad-RMHD module of the PLUTO code. Furthermore, radiation affects the flow dynamics through interactions between gas and radiation, particularly when the radiation luminosity is between sub-Eddington and Eddington luminosity levels (\(L_{\rm Edd}\)). We find that the radiation luminosity reaches the Eddington level (\(L_{\rm rad} \gtrsim 10^{46}\) erg s\(^{-1}\)) across all spin cases, as illustrated in Figure \ref{Figure_10}a. Now, we calculate and compare the synchrotron disk luminosity (\(L_{\rm syn}\)) and bremsstrahlung disk luminosity (\(L_{\rm br}\)) using equations (\ref{sync_lumi}) and (\ref{brem_lumi}), respectively. To determine these luminosities, we analyze the frequency-dependent radiative flux by varying the frequency from \(10^8\) to \(10^{24}\) Hz, as indicated in equations (\ref{sync_lumi}) and (\ref{brem_lumi}). The variation of synchrotron luminosity (\(L_{\rm syn}\)) is depicted in Figure \ref{Figure_10}b for all spin models. We observe that synchrotron luminosity increases by approximately two orders of magnitude when the flow enters the MAD state, after which it saturates. Notably, we do not observe significant differences in synchrotron luminosity across varying black hole spins, as shown in Figure \ref{Figure_10}b. This lack of variation occurs because black hole spin does not directly affect flow dynamics or the synchrotron radiation mechanism. Additionally, we present the variation of bremsstrahlung luminosity (\(L_{\rm br}\)) in Figure \ref{Figure_10}c. The bremsstrahlung luminosity shows minimal variability across all spin cases. This is primarily because bremsstrahlung luminosity predominantly originates from the highly dense torus region, where luminosity is proportional to the square of the density (\(\propto \rho^2\)), and density fluctuations are minimal in this area \citep{Aktar-etal24b}. Furthermore, bremsstrahlung luminosity is generally higher than synchrotron luminosity, with \(L_{\rm br} \gtrsim 10 L_{\rm syn}\), as demonstrated in our 2D simulations \citep{Aktar-etal24b}. We find that bremsstrahlung luminosity slightly increases with higher black hole spins, likely due to a marginal increase in torus size associated with these higher spins \citep{Jiang-etal-23, Aktar-etal24a, Aktar-etal24b}. Interestingly, we observe that the total radiative luminosity is significantly higher than both the bremsstrahlung and synchrotron luminosities, with the relationship \(L_{\rm rad} >> L_{\rm br} > L_{\rm syn}\) as illustrated in Figures \ref{Figure_8}a, b, and c. This suggests that the radiation mechanism has a considerable impact on the dynamics of the accretion flow. Additionally, we observe that the radiative luminosities (\(L_{\rm rad} \sim 10^{46}\) erg s\(^{-1}\)) are one order of magnitude higher than the bremsstrahlung luminosities (\(L_{\rm br} \sim (4 - 8) \times 10^{44}\) erg s\(^{-1}\)) at the final stage of the simulation for all spin values, as illustrated in Figures \ref{Figure_10}a and \ref{Figure_10}c. This difference occurs because there is a dominant free-free emission region located outside the disk. If the bremsstrahlung emission were to dominate only within the disk, \(L_{\rm rad}\) would be consistent with \(L_{\rm br}\). The significant bremsstrahlung emission outside the disk region is further supported by the high radiation energy density and electron temperature found in the jet area, as shown in Figures \ref{Figure_4}, \ref{Figure_6}b, and \ref{Figure_8}.

We further examine the spectral energy spectrum using equations (\ref{spec_flux}) - (\ref{spec_sync_emisi}) and the Compton scattering expression as outlined by \citet{Coppi-Blandford-90}. Our analysis compares the energy spectrum with black hole spin for time average between \( t = 5000t_g \) to \(6000t_g \), as shown in Figure \ref{Figure_11}. We observe two prominent peaks: one from synchrotron radiation and the other from bremsstrahlung. These peaks are present across all spin models with values of \( a_k = 0.0, 0.20, 0.50, 0.80, 0.98 \). The synchrotron peak occurs around \( \sim 10^{13} - 10^{14} \) Hz, while the bremsstrahlung peak is found at \( \sim 10^{19} - 10^{20} \) Hz. The overall shape of the spectral energy distribution (SED) is quite similar across all spin models. However, we note that the synchrotron peaks tend to shift to lower frequencies for lower spin values (\( a_k = 0.0, 0.20 \)) compared to higher spin values (\( a_k = 0.50, 0.80, 0.98 \)), as depicted in Figure \ref{Figure_11}. Additionally, the bremsstrahlung peaks maintain a similar frequency range for all spin values. We also find that the synchrotron peak is as bright as the bremsstrahlung peak for higher spin values (\( a_k \gtrsim 0.50 \)). In contrast, for lower spin values (\( a_k = 0.0, 0.20 \)), the bremsstrahlung peak is over 10 times brighter than the synchrotron peak, as illustrated in Figure \ref{Figure_11}. Finally, we do not observe any Compton enhancement in our 3D model, consistent with our previous findings in the 2D model \citep{Aktar-etal24b}.

It is important to note that the general consensus is that the radiative properties of black holes depend on their spin. Pioneering work by \citet{Novikov-Thorne-1973} demonstrated that the radiative efficiency of accretion is determined by the binding energy of matter at the innermost stable circular orbit (ISCO). They observed a monotonic increase in radiative efficiency with black hole spin, based on a fully relativistic model of geometrically thin, optically thick accretion disks around Kerr black holes. Subsequent GRMHD simulations also indicated a spin-dependent outflowing energy efficiency around black holes \citep{Tchekhovskoy-etal11, McKinney-etal-12, Tchekhovskoy-McKinney-12, Narayan-etal-22, Zhang-etal-24}. However, in our simulation model, we do not find clear evidence of a dependence of black hole spin on radiative properties, such as total radiative luminosities and spectral energy distribution, as illustrated in Figure \ref{Figure_10}a and Figure \ref{Figure_11}. This lack of correlation is likely due to the imprecise modeling of a spin-independent inner boundary (see subsection \ref{boundary_cond}), which is a limitation of the effective-Kerr potential.

%%%%%%%%%%%%%%%%%%%%%%%%%%%%%%%%%%%%%%%%%%%%%%%%%%
\section{Summary and Discussions}
\label{conclusion}
%%%%%%%%%%%%%%%%%%%%%%%%%%%%%%%%%%%%%%%%%%%%%%%%%%

In this paper, we examine the effect of black hole spin on the radiative, relativistic, and highly magnetized (Rad-RMHD) accretion flow around AGNs. To investigate the accretion properties, we utilize an equilibrium magnetized torus model, following our previous works \citep{Aktar-etal24b, Aktar-etal25}. We apply similar floor conditions as in our previous study \citep{Aktar-etal25}. For the simulation of the accretion flow, we adopt the radiation, relativistic magnetohydrodynamics (MHD) module in the PLUTO code \citep{Fuksman-Mignone-19}. It is important to note that our simulation model achieves much higher spatial resolution compared GRMHD simulations, especially in three dimensions and with the radiation mechanism, given the available computing resources (see Table \ref{Table-2}).

We observe that the flow attains the MAD state for all spin values, specifically for \(a_k = 0.0, 0.20, 0.50, 0.80, 0.98\), as depicted in Figure \ref{Figure_2}b. It also indicate the flux eruption events for all spin models. Additionally, we investigate the MAD state by analyzing physically motivated spatial average of plasma beta (\(\beta_{\rm ave}\)) in Figure \ref{Figure_2}c, in a manner similar to our previous work \citep{Aktar-etal25}. We find that the MAD state is achieved when the flow reaches \(\beta_{\rm ave} \lesssim 1\), meaning that the magnetic pressure is greater than or comparable to the gas pressure. 

Furthermore, we compare the distribution of various flow variables in the 2D (\(r-z\)) plane as shown in Figure \ref{Figure_3}. We find that the magnetization parameter is very high (\(\sigma_{\rm M} \gtrsim 10\)) near the black hole horizon and in the jet region for all spin models. Consequently, we observe that the Lorentz factor \(\Gamma \gtrsim 2.5\) in the jet spine and \(\Gamma \sim 1.0\) in the disk region across all cases. We also compare the radiation energy density (\(E_{\rm rad}\)) in both the 2D (\(r-z\)) cylindrical and (\(x-y\)) Cartesian planes. Our results indicate that the radiation energy density is significantly higher ($>10^{10}$) near the horizon and also along the jet compared to the disk region for the MAD state, as illustrated in Figure \ref{Figure_4}.

Finally, we utilize a two-temperature model, based on our previous works \citep{Okuda-etal23, Aktar-etal24b}, to separately calculate the electron and ion temperatures in the flow using post-processing simulation data. Our observations show that the electron temperature is significantly high near the horizon and jet region across all spin models, as illustrated in Figure \ref{Figure_7}. In Figure \ref{Figure_8}, we compute and compare the time variation of various luminosities, including radiative, synchrotron, and bremsstrahlung luminosities. We find that there is no noticeable difference in radiative and synchrotron luminosities concerning the spin of the black hole. However, the bremsstrahlung luminosity does exhibit a slight decrease as the black hole spin decreases, as shown in Figure \ref{Figure_8}c. Additionally, we analyze the spectral energy distribution by varying the black hole spin, as depicted in Figure \ref{Figure_9}. Our findings indicate that the SED is qualitatively similar for MAD state for all spin models. 

Although our simulations indicate that black hole spin has minimal impact on the global, time-averaged dynamics of the accretion flow, it plays a crucial role in shaping both the jet power and the magnetic field structure in the near-horizon region, as demonstrated by numerous previous studies \citep{Gammie-Popham-98, Tchekhovskoy-etal11, Tchekhovskoy-McKinney-12, Narayan-etal-22, Dhang-etal-25, Lowell-etal-25}. It is important to note that we are not focusing on the relationship between jet power and black hole spin in this work; we plan to investigate this topic in the future.

In this paper, we focus on a single temperature model and consider bremsstrahlung opacity as the only radiation opacity source in the flow. However, to accurately explore luminosity variations and spectral evolution, a proper two-temperature model is essential. In the future, we plan to develop a comprehensive three-dimensional two-temperature radiation model around a black hole to investigate the flow properties and spectral energy distribution in a MAD state.

%%%%%%%%%%%%%%%%%%%%%%%%%%%%%%%%%%%%%%%%%%%%%%%%%%%%%%%%%%%%%%%%%%%%%%%

\section*{Acknowledgments}
 We sincerely thank the anonymous referee for their valuable suggestions and comments that have enhanced the quality of the manuscript. This work is supported by the National Science and Technology Council of Taiwan through grant NSTC 112-2811-M-007-038, 112-2112-M-007-040 and 113-2112-M-007-031, and by the Center for Informatics and Computation in Astronomy (CICA) at National Tsing Hua University through a grant from the Ministry of Education of Taiwan. The simulations and data analysis have been carried out on the CICA Cluster at National Tsing Hua University.
%\end{acknowledgments}

%%%%%%%%%%%%%%%%%%%%%%%%%%%%%%%%%%%%%%%%%%%%%%%%%%%%%%%%%%%%%%%%%%%%%%

\appendix

%%%%%%%%%%%%%%%%%%%%%%%%%%%%%%%%%%%%%%%%%%%%%%%%%%%%%%%%%%%%%%%%%%%%%%
\section{Inflow equilibrium conditions}
\label{appendix-A}

\begin{figure*}
	\begin{center}
        \includegraphics[width=0.80\textwidth]{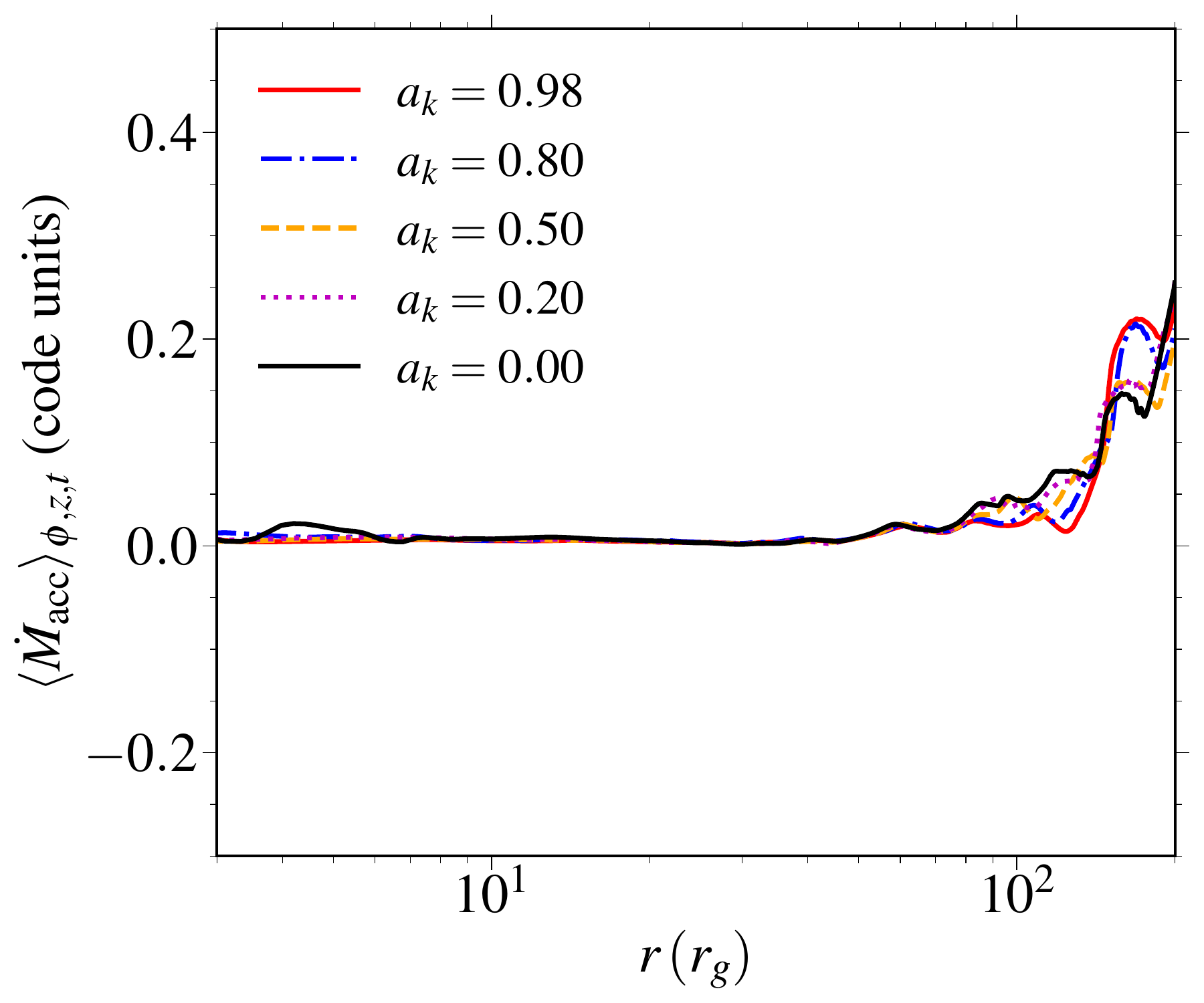} 
	\end{center}
	\caption{Radial variation of vertically, azimuthally integrated time-averaged mass accretion rate between $t = (5000-6000)t_g$ for all the spin models.} 
	\label{Figure_12}
\end{figure*}

One important consideration in simulation studies is to assess the inflow equilibrium or quasi-steady state at a given radius of the accretion flow. The inflow equilibrium radius is defined as the region within which the accretion flow has had sufficient time to settle into a quasi-steady state. In this region, matter flows inward at a roughly constant rate \citep{Penna-etal-10, Narayan-etal12, Zhang-etal-24, Dihingia-etal-25}. Formally, this radius is identified as the one within which the time-averaged mass accretion rate remains approximately independent of radius, i.e., $\dot{M}_{\rm acc} \simeq {\rm constant}$. In Figure \ref{Figure_12}, we present the variation of the azimuthally, vertically, and time-averaged mass accretion rate with radial distance for all spin models. We observe that the mass accretion rate remains nearly constant at radius $r \lesssim 50 r_g$ during the time interval from $t = (5000-6000)t_g$. The quasi-steady state is further supported by the time variation of the total radiative luminosity ($L_{\rm rad}$), as shown in Figure \ref{Figure_10}a. We find that $L_{\rm rad}$ reaches a quasi-steady state after $t \gtrsim 4000 t_g$ for all spin models. 

%Additionally, we note that the inflow equilibrium in our model is achieved much faster compared to various GRMHD simulations \citep{Penna-etal-10, Narayan-etal12, Dihingia-etal-25}, primarily because our model has a much shorter radial extent.   

%%%%%%%%%%%%%%%%%%%%%%%%%%%%%%%%%%%%%%%%%%%%%%%%%%%
%%                        Figure 12
%%%%%%%%%%%%%%%%%%%%%%%%%%%%%%%%%%%%%%%%%%%%%%%%%%%
%%%%%%%%%%%%%%%%%%%%%%%%%%%%%%%%%%%%%%%%%%%%%%%%%%%%

%%%%%%%%%%%%%%%%%%%%%%%%%%%%%%%%%%%%%%%%%%%%%%%%%%%%%%%%%%%%%%%%%%%%%%


%%%%%%%%%%%%%%%%%%%%%%%%%% References %%%%%%%%%%%%%%%%%%%%%%%%%%%%%%%%%%

\bibliography{refs}{}
\bibliographystyle{aasjournal}



\end{document}